\documentclass[fleqn,fleqn,usenatbib]{mnras}
\usepackage[T1]{fontenc}
\usepackage{ae,aecompl}
\usepackage{graphicx}
\usepackage{epsfig}
\usepackage{amsmath}
\usepackage{amssymb}
\usepackage{natbib}
\usepackage{epstopdf}
\usepackage{xurl}
\usepackage{float}

\usepackage{times}
\usepackage{amsmath}
\usepackage{amssymb}
\usepackage{amsfonts}
\usepackage{mathrsfs}
\usepackage{hyperref}
\title [Estimating ages from lithium depletion]{The Gaia-ESO Survey: Empirical estimates of stellar ages from lithium equivalent widths ({\sc eagles})}
\author[R. D. Jeffries et al.]
  {R.~D.~Jeffries$^{1}$, R. J.~Jackson$^1$, Nicholas~J.~Wright$^1$, G.~Weaver$^1$, G.~Gilmore$^2$, S.~Randich$^3$, A.~Bragaglia$^4$,\newauthor A.~J.~Korn$^5$, R.~Smiljanic$^6$, K.~Biazzo$^7$,  A.~R.~Casey$^{8,9}$, A.~Frasca$^7$, A.~Gonneau$^2$, G.~Guiglion$^{10}$, \newauthor L.~Morbidelli$^3$, L.~Prisinzano$^{11}$, G.~G.~Sacco$^3$, G.~Tautvai\v{s}ien\.{e}$^{12}$, C.~C.~Worley$^3$, S.~Zaggia$^{13}$ \\
   $^1$ Astrophysics Group, Keele University, Keele, 
      Staffordshire ST5 5BG\\
$^2$ Institute of Astronomy, University of Cambridge, Madingley Road, Cambridge CB3 0HA, United Kingdom\\
$^3$ INAF - Osservatorio Astrofisico di Arcetri, Largo E. Fermi 5, 50125, Florence, Italy\\
$^4$ INAF - Osservatorio di Astrofisica e Scienza dello Spazio di Bologna, via Gobetti 93/3, 40129, Bologna, Italy\\
$^5$ Observational Astrophysics, Division of Astronomy and Space Physics, Department of Physics and Astronomy, Uppsala University, Box 516,\\ SE-751 20 Uppsala, Sweden\\
$^6$ Nicolaus Copernicus Astronomical Center, Polish Academy of Sciences, ul. Bartycka 18, 00-716, Warsaw, Poland\\
$^7$ INAF - Osservatorio Astrofisico di Catania, via S. Sofia 78, 95123, Catania, Italy\\
$^8$ Monash Centre for Astrophysics, School of Physics \& Astronomy, Monash University, Clayton 3800, Victoria, Australia\\
$^9$ Monash Faculty of Information Technology, Monash University, Clayton 3800, Victoria, Australia\\
$^{10}$ Leibniz-Institut für Astrophysik Potsdam (AIP), An der Sternwarte 16, 14482, Potsdam, Germany\\
$^{11}$ INAF - Osservatorio Astronomico di Palermo, Piazza del Parlamento 1, 90134, Palermo, Italy\\
$^{12}$ Institute of Theoretical Physics and Astronomy, Vilnius University, Sauletekio av. 3, 10257 Vilnius, Lithuania\\
$^{13}$ INAF - Padova Observatory, Vicolo dell'Osservatorio 5, 35122 Padova, Italy
\\
}

\date{December 2022}

\pagerange{\pageref{firstpage}--\pageref{lastpage}} \pubyear{2020}

\def\LaTeX{L\kern-.36em\raise.3ex\hbox{a}\kern-.15em
    T\kern-.1667em\lower.7ex\hbox{E}\kern-.125emX}

\begin{document}
\label{firstpage}
\maketitle

\begin{abstract}
We present an empirical model of age-dependent photospheric lithium depletion, calibrated using a large, homogeneously-analysed sample of 6200 stars in 52 open clusters, with ages from 2--6000\,Myr and $-0.3<[{\rm Fe/H}]<0.2$, observed in the {\it Gaia}-ESO spectroscopic survey. The model is used to obtain age estimates and posterior age probability distributions from measurements of the Li\,{\sc i}~6708\AA\ equivalent width for individual (pre) main sequence stars with $3000 < T_{\rm eff}/{\rm K} <6500$, a domain where age determination from the HR diagram is either insensitive or highly model-dependent. 
In the best cases, precisions of 0.1 dex in log age are achievable; even higher precision can be obtained for coeval groups and associations where the individual age probabilities of their members can be combined. The method is validated on a sample of exoplanet-hosting young stars, finding agreement with claimed young ages for some, but not others. We obtain better than 10 per cent precision in age, and excellent agreement with published ages, for seven well-studied young moving groups. The derived ages for young clusters ($<1$ Gyr) in our sample are also in good agreement with their training ages, and consistent with several published, model-insensitive lithium depletion boundary ages. For older clusters there remain systematic age errors that could be as large as a factor of two. There is no evidence to link these errors to any strong systematic metallicity dependence of (pre) main sequence lithium depletion, at least in the range $-0.29 < {\rm [Fe/H]} < 0.18$. Our methods and model are provided as software -- "Empirical AGes from Lithium Equivalent widthS" ({\sc eagles}).

\end{abstract}


\section{Introduction}
\label{1}

Low-mass stars ($\leq 1\,M_{\odot}$) form the bulk of the Galactic population. Their long lifetimes make them witnesses and tracers of the assembly of the Galaxy, and the dynamical and chemical evolution of its various component structures. Knowing the ages of stars is an essential part of these investigations and also critical for exploring the formation and development of their exoplanetary systems. However, stellar age is not a directly observable parameter; age estimations are made in various ways and each technique has its own advantages, disadvantages and range of applicability \citep[e.g,][]{Soderblom2010a}. 

Most methods rely either on comparing the predictions of stellar evolutionary models with age-dependent observables, where it is assumed the relevant stellar physics is understood well-enough to yield reliable ages; or using secondary, empirical relationships describing how phenomena like rotation, stellar activity or chemical abundances change with age, but with an incomplete understanding of the stellar or Galactic physics driving the changes. In both cases, star clusters play a key role.
Firstly, they offer coeval stellar samples with similar initial composition but a range of masses, that can be used to test and improve stellar models. Secondly, they are the principal calibration samples for age-sensitive empirical diagnostics. 

For low-mass main sequence stars, stellar evolution models are of limited use in estimating age. Observables like effective temperature ($T_{\rm eff}$) and luminosity ($L$) change slowly and asteroseismological observations that give insight into the progress of core fusion are not available for most stars \citep{Epstein2014a}. Better opportunities are presented by low-mass pre main sequence (PMS) stars, where more rapid evolution is predicted; but even in these cases there are significant uncertainties in the model-dependent absolute ages \citep{Hillenbrand2009a, Bell2013a,Soderblom2014a}. These problems have led to extensive exploration of empirical age indicators such as rotation \citep[``gyrochronology",][]{Barnes2003a}, magnetic activity manifested as chromospheric and coronal emission \citep[][and references therein]{Mamajek2008a, Wright2011a}, stellar kinematics \citep{Wielen1977a} and the abundances of various chemical elements in the photospheres of low-mass stars \citep{Nissen2015a, Spina2016a, TucciMaia2016a, Magrini2018a, Casali2019a}. Of the age-dependent chemical abundance indicators, some are extrinsic -- arising from Galactic chemical evolution gradually changing the abundances in the interstellar medium from which a star was born (e.g., overall metallicity, the ratio of alpha elements to iron or the abundance of s-process elements). Others are intrinsic -- the photospheric abundances evolve due to nucleosynthesis and chemical mixing within the star (e.g., light element abundances).

This paper focuses on photospheric lithium\footnote{In this paper, "lithium" and "Li" refer to the $^7$Li isotope. The $^6$Li isotope is produced in far lower quantities, is far more fragile in stellar interiors and is not thought to contribute significantly to measured Li abundances.}, which has a long history as a, mostly intrinsic, age indicator and probe of stellar interiors  \citep{wallerstein65, herbig65a, Skumanich1972a}. More recent reviews of Li as an age indicator are provided by \cite{Jeffries2014b}, \cite{Barrado2016b} and \cite{Randich2021a}. Whilst some is produced in the Big Bang,  most Li in the present-day interstellar medium was likely produced in novae or asymptotic giant branch stars \citep[e.g.,][and references therein]{Romano2021a}.  However, Li is also destroyed in stellar interiors by $p,\alpha$ reactions at modest temperatures of $\sim 2.5\times 10^6$K and if mixing processes penetrate from regions at these temperatures to the surface, then photospheric Li depletion proceeds. Consequently, measuring Li in stars could be a good way to estimate their ages.

Li burning takes place in the cores of contracting PMS stars on timescales that decrease with mass \citep{Bildsten1997a}. If the PMS star remains fully convective, total Li depletion will be rapid. However, stars with mass $M>0.35 M_\odot$ will develop radiative cores on a timescale that also decreases with mass and once the base of any convection zone falls below the Li-burning temperature, then Li-depleted material is no longer rapidly mixed to the photosphere. This produces a complex, mass- and hence $T_{\rm eff}$-dependent pattern of Li depletion at the end of the PMS phase. Predictions of PMS Li depletion are highly sensitive to opacities and the uncertain treatment of convection \citep{Pinsonneault1997a, Dantona2000a, Piau2002a, Tognelli2021a}. Observations of young stars at the zero-age main sequence (ZAMS) also reveal a 1-2 orders of magnitude Li abundance dispersion, particularly among young K-stars. This is not predicted by standard models featuring only convective mixing and suggest another parameter, possibly rotation or dynamo-induced magnetic activity, is also important \citep{Soderblom1993a, Barrado2016a, Bouvier2018a, Jeffries2021a}.

Standard models also cannot explain why the Sun's Li abundance is two orders of magnitude lower than solar-type ZAMS stars. This points to slow mixing mechanisms that operate on the main sequence and gradually deplete photospheric Li further. Many candidate mechanisms have been proposed and it is not clear which are the most effective \citep[e.g.,][]{Garcia1991a, Chaboyer1995a, Charbonnel2005a, Denissenkov2010a}. Observations suggest that Li is gradually depleted in solar-type ZAMS stars by another factor of 3-4 over a billion years \citep[e.g.,][]{Sestito2005a, Randich2010a}. After that, observations of older star clusters again suggest a dispersion of unknown origin in some, but not all, clusters \citep[e.g.,][]{Randich2003a, pasquini2008a, Pace2012a}, possibly associated with rotation, binarity or planetary systems \citep[e.g.,][]{Israelian2004a, Bouvier2008a, Gonzalez2015a}. Others argue that any scatter is much reduced with better quality data and by confining comparisons to stars with very similar $T_{\rm eff}$, there may be a monotonic decline in Li with age from 1-10 Gyr \citep{Carlos2019a, Carlos2020a}.

Li depletion through the PMS and main sequence phases is therefore far from understood; the present generation of stellar models have significant uncertainties, and possibly missing physics, that greatly affect their predictions of Li depletion as a function of age, temperature and composition. This means they cannot be used to provide reliable or accurate age estimation and points to using Li as an empirical age estimator by calibrating its depletion with observations of coeval stars in clusters. Such datasets have been assembled previously to some extent \citep[e.g,][]{Sestito2005a, Gutierrez2020a}; most recently, \cite{Stanford-Moore2020a} provided quantitative age estimates using an empirical calibration constructed from literature measurements of Li in 10 open clusters and associations.

In this paper we exploit the very large, homogeneous set of spectroscopic observations of low-mass stars in open clusters, that were obtained as part of the {\it Gaia}-ESO Survey \citep[GES,][]{Gilmore2012a, Gilmore2022a, Randich2013a, Randich2022a}. These provide a much improved calibration of the relationship between lithium, $T_{\rm eff}$ and age that is used to construct an empirical model of Li depletion.
Our objectives are (i) to quantify how precisely age can be determined for stars of various ages and $T_{\rm eff}$; (ii) to find how much more precision can be obtained by fitting groups of stars that are assumed to be coeval; (iii) to validate the method by finding the ages for a selection of young stars and associations that were not observed as part of GES; and (iv) to explore to what extent Li depletion is determined only by age and $T_{\rm eff}$ or whether third parameters such as chemical composition might be confounding or contributing factors.

The dataset used in this project, the empirical model and how it is fitted to the data are described in \S~\ref{2} and includes a catalogue of lithium equivalent widths and $T_{\rm eff}$ for 6200 kinematically selected members of 52 open clusters taken from \cite{Jackson2022a}. In \S\ref{3} the age estimation performance of the model is analysed for single stars and groups and the sensitivity to the adopted age and temperature scales are discussed. \S\ref{4} applies the model to estimate the ages of young field stars that host exoplanets and to coeval ``moving groups" in the solar neighbourhood. In \S\ref{5} we discuss systematic uncertainties and compare the model with previous work. A summary is provided in \S\ref{6} and
an implementation of the technique, written in Python code\footnote{\protect\url{https://github.com/robdjeff/eagles}}, called ``Empirical AGes from Lithium Equivalent widthS" ({\sc eagles}) is described in Appendix~\ref{eagles}.

\section{An Empirical relationship between Lithium equivalent width, temperature and age}
\label{2}

The GES data used in this paper come from the sixth internal data release \citep[GESiDR6,][]{Gilmore2022a, Randich2022a} which includes a library of stacked spectra of cluster targets obtained with the FLAMES-GIRAFFE spectrograph (with HR15N order-sorting filter) and FLAMES-UVES spectrograph (centred at 580\,nm) \citep{Pasquini2002a} on the 8-m UT2-{\it Kueyen} telescope of the Very Large Telescope \footnote{The reduced spectra are available in the public DR4 data release of the Gaia-ESO survey in the ESO archive, \protect\url{http://archive.eso.org/cms.html}.}, together with  the GESiDR6 Parameter Catalogue\footnote{The parameter catalogue is available in the public DR5 release of the Gaia-ESO survey in the ESO archive.}. The latter contains values of $T_{\rm eff}$, metallicity ([Fe/H]), radial velocity (RV), gravity ($\log g$) and $T_{\rm eff}$- and gravity-sensitive spectral index \citep[$\gamma$,][]{Damiani2014a} (plus other parameters not used here) for a large proportion of targets, which were derived by the GES Working Groups (WGs, Hourihane et al. submitted). In this section we select a subset of 6200 stars that were identified as very probable members of the open clusters and associations observed by GES \citep[see][]{Jackson2022a, Bragaglia2022a} and estimate the equivalent width of their Li~{\sc i}~6708\AA\ feature (EW$_{\rm Li}$) from the GES spectra. These data, referred to as the training data, are used to define an empirical relationship between EW$_{\rm Li}$, $T_{\rm eff}$ and age.     

\subsection{The training data}
\label{2.1}
The training data were drawn from 52 clusters with metallicities in the range $-0.3 < {\rm [Fe/H]} < 0.2$ listed in Table \ref{clusters}. The metallicities in Table \ref{clusters} were taken from \cite{Randich2022a} with the exception of NGC\,6649, where the value is the median [Fe/H] of its cluster members. Also shown in Table \ref{clusters} are cluster ages from three different sources, which range from 2 Myr to 6 Gyr. Ages in the column headed "LIT" are representative cluster ages drawn from the literature listed in tables 1 and 2 of \cite{Jackson2022a}, except for the clusters 25\,Ori, NGC\,2451b, NGC\,2547 and NGC\,2516, where the age has been updated according to \cite{Franciosini2022a}. The column headed "GES" lists ages from tables~3 and~4 of \cite{Randich2022a}. The column headed "DIAS" lists ages from homogeneous determinations (by isochrone fitting) in \cite{Dias2021b}, where matches were found according to cluster RA, Dec and proper motion.  The final column headed "MEAN" lists the geometric mean of the three (or two) reported cluster ages. These mean values were used to define target ages for the training set. It is recognised that this provides an inhomogeneous scale and that there could be significant uncertainties in the ages of individual clusters. The effect of variations in the calibration age-scale is discussed in \S\ref{3.4}.   

Targets from the designated clusters were selected with $2900 < T_{\rm eff}/{\rm K} < 6600$, a reported value of RV and a probability of cluster membership $>0.9$ in \cite{Jackson2022a}. Note that these members were selected on the basis of their kinematics, not of their chemical (including Li abundance) or photometric properties. Any star with $T_{\rm eff}>4000$\,K and $\log g<3.4$ \citep[or its equivalent calculated from the $\gamma$ and $\tau$ indices,][]{Jackson2022a} were rejected as probable giant stars, and anything with EW$_{\rm Li} < -300$\,m\AA, EW$_{\rm Li}>800$\,m\AA\ or an uncertainty in EW$_{\rm Li}>300$\,m\AA\ were rejected as poor data. These selections left us with 6200 individual targets, of which 6106 have GIRAFFE spectra, 203 have UVES spectra and 109 have both.
Details of targets in the training set are shown in Table \ref{targets} where the membership probability and $T_{\rm eff}$ are taken from columns headed $T_{\rm eff}^p$ and $P_{\rm 3D}$ in table~3 of \cite{Jackson2022a}. For the majority (93 per cent) of targets, $T_{\rm eff}^p$ is the effective temperature reported in the GESiDR6 Parameter Catalogue,  otherwise it was inferred from the spectroscopic temperature index $\tau$ \citep{Damiani2014a} measured from the target spectra. The acceptance criteria of $P_{\rm 3D} > 0.9$ gives an average cluster membership probability of $P_{\rm 3D} > 0.994$ and the expected contamination level is only about 37 in the total sample of 6200 cluster stars.

%Ideally the training data would be evenly distributed over the chosen range of $T_{\rm eff}$ and $\log$age. However because the older stars in our cluster are generally more distant there is a paucity of data for M-dwarf targets and ages >1\,Gyr. These 

\begin{table}
\caption{Clusters ages. Columns 1 and 2 show the name and average metallicity of the cluster from \protect\cite{Randich2022a} and column 3 gives the number of members in each cluster used in our analysis. Columns 4 to 6 show cluster ages from three different sources. "Lit" ages are from \protect\cite{Jackson2022a} and \protect\cite{Franciosini2022a}, GES ages are from table~3 in Randich et al. (2022) and "Dias" ages are from \protect\cite{Dias2021b}.  Column 6 shows the geometric mean age for each cluster (see \S \ref{2.1}).}
\begin{tabular}{lrrrrrr} 
\hline
Cluster name	&	[Fe/H]	&	No.	    &	Lit  	&	GES	    &	Dias  & Mean	\\
                &       &   stars   &   ages    &   ages    &   ages  & ages     \\
            	&		&	    	&	(Myr)	&	(Myr)	&	(Myr) & (Myr)	\\
\hline	
NGC 6530	&	-0.02	&	285	&	1	&	2	&	5.3	&	2.2	\\
Trumpler 14	&	-0.01	&	79	&	1.5	&	2.8	&	4.8	&	2.7	\\
Chamaeleon I	&	-0.03	&	78	&	2	&	1.6	&	---	&	1.8	\\
Rho Ophiuchus	&	0.03	&	41	&	3	&	4.5	&	---	&	3.7	\\
NGC 2264	&	-0.10	&	460	&	4	&	3.2	&	7	&	4.5	\\
NGC 2244	&	-0.04	&	72	&	4.1	&	4	&	12.9	&	6.0	\\
Lambda Ori	&	-0.09	&	187	&	6	&	12.6	&	8.8	&	8.7	\\
Lambda Ori B35	&	-0.09	&	44	&	6.1	&	12.6	&	---	&	8.8	\\
25 Ori	&	0.00	&	159	&	19	&	13.5	&	12.2	&	14.6	\\
ASCC 50	&	-0.02	&	166	&	8	&	11.5	&	5.8	&	8.1	\\
Collinder 197	&	0.03	&	86	&	13	&	14.1	&	9.1	&	11.9	\\
Gamma Velorum	&	-0.02	&	201	&	18	&	20	&	12.2	&	16.4	\\
IC 4665	&	0.01	&	32	&	23	&	33.1	&	53	&	34.3	\\
NGC 2232	&	0.02	&	74	&	38	&	17.8	&	30.6	&	27.5	\\
NGC 2547	&	-0.03	&	147	&	35	&	32.4	&	39.1	&	35.4	\\
IC 2602	&	-0.06	&	53	&	44	&	36.3	&	47	&	42	\\
NGC 2451b	&	-0.02	&	57	&	30	&	40.7	&	45.6	&	38	\\
NGC 6649	&	-0.05	&	4	&	50	&	70	&	40.8	&	52	\\
IC 2391	&	-0.06	&	33	&	51	&	28.8	&	48.8	&	42	\\
NGC 2451a	&	-0.08	&	40	&	65	&	35.5	&	54	&	50	\\
NGC 6405	&	-0.02	&	52	&	94	&	34.7	&	78	&	63	\\
NGC 6067	&	0.03	&	21	&	120	&	125	&	127	&	124	\\
NGC 2516	&	-0.04	&	450	&	138	&	239	&	276	&	209	\\
Blanco 1	&	-0.03	&	126	&	125	&	104	&	102	&	110	\\
NGC 6709	&	-0.02	&	43	&	150	&	190	&	160	&	166	\\
NGC 6259	&	0.18	&	15	&	210	&	269	&	328	&	265	\\
NGC 6705	&	0.03	&	119	&	280	&	309	&	294	&	294	\\
Berkeley 30	&	-0.13	&	22	&	300	&	295	&	454	&	343	\\
NGC 3532	&	-0.03	&	397	&	300	&	398	&	413	&	367	\\
NGC 6281	&	-0.04	&	23	&	314	&	512	&	328	&	375	\\
NGC 4815	&	0.08	&	19	&	560	&	371	&	441	&	451	\\
NGC 6633	&	-0.03	&	17	&	575	&	691	&	606	&	622	\\
Trumpler 23	&	0.20	&	10	&	800	&	707	&	292	&	549	\\
NGC 2355	&	-0.13	&	59	&	900	&	1000	&	1235	&	1036	\\
NGC 6802	&	0.14	&	26	&	900	&	660	&	645	&	726	\\
Pismis 15	&	-
0.02	&	30	&	1300	&	871	&	1811	&	1270	\\
Trumpler 20	&	0.13	&	94	&	1400	&	1862	&	1510	&	1579	\\
NGC 2141	&	-0.04	&	507	&	1800	&	1862	&	2897	&	2133	\\
Czernik 24	&	-0.11	&	33	&	2000	&	2691	&	1342	&	1933	\\
Haffner 10	&	-0.10	&	194	&	2000	&	3801	&	5058	&	3375	\\
NGC 2158	&	-0.15	&	189	&	2000	&	1548	&	---	&	1760	\\
NGC 2420	&	-0.15	&	312	&	2200	&	1737	&	2223	&	2040	\\
Berkeley 21	&	-0.21	&	70	&	2200	&	2138	&	---	&	2169	\\
Berkeley 73	&	-0.26	&	14	&	2300	&	1412	&	2223	&	1933	\\
Berkeley 22	&	-0.26	&	97	&	2400	&	2454	&	---	&	2427	\\
Berkeley 31	&	-0.29	&	96	&	2900	&	2818	&	3689	&	3112	\\
NGC 6253	&	0.16	&	86	&	3000	&	3000	&	3539	&	3170	\\
Messier 67	&	0.00	&	79	&	3500	&	3500	&	3758	&	3584	\\
NGC 2425	&	-0.13	&	98	&	3600	&	2398	&	4017	&	3261	\\
Berkeley 36	&	-0.15	&	143	&	4000	&	6760	&	3411	&	4518	\\
Berkeley 39	&	-0.14	&	455	&	6000	&	5623	&	6471	&	6021	\\
\hline
\end{tabular}
\label{clusters}
\end{table}

\begin{table*}
\caption{Training data used to calibrate the empirical relations of EW$_{\rm Li}$ as a function of $T_{\rm eff}$ and age (see \S\ref{2.1}). Details are shown for 6200 cluster members observed as part of the {\it Gaia}-ESO survey. 109 targets have both GIRAFFE (Filter $=$ 665.0 nm) and UVES (Filter $=$ 580.0 nm) measurements. Also shown is data for 1503 GES targets with low ($<$0.01) probabilities of cluster membership classified as field stars (see \S\ref{2.3}). A sample of the table is shown here. The full version is available online as supplementary material.} 
\label{training}
\begin{tabular}{lllllllllll}
\hline
Cluster & Target & Filter & RA    & DEC   & Age   & Probability   & $(G_{\rm BP}$-$G_{\rm RP})_0$ & $T_{\rm eff}$ & EW$_{\rm Li}$ & eEW$_{\rm Li}$\\
        &        & centre $\lambda$ &(deg) & (deg) & (Myr) & member & (mag) & (K) & (m\AA) & (m\AA)\\
\hline
25 Ori              &05224842+0140439&665.0nm &  80.70175&   1.67886&    14.6&   0.998&   3.013&   3148.&   633.7&    38.6\\
25 Ori              &05225186+0145132&665.0nm &  80.71608&   1.75367&    14.6&   1.000&   3.173&   3203.&   544.4&    41.2\\
25 Ori              &05225609+0136252&665.0nm &  80.73371&   1.60700&    14.6&   1.000&   2.727&   3333.&    12.0&    19.5\\
25 Ori              &05225678+0147404&665.0nm &  80.73658&   1.79456&    14.6&   0.992&   2.821&   3299.&   296.4&    26.8\\
25 Ori              &05225889+0145437&665.0nm &  80.74538&   1.76214&    14.6&   1.000&   2.857&   3320.&   463.6&    34.6\\
25 Ori              &05230387+0134335&665.0nm &  80.76613&   1.57597&    14.6&   1.000&   1.643&   4158.&   499.6&    10.6\\
\hline
 \label{targets}
 \end{tabular}
\end{table*}

\begin{figure}
	\includegraphics[width = 85mm]{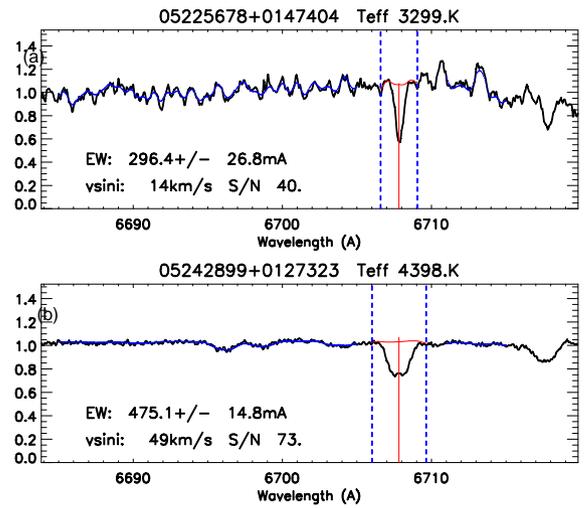}
	\caption{Typical spectra observed in open cluster 25\,Ori. Plot (a) shows the spectrum of a cooler star with the smoother, blue lines showing sections of the empirical continuum spectra scaled to match the observed spectrum. Blue vertical dashed lines indicate the extent of the top hat profile used to measure EW$_{\rm Li}$ (the integral pf the difference between the black line and the red line). Plot (b) shows the spectrum of a hotter and faster rotating star compared  with sections of the rotationally broadened empirical continuum (see \S\ref{2.2}).}  
	\label{Li_spectra}	
\end{figure}

\subsection{Lithium equivalent widths}
\label{2.2}

At an early stage, we decided to base our analysis on EW$_{\rm Li}$, rather than a lithium abundance derived from EW$_{\rm Li}$ or from a spectral synthesis. The reasons were twofold. First, EW$_{\rm Li}$ is independent of the temperature scale or the adoption of any particular set of stellar atmosphere models or NLTE corrections. Our results are then in principle applicable, without scaling or offset, to other datasets outside the GES project. Second, the uncertainties in Li abundance and temperature are highly correlated and would be difficult to disentangle when it comes to estimating ages for individual stars, whereas uncertainties in EW$_{\rm Li}$ and $T_{\rm eff}$ should be close to orthogonal.

The adoption of EW$_{\rm Li}$ as the main dependent variable does bring its own problem of defining the lithium equivalent width. This is rarely a problem in high resolution spectra of warmer, slowly rotating stars where the continuum can be readily identified. It is more problematic in cooler stars with molecular bands and those with rapid rotation. One third of our targets are late K or M-dwarfs with $T_{\rm eff}<4250$\,K, where measurements of EW$_{\rm Li}$ are complicated by molecular absorption features \citep[e.g.][]{Pavlenko1996a, Rajpurohit2014a}, making the pseudo-continuum highly sensitive to small changes in temperature. This can easily lead to systematic shifts in EW$_{\rm Li}$. 

EW$_{\rm Li}$ values are available for the majority of our training sources in the released  GESiDR6 parameter catalogue \citep{Franciosini2022b}. However, the use of these EWs causes difficulties in three ways: (i) EWs were unavailable for 687 of the training sources and 2859 lacked an EW that was corrected for the blend with the nearby Fe~{\sc i}\,6707.4\AA\ line (mostly the cool stars). (ii) The EWs reported for cool stars ($T_{\rm eff}<4250$\,K) are pseudo-equivalent widths that still include a significant contribution from molecular absorption and other blends. Whilst this choice is justified and accounted for in estimating GES Li abundances \citep[see][]{Franciosini2022b}, there is a lack of comparability with literature EWs, where EW$_{\rm Li} \sim 0$ in older M-dwarfs (see \S\ref{4.2}). (iii) 354 targets have only reported upper limits to EW$_{\rm Li}$, which is problematic for our fitting methods.

\begin{figure*}
	\begin{minipage}[t]{1\textwidth}
	\includegraphics[width = 170mm]{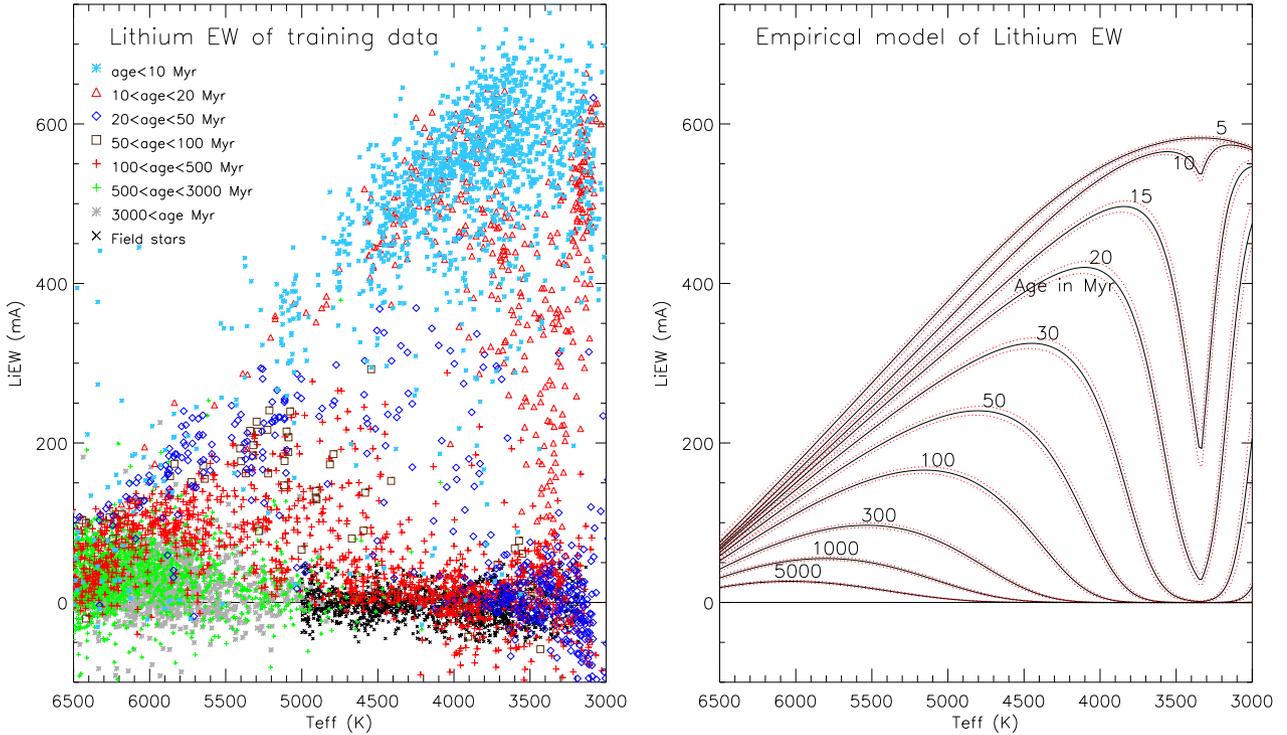}
        \end{minipage}
	\caption{Equivalent width of the of the Li\,{\sc i}~6708\AA\ line as a function of $T_{\rm eff}$ and age. The left-hand panel shows measured values of equivalent width for the training data (EW$_{\rm Li}$) colour-coded in broad bins of age (see \S\ref{2.2} and \S\ref{2.3}). The right-hand panel shows model isochrones (EW$_{\rm m}$ from Eqns.~\ref{EW_m} and~\ref{ABC}) derived by fitting the training data with the model described in \S\ref{2.3} and \S\ref{2.4}. Dotted lines show the the $\pm 1 \sigma $ uncertainties in the model values due to the RMS uncertainties in the expectation values of the constants defining EW$_{\rm m}$ (see Table~3). This uncertainty does not include any systematic uncertainties in the adopted $T_{\rm eff}$ and age scales in the training data, which are a larger source of uncertainty.}  
	\label{lithium_at3}	
\end{figure*}

To address all these issues, the original reduced spectra were retrieved and EW$_{\rm Li}$ was estimated using a direct flux integration of the spectrum over a top-hat profile, after subtracting the normalised median spectra defined by sets of field stars of comparable temperature observed in GES cluster fields and which are assumed to have minimal Li. This method estimates EW$_{\rm Li}$ even in low signal-to-noise spectra; is corrected for blends to first-order; and was used by \cite{Binks2021a} to measure the  EW$_{\rm Li}$ of stars in the temperature range $3000 < T_{\rm eff}/{\rm K} < 5000$, where the majority of field stars are expected to be fully lithium depleted\footnote{Models of PMS Li depletion alone show that Li has been entirely depleted in these stars by the ZAMS \citep{Piau2002a} and this is empirically confirmed in clusters like the Hyades \citep{Thorburn1993a}.}. For the current analysis the catalogue of template continuum spectra was extended to cover the full temperature range of the training data. In the case of warmer stars this necessitates accounting for the finite level of median EW$_{\rm Li}$ observed for field stars at $T_{\rm eff} > 5500$\,K, since Li depletion is not completed in these stars. Hence the measured equivalent width is now defined as
\begin{equation}
{\rm EW_{Li}} = \Delta{\rm EW_{Li}} + {\rm EW}_0 
\label{EW_0}
\end{equation}
where $\Delta$EW$_{\rm Li}$ is the equivalent width measured by comparing the target spectra to a continuum defined by the median field star spectra in the appropriate temperature range and EW$_0$ is the equivalent width of the template median field star spectrum which is assumed to be zero below $T_{\rm eff}$=5550\,K and peaks at 34\,m\AA~at $T_{\rm eff}=6100$\,K (see Appendix~\ref{AppA}). 

Figure~\ref{Li_spectra} shows heliocentrically corrected spectra for two typical targets in the region of the Li\,{\sc i}~6708\,\AA~feature. Vertical dashed lines show the top hat profile used to define the limits of the integral that measures EW$_{\rm Li}$. The template continuum spectra were rotationally broadened according to the target $v\sin i$, which was calculated from the parameter {\sc vrot} of the GES spectrum meta-data \citep[see appendix B2 of][]{Randich2022a}. The width of the top hat profile scales with $v\sin i$ as $\pm(1+v\sin i/60\,{\rm km\,s}^{-1})$. In cases where {\sc vrot} is not defined in some low S/N spectra, the targets were assigned a default $v\sin i$ of 30\,km\,s$^{-1}$. 

The measured EW$_{\rm Li}$ for the training data are shown in Fig.~\ref{lithium_at3}~and listed in Table \ref{targets}. Measurement uncertainties, $\sigma_{\rm Li}$ in Table \ref{targets}, were estimated as the RMS value of the EWs measured using the same procedure and top hat profile, but centred at a set of 20 wavelengths on either side of the Li feature (see Fig.~\ref{Li_spectra}).  

\subsection{Lithium EW as a function of age and temperature}
\label{2.3}
\begin{figure*}
	\begin{minipage}[t]{1\textwidth}
	\includegraphics[width = 170mm]{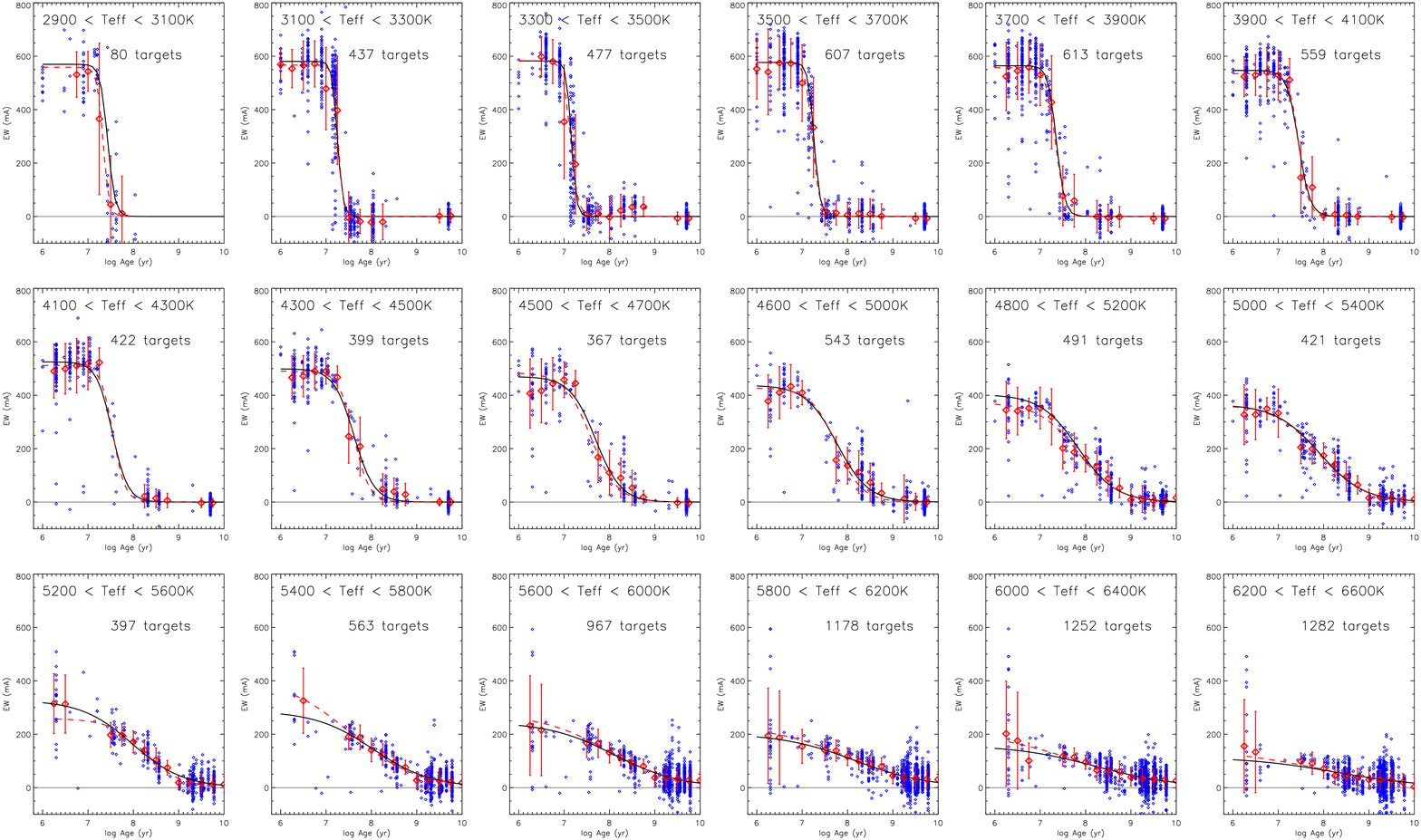}
    \end{minipage}
	\caption{EW$_{\rm Li}$ as a function of age and $T_{\rm eff}$. Plots show measured values of EW$_{\rm Li}$ as a function of log age in bins of temperature over ranges indicated on each plot. Red diamonds and error bars show the average and RMS uncertainty of  EW$_{\rm Li}$  in bins of log age. The red dashed curve shows an empirical model (Eqn.~\ref{EW_m}) fitted to the data in each temperature bin; black curves show the full maximum likelihood model evaluated at the central $T_{\rm eff}$ of each bin.} 
	\label{lithium_at1_rob}	
\end{figure*}

Figure \ref{lithium_at1_rob} shows EW$_{\rm Li}$ as a function of $\log{\rm age/yr}$ (base 10) for $2900< T_{\rm eff}/{\rm K} < 6600$ in 200\,K steps. Bins are $\pm$100\,K wide for  $T_{\rm eff} < 5000$\,K and  $\pm$200\,K wide at higher temperatures. Where there are $\geq 10$ stars in $\pm$0.25\,dex bins of $\log{\rm age}$, red triangles and error bars show the mean and  standard deviation of EW$_{\rm Li}$ .  Visual inspection of these  plots suggests that the mean EW$_{\rm Li}$ in a given temperature bin can be well-described by a $\tanh$-type function of $\log$ age;
\begin{eqnarray}  
{\rm EW}_m & =  A (1-\tanh[(\log{\rm Age/yr}-C)/B])\, ,
\label{EW_m}
\end{eqnarray}   
which transitions from an undepleted value at small $\log{\rm age}$ to a value close to zero for the oldest stars. EW$_{\rm Li}$($\log$ Age) is then defined by the constants $A$, $B$ and $C$, which must be empirically determined. This becomes interesting if the constants $A$, $B$ and $C$ are themselves simple functions of $T_{\rm eff}$ which can be modelled using the full training dataset.

\begin{figure}
	\includegraphics[width = 80mm]{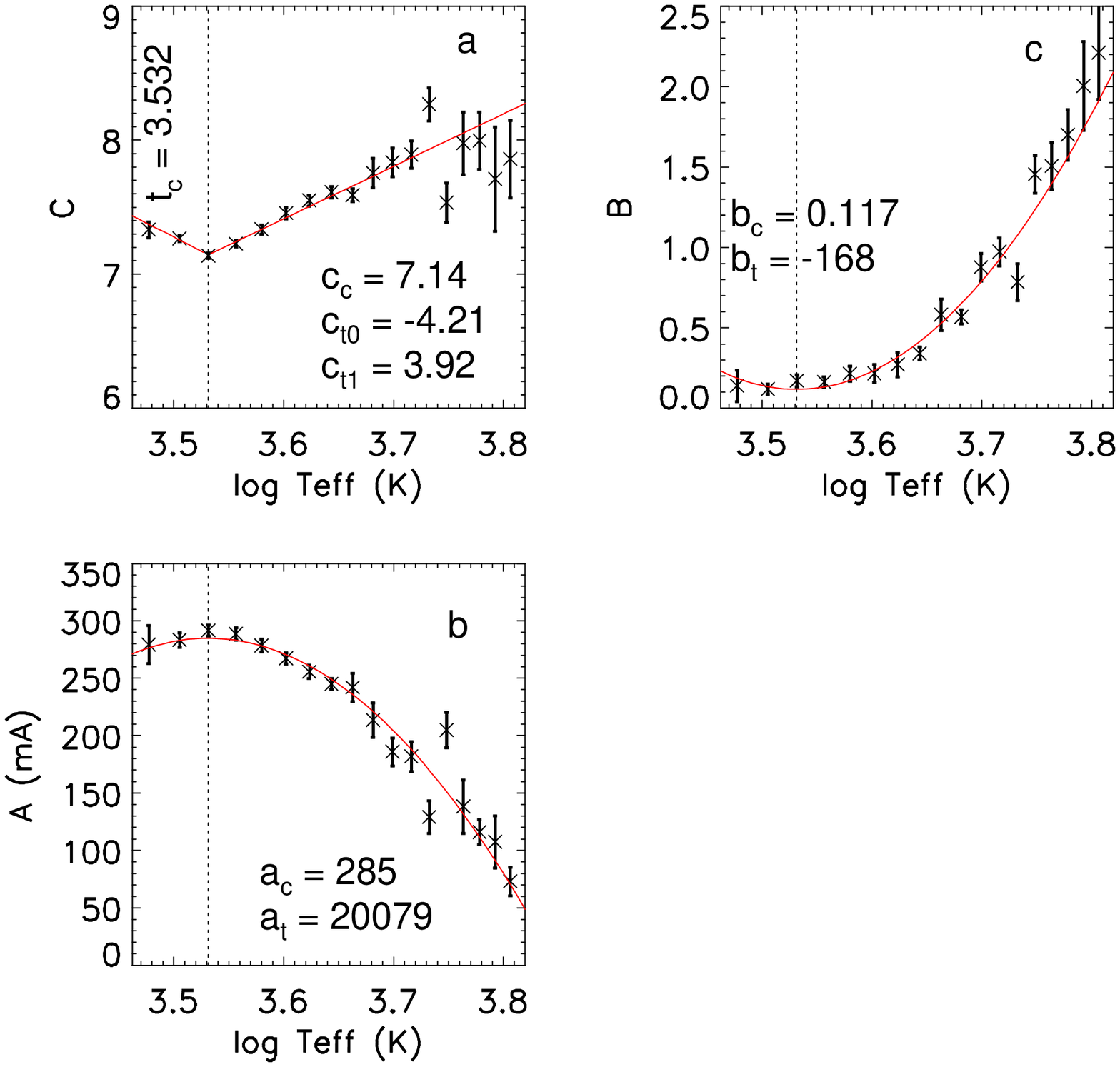}
	\caption{The results of fitting data in individual $T_{\rm eff}$ bins  in Fig.~\ref{Li_spectra} to an empirical model of EW$_{\rm Li}$ as a function of $\log$ age (Eqn.~\ref{EW_m}). Points show the mean value and rms uncertainty of  $A$, $B$ and $C$ as a function of $T_{\rm eff}$,  with  curves showing least square fits of the expressions defining $A$, $B$ and $C$ in Eqn.~\ref{ABC}. Text on the plot shows the fitted values of the constant terms in Eqn.~\ref{ABC}}.  
	\label{lithium_at1_fit}	
\end{figure}

To investigate this possibility a three-parameter maximum likelihood analysis was used to estimate $A$, $B$ and $C$ (and their uncertainties) in each $T_{\rm eff}$ bin separately. For this analysis the total uncertainty in EW$_{\rm Li}$ for individual targets was estimated by interpolating the standard deviation of data in 0.25\,dex bins of $\log{\rm age}$ (shown as error bars in Fig.~\ref{lithium_at1_rob}). There is a paucity of data for M-dwarfs with age $>1$\,Gyr at $T_{\rm eff}<4200$\,K (because the older clusters in the sample are more distant). This has no great effect on estimates of $A$, $B$ and $C$ because there are sufficient data to see that Li has already been fully depleted at ages $<1$\,Gyr at these spectral types. However, in order to better constrain the intrinsic dispersion of EW$_{\rm Li}$ at older ages and cooler temperatures (see \S\ref{2.4}), additional data were added to the training set at an assumed age of 5 Gyr using $EW_{\rm Li}$ measured for 1503 field stars with $T_{\rm eff}<5000$\,K that came from the sample used to construct the Li-depleted template spectra in Appendix~\ref{AppA}. The properties of these additional targets are shown in the left hand panel of Fig.~\ref{lithium_at3}~and in Table \ref{targets}.

The results of fitting the 3 parameter model of EW$_{\rm Li}$ for the 18 $T_{\rm eff}$ bins in Fig.~\ref{lithium_at1_rob}~are plotted as a function of $\log T_{\rm eff}$ in Fig.~\ref{lithium_at1_fit}. These plots indicate the following:
\begin{itemize}
    \item Parameter $C$, which defines the $\log$ age at which the transition from high EW$_{\rm Li}$ to low EW$_{\rm Li}$ occurs, can be described by linear functions of $\log T_{\rm eff}$ above and below some critical $\log{T_{\rm eff}}$, which we define as $t_c$ (the red lines in Fig.~\ref{lithium_at1_fit}a).
    \item Parameter $A$ that is the limiting value of EW$_{\rm Li}/2$ at young ages is reasonably described by a second order polynomial of $\log{T_{\rm eff}}$ with its maximum value fixed at $\log{T_{\rm eff}} = t_c$ (the red curve in Fig.~\ref{lithium_at1_fit}b).
   \item Parameter $B$, which determines the rate of transition from high to low EW$_{\rm Li}$ can also be  described with a second order polynomial with its minimum at $\log{T_{\rm eff}}=t_c$ (the red curve in Fig.~\ref{lithium_at1_fit}c).
   
\end{itemize}

These empirical observations lead to the following definitions of $A$, $B$ and $C$ as a function of $\log T_{\rm eff}$; 
\begin{eqnarray}
\label{ABC}
A & = & a_c - a_t(\log{T_{\rm eff}} - t_c)^2/(2t_c)\, ,                \\
B & = & b_c - b_t(\log{T_{\rm eff}} - t_c)^2/(2t_c)\, ,                \nonumber \\
C & = & c_c + c_{t0}(\log{T_{\rm eff}} - t_c)\ \ \ {\rm for}\ \log{T_{\rm eff}} < t_c\, ,   \nonumber \\
C & = & c_c + c_{t1}(\log{T_{\rm eff}} - t_c)\ \ \ {\rm for}\ \log{T_{\rm eff}} > t_c\, , \nonumber    
\end{eqnarray}
where $t_c, a_c, a_t, b_c, b_t, c_c, c_{t0}$ and $c_{t1}$ are empirical constants.

Equations~\ref{EW_m} and~\ref{ABC} define an empirical relationship for EW$_{\rm Li}$, as a function of $T_{\rm eff}$ and $\log$ age that can be calibrated using the whole training dataset. Approximate values for these constants, calculated from a least-squares fit to the values of $A$, $B$ and $C$ measured in $T_{\rm eff}$ bins are shown in Fig.~\ref{lithium_at1_fit}. A more optimal approach uses a multi-dimensional maximum likelihood analysis to simultaneously fit equation~\ref{ABC} to the full training dataset (\S\ref{2.5}). There is an explicit assumption here that EW$_m$ only depends on $T_{\rm eff}$ and age. Any dependence on additional parameters might be manifested as increased dispersion around the model (\S\ref{2.4}).

 \begin{figure*}
    \centering
	\begin{minipage}[]{1\textwidth}
	\includegraphics[width = 170mm]{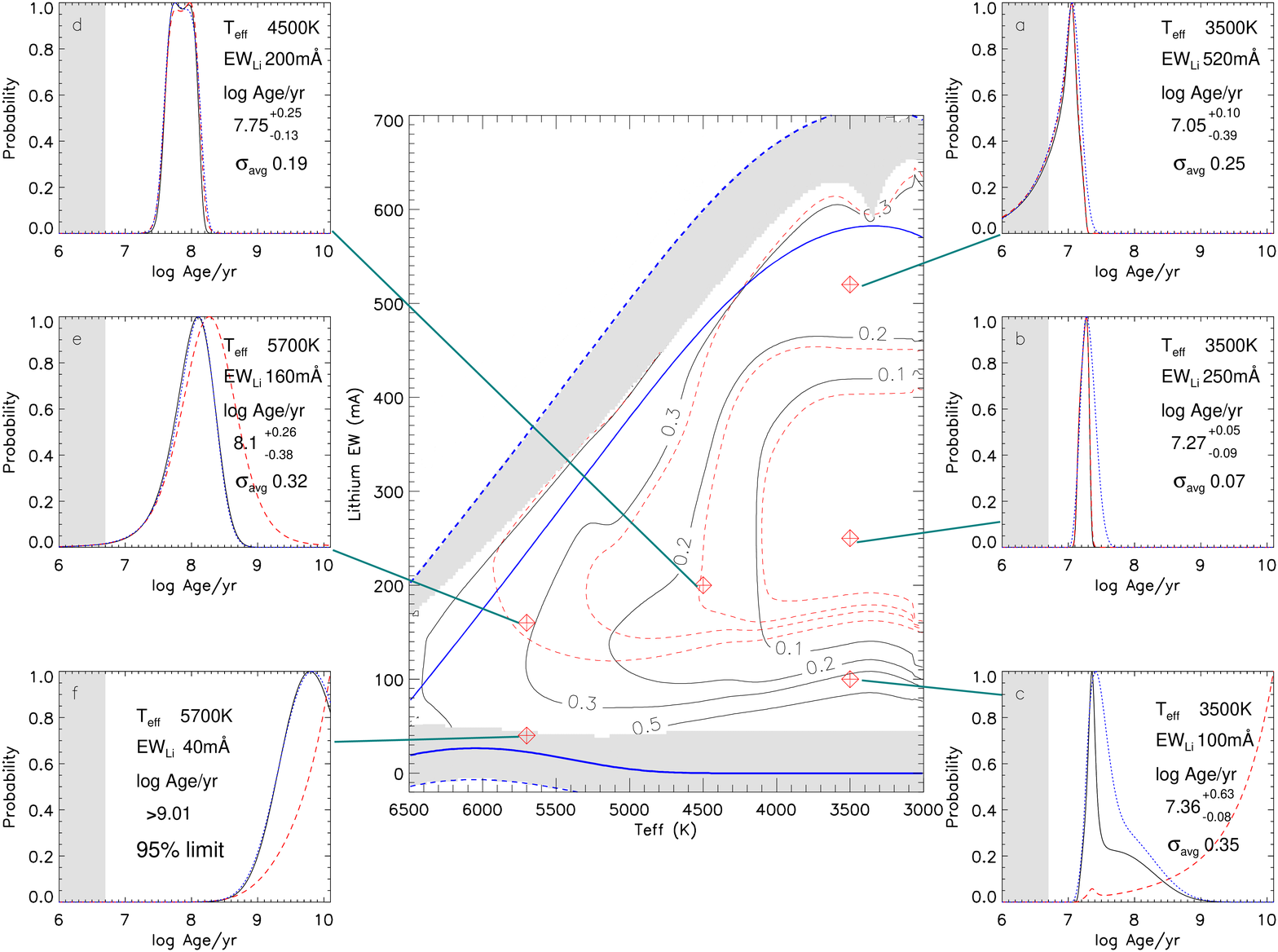}
    \end{minipage}
	\caption{The sensitivity of $\log$ age estimates in the of EW$_{\rm Li}$, $T_{\rm eff}$ plane. The central plot shows contours (black solid lines) of the mean error bars (in dex) of a log age determination assuming data with no measurement uncertainties and a prior probability distribution that is flat in age (appropriate for estimating the ages of single stars). Heavy blue lines show model values of EW$_{\rm m}$ at ages of 5\,Myr and 5\,Gyr. The blue dashed lines mark the boundaries beyond which the model cannot sensibly be applied. The shaded grey region marks the region where even with perfect data, only upper or lower limits on the age can be obtained. The surrounding plots (a-f) show the log age probability distributions derived at representative points in the EW$_{\rm Li}$, $T_{\rm eff}$ plane, labelled with the most probable log age and (asymmetric) 68 per cent confidence intervals. The red dashed contours in the plots show the effects of adopting measurement errors in EW$_{\rm Li}$ of $\pm 30$\,m\AA, the median value for the training data. The blue dotted curves show the effects of assuming an additional $T_{\rm eff}$ error of $\pm 200$\,K (see \S\ref{3.1}).} 
	\label{one_target}	
\end{figure*}

\subsection{Dispersion of Lithium EW as a function of temperature and age}
\label{2.4}
The full maximum likelihood analysis requires an estimate of the total uncertainty, $\delta_{\rm Li}$, in the measured values of EW$_{\rm Li}$ relative to the model values, EW$_{\rm m}$, at a given $T_{\rm eff}$ and age . This uncertainty arises from two sources which (we assume) add in quadrature, firstly the measurement uncertainty $\sigma_{\rm Li}$ (see \S\ref{2.2} and Table~\ref{targets}) and secondly what we label as an intrinsic dispersion around the model, $\rho_m$, which will incorporate genuine astrophysical dispersion, perhaps associated with differences in rotation and composition, a contribution from individual uncertainties in $T_{\rm eff}$ and age in the training dataset and any deficiencies in the simple model for Li depletion described by Eqns.~\ref{EW_m} and~\ref{ABC}. 

Equation \ref{delta_m} shows the expression used for $\rho_m$. The first component, $E$, is a function of $\log$~age, being highest for the youngest clusters and declining with age to a uniform low level for the older clusters. The second component, $F$, is proportional to the rate of change of EW$_{\rm m}$ with $\log$~age (see Eqns.~\ref{EW_m}) and is hence dependent on both $T_{\rm eff}$ and age. This term has two elements; the first applies over the full range of $T_{\rm eff}$  whereas the second applies over a restricted range of ages and temperatures and is intended to represent approximately the effect of the additional rotation-dependent dispersion observed in late G- and K-stars at the ZAMS (see \S\ref{1}). 
\begin{eqnarray}
   \rho_{\rm m} &=& \sqrt{E^2 + F^2}\, , \nonumber \\
    E &=& e_0\exp\left[-e_2\left( \log {\rm age/yr} -6\right)\right] +e_1\ ,  \nonumber \\
    F &=& \left( f_0 + f_1 \exp{\tfrac{-(\log ({\rm age}/{\rm yr})-8)^2}{2(0.2)^2}}\right)\left|\tfrac{\partial {\rm EW}_m}{\partial \log {\rm age}}\right|\, ,  \nonumber \\ 
      &  &{\rm for}\ 4200 < T_{\rm eff}<5200\,{\rm K}\, , \nonumber \\   
    F &=& f_0\left|\tfrac{\partial {\rm EW}_m}{\partial \log {\rm age}}\right|\,\ {\rm elsewhere}\ , \nonumber \\  
     \label{delta_m}
\end{eqnarray}
  where $e_0$, $e_1$,$e_2$, $f_0$ and $f_1$ are constants fitted to the training dataset.

\subsection{A maximum likelihood fit of the model to the training dataset}
\label{2.5}
Best-fit values for the eight parameters that determine EW$_{\rm m}(\log T_{\rm eff}, \log {\rm age})$ (Eqns.~\ref{EW_m} and~\ref{ABC}) and the five additional parameters that determine $\rho _{\rm m}$ (Eqn.~\ref{delta_m}) were evaluated in two steps. EW$_{\rm m}$ values were  evaluated at the $T_{\rm eff}$ and age of the training dataset stars and the results used to measure the difference, $\delta_{\rm Li}$, between the measured and model EW for individual targets. These data were binned in 100\,K bins of $T_{\rm eff}$ and 0.25\,dex bins of $\log{\rm age}$ to determine the RMS total uncertainty per bin, $\delta_{\rm bin}$. The RMS measurement error per bin $\sigma_{\rm bin}$ was then subtracted (in quadrature) to estimate the RMS intrinsic dispersion per bin $\rho_{\rm bin} = \sqrt{\delta_{\rm bin}^2 - \sigma_{\rm bin}^2}$ and its uncertainty $\rho_{\rm bin}/\sqrt{n}$ for bins containing $n \geq 8$ targets.
At this stage we discarded the one star with $T_{\rm eff}<3000$\,K and also stars with $T_{\rm eff}>6500$\,K because their EW$_{\rm Li}$ was too small compared with their uncertainties and scatter to give useful constraints. 

The remaining 5944 stars spanning $3000<T_{\rm eff}/{\rm K}<6500$
were used to compute the summed $\log$ likelihood of fit of the model function $\rho_{\rm m}$ over a grid of trial values for $e_0, e_1, e_2$, $f_0$ and $f_1$. The likelihood is defined (as a first approximation, using the parameters from Fig.~\ref{lithium_at1_fit} to determine $\rho_{\rm bin}$) as;
\begin{equation}
\mathscr{L} = \prod \frac{1}{\sqrt{2\pi\rho_{\rm bin}^2/n}} \exp{\left( \frac{-(\rho_{\rm bin} - \rho_{\rm m} )^2}{2\rho_{\rm bin}^2/n} \right)} + z\, .
\label{eq5}
\end{equation}
When fitting a group of stars at the same age within a single training cluster, a regularisation constant, $z = 10^{-12}$, was added to the individual likelihoods to avoid numerical problems due to any extreme outliers in individual measurements of EW$_{\rm Li}$ and $\sigma_{\rm Li}$. 

Expectation values of $e_0,e_1,e_2$, $f_0$ and $f_1$ were then used to calculate the model dispersion $\rho_{\rm m}$ at the $T_{\rm eff}$ and age of the training data and the results used to calculate the likelihood of fit of EW$_{\rm m}$ over a uniform grid of values of $t_c,a_c,a_t,b_c,b_t,c_c,c_{t0}$ and $c_{t1}$. The summed $\log$ likelihood was calculated at each point in the grid, with the likelihood defined as:
\begin{equation}
\mathscr{L} = \prod \frac{1}{\sqrt{2\pi(\sigma_{\rm Li}^2 + \rho_{\rm m}^2})} \exp{\left( \frac{-( {\rm EW_{\rm Li}} - {\rm EW} _m)^2}{2(\sigma_{\rm Li}^2+ \rho_{\rm m}^2)} \right)} + z
\label{eq6}
\end{equation}

The adopted values for the constants were then computed as expectation values over the parameter grid. The two stages of calculation (Eqns.~\ref{eq5} and~\ref{eq6}) were then repeated until the model parameters converged. Expectation values of the constants presented in Table~\ref{constants} are used to calculate EW$_{\rm m}$ and $\rho_{\rm m}$. Results (shown as black curves) are compared with the binned training data in Fig.~\ref{lithium_at1_rob}, whilst Fig.~\ref{lithium_at3} shows a graphical representation of the model in the form of isochrones of EW$_{\rm m}$ over the constrained $T_{\rm eff}$ range at ages of 5--5000\,Myr. Dotted lines indicate uncertainty in the calculation of EW$_{\rm m}$ produced by combining (in quadrature) the effects of the uncertainties in the expectation values of the constants defining EW$_{\rm m}$ (see Table~\ref{constants}). 

\begin{table}
\caption{Results of the maximum likelihood fit of the empirical model of lithium equivalent width as a function of $T_{\rm eff}$ and age to the training data in Table 2. Columns 2 and 3 show the range explored for each parameter with the final column showing the expectation value and uncertainty of the constants defining the empirical model in Eqns. 2--4.}
\centering
\begin{tabular}{llll} \hline
parameter & lower  &  upper  & expectation \\
          & limit  &  limit  & value \\
\hline          
\multicolumn{4}{l}{constants defining EW$_{\rm m}$ (see Eqns.~\ref{EW_m} and \ref{ABC})} \\
$t_c$	&	3.518	&	3.53	&	3.524	$\pm$	0.001	\\
$a_c$	&	285	&	300	&	291.3	$\pm$	1.1	\\
$a_t$	&	19500	&	22500	&	20687	$\pm$	239	\\
$b_c$	&	0.08	&	0.14	&	0.111	$\pm$	0.006	\\
$b_t$	&	-180	&	-150	&	-164.7	$\pm$	3.5	\\
$c_c$	&	7.08	&	7.18	&	7.131	$\pm$	0.007	\\
$c_t0$	&	-10	&	-4	&	-6.44	$\pm$	0.50	\\
$c_t1$	&	3.6	&	4.5	&	4.040	$\pm$	0.095	\\
\hline
\multicolumn{4}{l}{constants defining $\rho_{\rm m}$ (see Eqn.~\ref{delta_m})} \\
$e_0$	&	60	&	110	&	84.3	$\pm$	1.9	\\
$e_1$	&	0	&	20	&	1.8	$\pm$	1.0	\\
$e_2$	&	0.3	&	1.2	&	0.47	$\pm$	0.02	\\
$f_0$	&	0.04	&	0.12	&	0.079	$\pm$	0.003	\\
$f_1$	&	0.1	&	0.4	&	0.219	$\pm$	0.028	\\
\hline
    \end{tabular}
  \label{constants}
\end{table}

\section{Model Performance for Single Stars and Clusters}
\label{3}

%The set of isochrones in Fig.~\ref{lithium_at3} defines an empirical age scale  that can be used to constrain the age of stars or a coeval cluster or association from measured  EW$_{\rm Li}$ and $T_{\rm eff}$ values. 
Examination of  Fig.~\ref{lithium_at3} immediately indicates the range of ages and temperatures over which an EW$_{\rm Li}$ measurement is likely to provide a strong age constraint.
\itemize
\item For stars aged $<$10\,Myr, changes in EW$_m$  occur only over a limited range of temperatures ($4000< T_{\rm eff} <5500$\,K), but even this probably reflects a lack of detail in the model rather than a genuine change in the observed EW$_{\rm Li}$ patterns. Hence  comparison of measured levels of EW$_{\rm Li}$  with the empirical model EW$_{\rm m}$  can only provide upper limits on the age in this region of parameter space.

\item For stars with ages of 10--1000\,Myr there is continuous reduction in EW$_{\rm m}$ with age over a broad range of $T_{\rm eff}$, which should allow the age of a star or a cluster of coeval stars to be estimated, with a precision that will depend on the number and $T_{\rm eff}$ distribution of stars with EW$_{\rm Li}$ values.

\item For stars aged $>$1\,Gyr the predicted level of EW$_{\rm m}$ falls to its base level for $T_{\rm eff} < 4500$\,K and the isochrones become closely bunched at higher $T_{\rm eff}$. The slow decline at higher temperatures might provide some age sensitivity (though see \S\ref{5.1}), but could easily be obscured by measurement errors or an intrinsic dispersion. Thus for stars older than a few Gyr, comparison of EW$_{\rm Li}$ with EW$_{\rm m}$ is likely to provide only a lower limit to the stellar age.

\subsection{Estimating the age of single stars}
\label{3.1}

The best-fit model derived in \S\ref{2.5} can be used to estimate the age probability distribution for stars that have measurements of EW$_{\rm Li}$, $\sigma_{\rm Li}$ and $T_{\rm eff}$ (and its uncertainty). This is done by calculating EW$_{\rm m}$ and $\rho_m$ at a given $T_{\rm eff}$ over a uniform grid of log age (between 1 Myr and 12.4 Gyr, see below) and computing the corresponding likelihood function using Eqn.~\ref{eq6}. For single field stars, the regularisation constant $z$ is not needed and is set to zero. Since we believe the model lacks any age discrimination for ages $<5$ Myr, then $\mathscr{L}$ is set to its value at 5 Myr in this range. 

To use this likelihood function to estimate an age probability distribution it is multiplied by an age probability prior. For single stars we assume that this is a flat distribution in age and hence $p({\rm log\ age}) \sim {\rm age}$, between 1 Myr (we assume younger stars would be embedded in dust and unobservable) and 12.4 Gyr (an assumed age for the oldest stars in the Galaxy). In specific applications the prior probability distribution could be more informative (see \S\ref{sec_prior}). The product of the likelihood function and prior probability function is referred to as the posterior probability distribution of log age
\begin{equation}
    \ln P(\log {\rm age}) = \ln \mathscr{L} + \ln p
    \label{prior}
\end{equation}
If the posterior shows a clear peak, with $\ln P_{\rm max}>0.5$ above its value at 5\,Myr and 12.4\,Gyr, then the peak of the distribution is adopted as the best estimate of log age, with (asymmetric) 68 per cent confidence limits estimated by integrating $P$, with respect to log age above and below the most probable log age to include $\pm 34$ per cent of the probability distribution\footnote{The software also returns the median of the posterior as an estimator of $\log$ age.}. If $\ln P$ is $<0.5$ below $\ln P_{\rm max}$ at 5 Myr/12.4 Gyr, then a 95 per cent upper/lower limit to the log age is calculated respectively.

The performance of the model is illustrated in Fig.~\ref{one_target}, which shows a contour plot of the 68 per cent confidence limits (half the difference between the upper and lower limit) in log age in the EW$_{\rm Li}$, $T_{\rm eff}$ plane. Initially (the black, solid lines), this is evaluated assuming perfect data - so that the uncertainty in the derived age is due only to the intrinsic dispersion present in the training data (\S\ref{2.4}). Posterior probability distributions of log age, normalised to a peak of 1, are shown for a set of six selected EW$_{\rm Li}$, $T_{\rm eff}$ points, that illustrate the variety of behaviours, and labelled with the estimated (most probable) log age and its uncertainties. Figure~\ref{one_target} demonstrates that the greatest age sensitivity is achieved where EW$_{\rm Li}$ is changing fastest with respect to log age (e.g., Figs.~\ref{one_target}b,d). The probability distributions become quite asymmetric as data points approach regions where EW$_{\rm Li}$ becomes less sensitive to age, either because it is almost undepleted (e.g., Fig.~\ref{one_target}a) or almost entirely depleted (Fig.~\ref{one_target}c). Grey shaded areas mark the regions where even a perfect measurement can only yield an upper or lower limit to log~age (e.g., Fig.~\ref{one_target}f). The choice of prior can be quite important. This is discussed further in \S\ref{sec_prior}, but we note here that a flat prior in age means that where the likelihood function is poorly constrained by the Li measurements and/or has a broad upper tail that extends to ages $>1$ Gyr (e.g., Figs.~\ref{one_target}c, f) then the resulting age estimate is pushed to larger values, has a larger upper error bar or is more likely to result in a lower limit to the age than if a flat prior in $\log$~age were adopted.

The influence of measurement uncertainties is of course to make the age determinations less precise. The red dashed lines show how the plot is modified if the EW$_{\rm Li}$ measurements have uncertainties of 30\,m\AA, a median value for the GES data\footnote{The GES EW$_{\rm Li}$ measurements are heteroscedastic -- the errors tend to be smaller in the older and hotter stars and larger in the cooler and younger stars, but for the purposes of illustrating the effects of an uncertainty, a fixed value is assumed in Fig.~\ref{one_target}}. In the main panel the contours move inward so that a smaller portion of parameter space yields ages to a given level of precision. In the subplots, the inferred age probability distributions are broadened. The effect can be dramatic at smaller EW$_{\rm Li}$ because these EW$_{\rm Li}$ values could then be consistent with complete depletion, leading to an unconstrained likelihood at older ages and the posterior becoming dominated by the exponentially rising prior (see Figs.~\ref{one_target}c, f). The effects are much smaller in other regions of the EW$_{\rm Li}$/$T_{\rm eff}$ plane because the intrinsic dispersion is larger than the assumed measurement uncertainty, particularly at young ages (e.g., Figs.~\ref{one_target}a, b, d).

Uncertainties in $T_{\rm eff}$ are already accounted for to some extent when the empirical model is fitted to the training data, since uncertainties in the GES temperatures contribute to the modelled intrinsic dispersion in EW$_{\rm m}$-EW$_{\rm Li}$. The effects of $T_{\rm eff}$ errors in making an age estimate should only be included explicitly if those errors significantly exceed those in the GES training data.
Estimated uncertainties in $T_{\rm eff}$ are available for most of the GES training dataset and amount to $\pm$80\,K.  The effects of larger $T_{\rm eff}$ errors can be demonstrated by marginalising $\mathscr{L}$ over a set of normally distributed $T_{\rm eff}$ values with standard deviation equivalent to that which should be added in quadrature to the GES uncertainties to obtain the measured $T_{\rm eff}$ errors.  For the purposes of illustration, Fig.~\ref{one_target} shows the effects on the age estimates for single stars of an {\it additional} $\pm 200$ K error in $T_{\rm eff}$ (i.e. {\it net} $T_{\rm eff}$ uncertainties of $\sim 215$ K). The revised probability distributions are shown with a blue dashed line in each of the sub-plots. As expected, the additional $T_{\rm eff}$ uncertainty broadens the age probability distributions, but by very little in regions where the isochrones in Fig.~\ref{lithium_at3} are flat. Even for these large $T_{\rm eff}$ errors, any shift in the age estimates is much less than the uncertainties that were already present due to the intrinsic dispersion in EW$_{\rm Li}$.

\begin{figure}
	\includegraphics[width = 85mm]{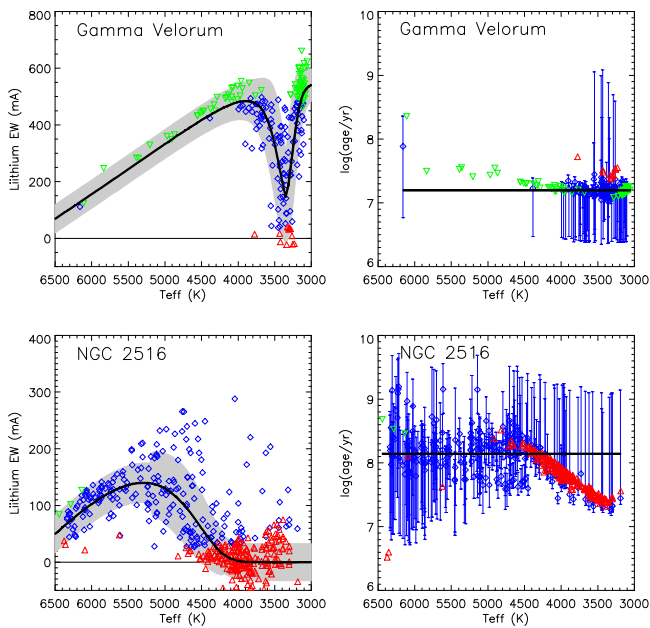}
	\caption{Left panels: Examples of the fits to the lithium depletion patterns of the Gamma Vel (top) and NGC 2516 clusters. The solid line shows the EW$_{\rm m}$ isochrone at the modelled most probable age of the cluster dataset. The shaded regions illustrate the modelled intrinsic dispersion $\pm\rho_{\rm m}$ at these ages. Blue diamonds mark stars for which an age and error bar were determined; red and green triangles mark those stars for which only (95 per cent) lower or upper limits to the age could be determined. Right panels: The corresponding ages determined for the individual stars in these clusters. The error bars on the blue diamonds are 68 per cent confident limits. The very narrow horizontal bands represent the overall most probable age and its 1-sigma uncertainty from the combined probability function (see \S\ref{3.2}).}
	
\label{bounds_targets}	
\end{figure}

\begin{figure*}
	\begin{minipage}[t]{1\textwidth}
	\includegraphics[width = 170mm]{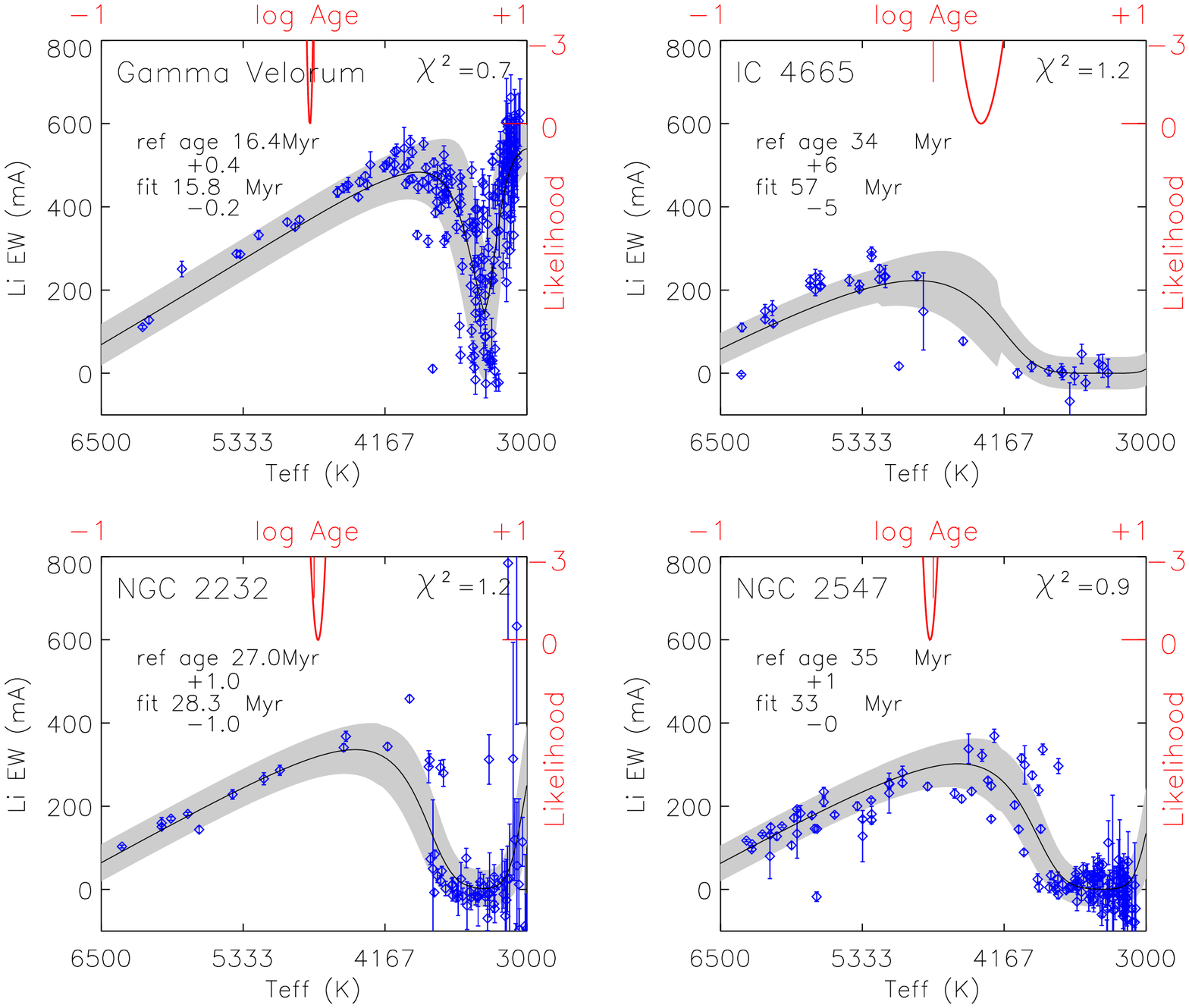} 
    \end{minipage}
	\caption{Four examples of best-fitting Model isochrones compared with the GES training data. Blue points show EW$_{\rm Li}$ as a function of $T_{\rm eff}$ (scales on the left hand and bottom axes). The black line shows the model value of EW$_{\rm Li}$ at the fitted most probable age (shown as text on the plot) and the shaded regions are the model intrinsic dispersion at the best-fit age or its upper limit. The reduced chi-squared value for the fit is shown in the top-right corner. The upper red paraboloid shows negative log likelihood (normalised to zero at the maximum-likelihood) as a function of $\log$~age relative to the age adopted in the training set (scales on the right hand and top axes). Equivalent plots for all the clusters in Table~\ref{ages} are given in Appendix~\ref{clusterplots} (available online as supplementary material).}
	\label{lithium_at5}	
\end{figure*}

\subsection{Estimating the ages of clusters and coeval groups}
\label{3.2}

The procedure described in \S\ref{3.1} can be adapted to estimating an age for groups of stars assumed to be coeval. $\mathscr{L}$ is estimated for each star over a uniform grid in age and the summed $\ln \mathscr{L}$ used to represent the overall log likelihood function of the cluster. A difference over the treatment of single, field stars is that we then assume a prior probability distribution that is flat in log age (i.e., $\ln p =0$ in Eqn.~\ref{prior}). The rationale for this is that older clusters are much rarer than younger clusters, such that the number of clusters $N(\log\ {\rm age}$) is roughly flat between 1\,Myr and 3\,Gyr in the \cite{Dias2021b} catalogue. The combined $P(\log\ {\rm age})$ can then be used to  estimate an age and uncertainties as for single field stars.

The summing of log likelihoods ensures that the contribution each star makes towards estimating the age of the cluster is appropriately weighted according to how precisely the age of the star could be determined and the shape of its likelihood function. 
The process is visualised in Fig.~\ref{bounds_targets} where we show the EW$_{\rm Li}$-$T_{\rm eff}$ relationship and the derived ages determined for individual stars in two clusters, Gamma Velorum and NGC~2516\footnote{We note that these clusters were themselves used to constrain the model. The derived ages  will be compatible with the mean age scale defined by all the clusters but the statistical uncertainties may be slightly  underestimated.}.

Gamma Velorum is a rich young cluster and our model yields a most probable cluster age of $15.8^{+0.4}_{-0.2}$ Myr. Some of the individual Li measurements yield reasonably well-determined ages because EW$_{\rm Li}$ is in transition from undepleted levels to being completely depleted. Stars with the highest EW$_{\rm Li}$ yield only upper limits to their age because their EW$_{\rm Li}$ is consistent with undepleted levels in the youngest clusters. There are also a small group of stars at the bottom of the Li dip at $T_{\rm eff} \sim 3300$\,K with very low EW$_{\rm Li}$ and where only age lower limits are found.
The tight cluster age constraints arise from a group of $\sim 50$ stars with small individual age error bars and $3200 < T_{\rm eff}/{\rm K}< 4000$, in the "sweet-spot" of Fig.~\ref{one_target}, but also from the tension between the age lower limits found for the almost entirely Li-depleted stars that define the "Li dip" and the age upper limits found in slightly warmer and cooler stars.

In contrast, NGC~2516 is an older cluster with a most probable age of $140^{+4}_{-4}$ Myr. Here there are a large group of low EW$_{\rm Li}$ stars at $T_{\rm eff} <4500$K that provide only age lower limits. The firm age constraints come mainly from the stars with moderate and high EW$_{\rm Li}$ at $4000 < T_{\rm eff}/{\rm K}< 5200$. An interesting feature seen here is a bimodality in the ages at $4500<T_{\rm eff}/{\rm K} < 5200$. This is caused by the additional intrinsic dispersion in EW$_{\rm m}$ that is introduced in Eqn.~\ref{delta_m} for ages around 100\,Myr. Depending on the measurement uncertainties, this can produce a local minimum in the likelihood around this age (see for example Fig.~\ref{one_target}d) and as a result the {\it most probable} age (for an individual star) transitions abruptly between a peak in the age probability function at $\sim 60$ Myr for higher EW$_{\rm Li}$ to an age of $\sim 150$\,Myr for lower EW$_{\rm Li}$. The median age transitions more smoothly, without this bimodality.  

\subsection{Comparing best fit ages to the training ages}
\label{3.3}

The procedure for estimating the age of a cluster was exercised on each of the training clusters in turn to examine the residuals between model predictions and training ages and to look for any trends or additional systematic error. A useful model-predicted age was obtained for 34 of the clusters such that the uncertainties in $\log$ age were below $\pm 0.3$ dex. Results for these clusters are listed in Table~\ref{ages} together with results for four well-populated young clusters where only a 95 per cent upper limit on age could be determined and two older clusters where a 95 per cent age lower limit is quoted. The remaining clusters had insufficient members for the process to provide a clear or usefully narrow peak in log likelihood.

Figure~\ref{lithium_at5} shows examples of fits to EW$_{\rm Li}$ as a function of $T_{\rm eff}$  on the left hand and lower axes, and $-\ln{\mathscr{L}}$ as a function of $\log{\rm age}$ on the right hand and upper axes (i.e. The minimum of the negative log likelihood marks the most likely age and has been normalised to zero)\footnote{Recall that for a prior that is flat in log age, the log likelihood is equivalent to $\ln P(\log$ age) (see Eqn.~\ref{prior}).}. The black curves show EW$_{\rm m}$ as a function of $T_{\rm eff}$ at the most probable age with text on the plot indicating the reference age (the "Mean" age in Table~\ref{clusters}) together with the best-fit age and its asymmetric 68 per cent confidence limits. The sharpness of the likelihood peak is directly related to the number of cluster members, their $T_{\rm eff}$ distribution and hence the extent to which EW$_{\rm Li}$ is a useful age indicator at the age of that cluster in Figs.~\ref{lithium_at3} and \ref{one_target}. Equivalent plots for all the clusters are included in Appendix~\ref{clusterplots} (available online as supplementary material).

The most probable values of $\log{\rm age}$ and their upper and lower 68 per cent bounds (or 95 per cent limiting values)  are listed in Table~\ref{ages}. Also shown in Table~\ref{ages} are the reduced $\chi^2$ values with respect to the isochrone at the most probable age. These range from 0.5 to 1.7 indicating that our modelling of the dispersion that is added to the measurement uncertainties provides reasonable goodness-of-fit statistics. For the clusters in Table~\ref{ages}, the probability distributions are sufficiently symmetric that the most probable and median ages are almost identical.

\begin{figure}
    \centering
	\includegraphics[width = 75mm]{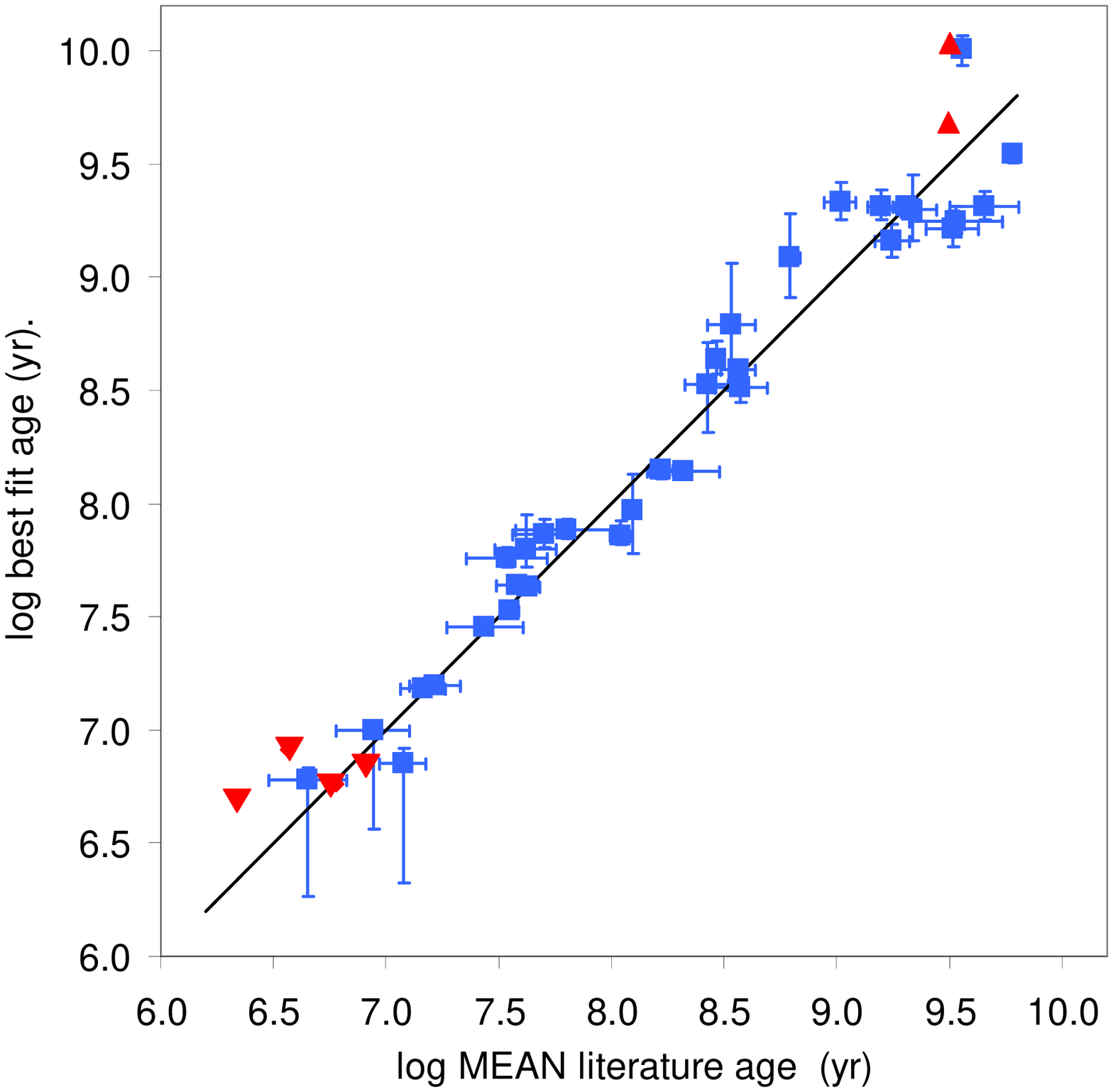} 
	\caption{Comparison of best fit ages predicted by our model with the assumed cluster age (i.e. the "Mean" training age in  Table~\ref{ages}). Blue squares show clusters with a resolved most probable age (see \S\ref{3.3}). Error bars show the RMS variation of $\log$ literature age and the asymmetric $\pm$68 percent limits of  best-fit age respectively (see Table~4). Red diamonds show ages of clusters where only a 95 per cent upper or lower limit could be determined.}
	\label{lithium_lit_ages}	
\end{figure}

Figure \ref{lithium_lit_ages} compares the model-predicted ages for individual clusters to the ages that were adopted for those clusters in the training data (i.e. the residuals between data and model). Both are listed in Table~\ref{ages}, where uncertainties in the training age are taken as the standard deviation of the ages drawn from the different sources in Table~\ref{clusters}. The RMS difference between the training data ages and best-fit ages of the 34 clusters with well-determined ages is 0.13~dex. This is comparable to the RMS variation of 0.11\,dex in $\log{\rm age}$ between the different age scales shown in Table \ref{clusters} for the same clusters.  Fitting the empirical model, EW$_{\rm m}$ to the full training set averages out local differences between the training age scale and best fit ages such that the two age scales align when averaged over the full set of clusters, with a mean offset 0.02$\pm$0.03\,dex and the slope of the least squares fit in Fig.~\ref{lithium_lit_ages} is 0.97$\pm$0.03. There are however indications that the model performs better for younger clusters than for older clusters. Several of the oldest clusters (e.g., NGC~2355, M67, NGC~6253, Berkeley 36) show significant discrepancies between training and best-fit ages, of a factor of two or more in either direction. These discrepancies and possible systematic errors in the model are discussed further in \S\ref{5.1}.

\subsection{Sensitivity to the training age scale}
\label{3.4}
\begin{figure}
	\includegraphics[width = 70mm]{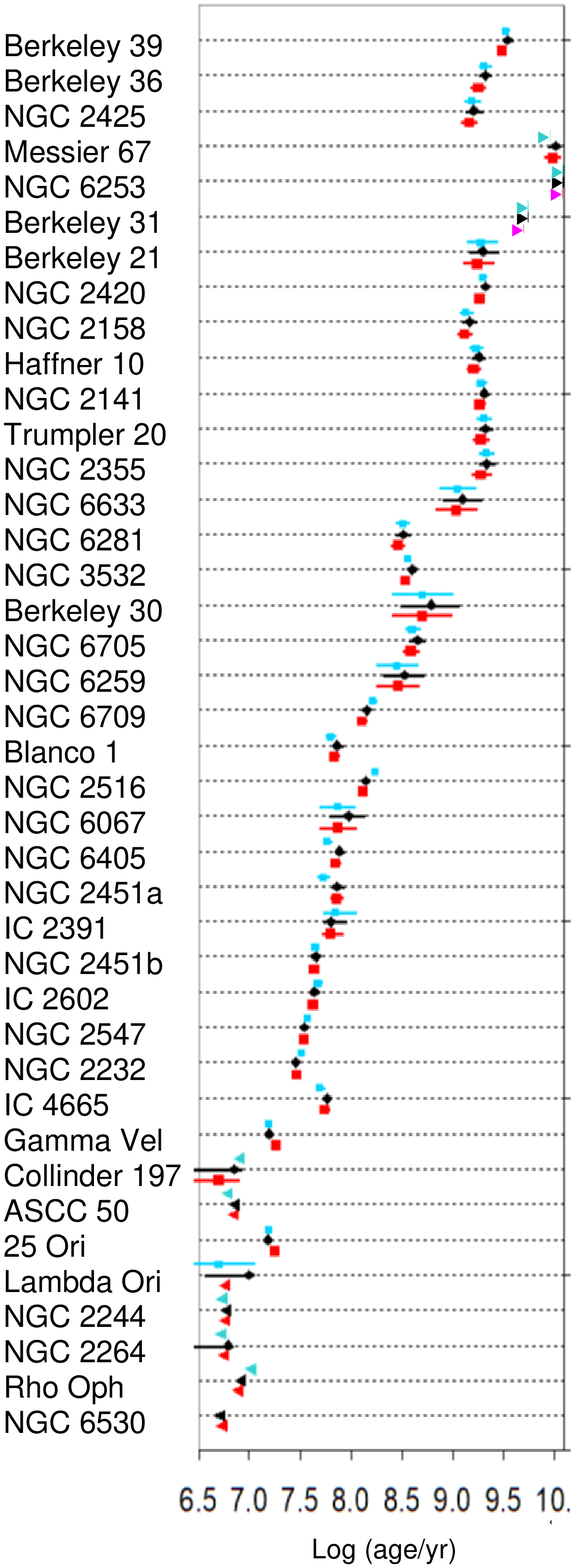} 
	\centering
	\caption{Comparison of EW$_{\rm Li}$ ages for GES clusters obtained using three different age scales for the training set. Black diamonds show EW$_{\rm Li}$ ages found using the MEAN age scale in Table 1. Red squares show results found using the LIT ages and the blue circles show results for the GES ages in Table 1. Error bars show the the 68 per cent limits on $\log{\rm Age}$ with triangles indicating 95 percent upper and lower bounds.}
	\label{compare_Li_ages}	
	\label{compare_ages}
\end{figure}

The ages derived from EW$_{\rm Li}$ are not absolute, they are scaled relative to the cluster ages adopted for the training data; if these are changed then so is EW$_{\rm m}$($\log$ age, $T_{\rm eff}$) and hence the ages determined for individual clusters. Figure~\ref{compare_Li_ages} shows the best fit ages calculated using three different scales to define the cluster ages of the training data: the LIT ages from \cite{Jackson2022a} and \cite{Franciosini2022a}; the GES age scale from \cite{Randich2022a}, and the geometric MEAN ages that we have adopted (see \S\ref{2.1} and Table\ref{clusters}).   

The RMS difference between the best fit ages calculated using the three age scales is 0.05\,dex. This is much lower than the RMS variation between the three age scales listed in Table~\ref{clusters} (0.11\,dex) indicating that whilst the various age scales show significant differences for individual clusters they are reasonably well aligned over the 5\,Myr--5\,Gyr age range of the training data indicating that the exact choice of training age scale has only a limited effect on the calculated values of best fit age and also that the sensitivity of the Li-derived ages to the ages adopted (or the errors in age) for individual clusters is very low, diluted by the large sample size.

\begin{table}
\caption{Training and most probable ages of GES clusters. Column 2 shows the mean $\log{\rm age}$ and it uncertainty from Table~\ref{clusters}, column 3 shows the best fit age (or limiting value) and column 4 shows the reduced $\chi^2$  of the best fit isochrone of EW$_{\rm m}$ relative to the the measured $EW_{\rm Li}$ using model values of dispersion in EW, $\rho_{\rm m}$.}

\centering
\begin{tabular}{llll} \hline
Cluster name	 & \multicolumn{2}{c}{$\log{\rm age/yr}$}&Reduced \\
                 &  training data & best fit value & $\chi^2$  \\
                 \hline
NGC 6530	&	6.34$\pm$0.36	&	<6.7	&	--\\
Rho Ophiuchus	&	6.57$\pm$0.12	&	<6.91	&	--\\
NGC 2264	&	6.65$\pm$0.18	&	6.78$^{+0.05}_{-0.52}$	&	1.24\\
NGC 2244	&	6.78$\pm$0.29	&	<6.76	&	--\\
Lambda Ori	&	6.94$\pm$0.16	&	7.00$^{+0.02}_{-0.44}$	&	1.73\\
25 Ori	&	7.17$\pm$0.10	&	7.18$^{+0.01}_{-0.01}$	&	1.12\\
ASCC 50	&	6.91$\pm$0.15	&	<6.85	&	--\\
Collinder 197	&	7.07$\pm$0.10	&	6.85$^{+0.07}_{-0.53}$	&	0.91\\
Gamma Velorum	&	7.21$\pm$0.11	&	7.20$^{+0.01}_{-0.01}$	&	0.71\\
IC 4665	&	7.54$\pm$0.18	&	7.76$^{+0.05}_{-0.04}$	&	1.25\\
NGC 2232	&	7.44$\pm$0.17	&	7.45$^{+0.02}_{-0.02}$	&	1.24\\
NGC 2547	&	7.55$\pm$0.04	&	7.53$^{+0.02}_{-0.01}$	&	0.93\\
IC 2602	&	7.63$\pm$0.06	&	7.63$^{+0.04}_{-0.03}$	&	1.04\\
NGC 2451b	&	7.58$\pm$0.09	&	7.64$^{+0.03}_{-0.03}$	&	1.53\\
IC 2391	&	7.62$\pm$0.14	&	7.80$^{+0.16}_{-0.08}$	&	0.53\\
NGC 2451a	&	7.70$\pm$0.13	&	7.86$^{+0.07}_{-0.06}$	&	1.46\\
NGC 6405	&	7.80$\pm$0.23	&	7.88$^{+0.05}_{-0.04}$	&	0.72\\
NGC 6067	&	8.09$\pm$0.01	&	7.97$^{+0.16}_{-0.2}$	&	0.5\\
NGC 2516	&	8.32$\pm$0.16	&	8.14$^{+0.02}_{-0.01}$	&	1.01\\
Blanco 1	&	8.04$\pm$0.05	&	7.86$^{+0.07}_{-0.04}$	&	0.9\\
NGC 6709	&	8.22$\pm$0.05	&	8.15$^{+0.04}_{-0.04}$	&	0.87\\
NGC 6259	&	8.42$\pm$0.10	&	8.53$^{+0.19}_{-0.21}$	&	1.17\\
NGC 6705	&	8.47$\pm$0.02	&	8.64$^{+0.08}_{-0.07}$	&	1.1\\
Berkeley 30	&	8.53$\pm$0.11	&	8.79$^{+0.27}_{-0.3}$	&	1.11\\
NGC 3532	&	8.56$\pm$0.08	&	8.59$^{+0.03}_{-0.02}$	&	0.89\\
NGC 6281	&	8.57$\pm$0.12	&	8.51$^{+0.07}_{-0.07}$	&	0.85\\
NGC 6633	&	8.79$\pm$0.04	&	9.09$^{+0.2}_{-0.18}$	&	0.49\\
NGC 2355	&	9.02$\pm$0.07	&	9.33$^{+0.09}_{-0.08}$	&	0.84\\
Trumpler 20	&	9.20$\pm$0.06	&	9.32$^{+0.07}_{-0.07}$	&	1.35\\
NGC 2141	&	9.33$\pm$0.12	&	9.30$^{+0.05}_{-0.04}$	&	1.3\\
Haffner 10	&	9.53$\pm$0.21	&	9.25$^{+0.07}_{-0.06}$	&	1.48\\
NGC 2158	&	9.25$\pm$0.08	&	9.16$^{+0.08}_{-0.07}$	&	1.16\\
NGC 2420	&	9.31$\pm$0.06	&	9.31$^{+0.04}_{-0.04}$	&	0.96\\
Berkeley 21	&	9.34$\pm$0.01	&	9.29$^{+0.16}_{-0.13}$	&	0.67\\
Berkeley 31	&	9.49$\pm$0.06	&	<9.69	&	--\\
NGC 6253	&	9.50$\pm$0.04	&	<10.04	&	--\\
Messier 67	&	9.55$\pm$0.02	&	10.01$^{+0.06}_{-0.07}$	&	1.61\\
NGC 2425	&	9.51$\pm$0.12	&	9.21$^{+0.08}_{-0.08}$	&	1.18\\
Berkeley 36	&	9.65$\pm$0.16	&	9.31$^{+0.07}_{-0.06}$	&	0.74\\
Berkeley 39	&	9.78$\pm$0.03	&	9.54$^{+0.04}_{-0.04}$	&	0.91\\

\hline
    \end{tabular}
\label{ages}
\end{table}

\subsection{Sensitivity to target temperature scales}
\label{3.5}

\begin{figure}
    \centering
	\includegraphics[width = 75mm]{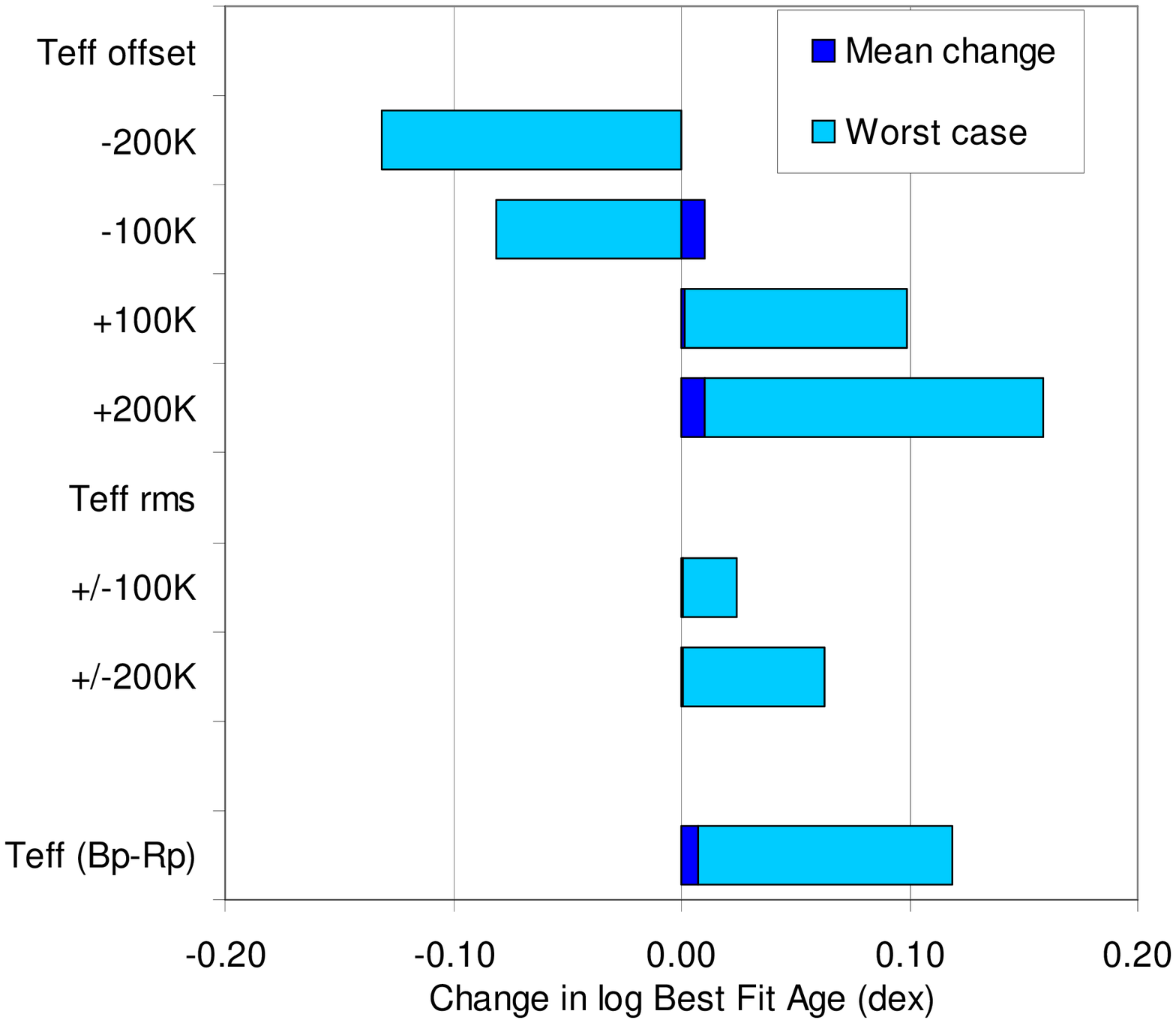} 
	\caption{Effects of changes stellar temperatures on estimated values of cluster age. Results are shown for a subset of 28 clusters from the training set with mean uncertainties in estimated age $<$0.2dex. The top four bars show the average and maximimum change caused by different levels of offset in T$_{\rm eff}$. The middle two bars show the effects of additional (Normally distributed) uncertainties in measured temperatures and the lower bar shows the effect of using T$_{\rm eff}$ values estimated from the $(G_{\rm Bp}-G_{\rm Rp})_0$ (see \S\ref{3})}.
	\label{teff_offsets}	
\end{figure}

In \S\ref{3.1} we showed that the effects of $T_{\rm eff}$ uncertainties are small, partly because some level of uncertainty is already incorporated when the empirical model was fitted to the training data, since the inherent uncertainties in $T_{\rm eff}$ contribute to the model dispersion, $\rho_{\rm Li}$ and partly because $\partial  {\rm age}/\partial T_{\rm eff}$ is small over most of the $EW_{\rm Li}/T_{\rm eff}$ plane. Nevertheless, the empirical model of EW$_{\rm m}$ was calibrated using the GES temperatures and if the model is used to estimate the age of other targets or clusters a systematic error will occur if the temperature scale is offset relative to that used in GES. The magnitude of this error will depend on where measured data lies in the EW$_{\rm Li}$ /T$_{\rm eff}$ plane. To illustrate the typical size of this error we selected 28 clusters that have an average Li-determined uncertainty in $\log{\rm age} < 0.2$\,dex and recalculated their ages for offsets in $T_{\rm eff}$ between $-200$\,K and $+200$\,K. The results shown in Fig.~\ref{teff_offsets} are that while the mean change in cluster age is small ($<$0.02\,dex) for offsets of $\pm 100$\,K, in the worst case cluster (i.e. when the distribution of EW$_{\rm m}/T_{\rm eff}$ in that cluster is most sensitive to an offset in temperature scale) the change in estimated age is $\pm 0.1$\,dex at $\pm 100$\,K, comparable with the RMS uncertainty in cluster age, rising to $\pm 0.15$ dex at $\pm 200$\,K, suggesting that offsets in the overall $T_{\rm eff}$ scale could become a significant source of systematic error if they are larger than $\pm 100$\,K from the GES scale.

Estimates of cluster age may also be sensitive to random uncertainties in $T_{\rm eff}$. In \S\ref{3.1}, we showed the effect on individual targets is mostly small if additional $T_{\rm eff}$ uncertainties are $<200$\,K.
The effects of introducing  additional (normally distributed) $T_{\rm eff}$ uncertainties of $100$\,K and $200$\,K on cluster ages are shown in Fig.~\ref{teff_offsets}. The impact is small indicating that any unmodelled additional uncertainty in measured $T_{\rm eff}$ of $<200$\,K  RMS will have only a limited effect on the estimated age even in the worst cases.

Finally we consider the case where there is no suitable temperature measurements available for  targets with reported EW$_{\rm Li}$ data. In this case a practical alternative is to estimate target temperatures using an empirical calibration curve of $\log T_{\rm eff}$ as a function of $\alpha = (G_{\rm Bp}-G_{\rm Rp})_0$ fitted to GES training data which yields, 
\begin{eqnarray}
\log T_{\rm eff} & = & 3.87456 -0.221311\, \alpha +  0.0676576\, \alpha^2 \nonumber\\
&& -0.0183638\, \alpha^3 +  0.00201383\,\alpha^4 \nonumber
\end{eqnarray}
for~$3500 < T_{\rm eff} < 4800$\,K and
\begin{equation}
\log T_{\rm eff} =  3.98419 -0.308435\, \alpha +  0.0452639\, \alpha^2
\label{BpRp2Teff}
\end{equation}
for~$4800 < T_{\rm eff} <6500$\,K.
Recalculating the cluster ages, using cluster reddening values from \cite{Jackson2022a}, results in offsets that are summarised in the lower bar in  Fig.~\ref{teff_offsets}. This shows that using $T_{\rm eff}$ values estimated from $G_{\rm Bp}-G_{\rm Rp}$ produces an average offset of 0.01\,dex in estimated age with a worst case value of 0.12~dex -- similar to the changes in $\log$ age caused by an offset of 100\,K in temperature scale.

\section{Application Examples}

\label{4}

\subsection{Estimating the ages of young exoplanets}
\label{4.1}

\begin{table*}

    \caption{The properties of the exoplanet host stars discussed in \S\ref{4.1} and their most probable ages (with 68 per cent confidence interval or a 95 per cent confidence lower limit, to 2 significant figures) and median ages estimated from the model in this paper. EW$_{\rm Li}$ values were corrected for a blend with a nearby Fe~{\sc i} line where necessary.}
    \centering
         \begin{tabular}{lcccccccc} \hline
  ID       & $T_{\rm eff}$ & EW$_{\rm Li}$ & Ref Age & Ref & Our Age & Median \\
           &    (K)        & (m\AA)        & (Myr)   &     & (Myr)   & (Myr) \\
\hline
 TOI-251   & $5875^{+100}_{-190}$          & $134 \pm 17$ & 40-320 & (1) &$230^{+290}_{-150}$ & 205 \\
 TOI-942   & $4928^{+125}_{-85}$           & $257 \pm 54$  & 20-160 & (1)& $57^{+69}_{-29}$ &65\\
 HD~110082  & $6200 \pm 100$                & $83 \pm 20$   & $250^{+50}_{-70}$& (2) & $ > 150$ & \\
 TOI-1807   & $4757^{+51}_{-50}$           &  $84.1 \pm 7.0$ &  $180 \pm 40$& (3) & $240^{+180}_{-90}$& 247         \\
 TOI-2076   & $5187^{+54}_{-53}$           &  $70.3 \pm 7.1$ &  $204 \pm 50$ & (3)   &  $610^{+740}_{-280}$ & 673\\
 KOI-7368   & $5241 \pm 100$               & $223\pm 15$ & $36^{+10}_{-8}$& (4) & $66^{+39}_{-33}$& 59\\
KOI-7913A  & $4324 \pm 70$                & $45 \pm 7 $ & $36^{+10}_{-8}$& (4) & $>180$ &\\
KOI-7913B  & $4038 \pm 70$                & $15 \pm 16$ & $36^{+10}_{-8}$& (4) & $>810$ &\\
Kepler 1643& $4916 \pm 110$               & $113 \pm 6$ & $46^{+9}_{-7}$ &(4) &  $220^{+100}_{-80}$& 205\\
HIP 94235  &  $5991 \pm 50$               & $141.3 \pm 9.2$ & 120 & (5) & $150^{+170}_{-100}$& 121\\
 \hline
 \multicolumn{7}{l}{(1) \cite{Zhou2021a}, (2) \cite{Tofflemire2021a}, (3) \cite{Hedges2021a}}\\
 \multicolumn{7}{l}{(4) \cite{Bouma2022a}, (5) \cite{Zhou2022a}} \\
% TOI-251   & $5875^{+100}_{-190}$          & $134 \pm 17, (3)$  & 40-320 &$227^{+294}_{-147}$ \\
% TOI-942   & $4928^{+125}_{-85}$           & $257 \pm 54$  & 20-160 & $57^{+69}_{-29}$ \\
% HD~110082  & $6200 \pm 100$                & $83 \pm 20$   & $250^{+50}_{-70}$ & $ > 150$ \\
% TOI-1807   & $4757^{+51}_{-50}$           &  $84.1 \pm 7.0$ &  $180 \pm 40$ & $235^{+179}_{-85}$         \\
% TOI-2076   & $5187^{+54}_{-53}$           &  $70.3 \pm 7.1$ &  $204 \pm 50$            &  $613^{+744}_{-276}$ \\
% KOI-7368   & $5241 \pm 100$               & $223\pm 15$ & $36^{+10}_{-8}$ & $66^{+39}_{-33}$\\
%KOI-7913A  & $4324 \pm 70$                & $45 \pm 7 $ & $36^{+10}_{-8}$ & $>178$\\
%KOI-7913B  & $4038 \pm 70$                & $15 \pm 16$ & $36^{+10}_{-8}$ & $>808$ \\
%Kepler 1643& $4916 \pm 110$               & $113 \pm 6$ & $46^{+9}_{-7}$ &  $215^{+99}_{-82}$\\
%HIP 94235  &  $5991 \pm 50$               & $141.3 \pm 9.2$ & 120 & $145^{+164}_{-99}$\\
% \hline      
   \end{tabular}
    \label{fieldstars_tab}
\end{table*}

\begin{figure}
	\includegraphics[width = 70mm]{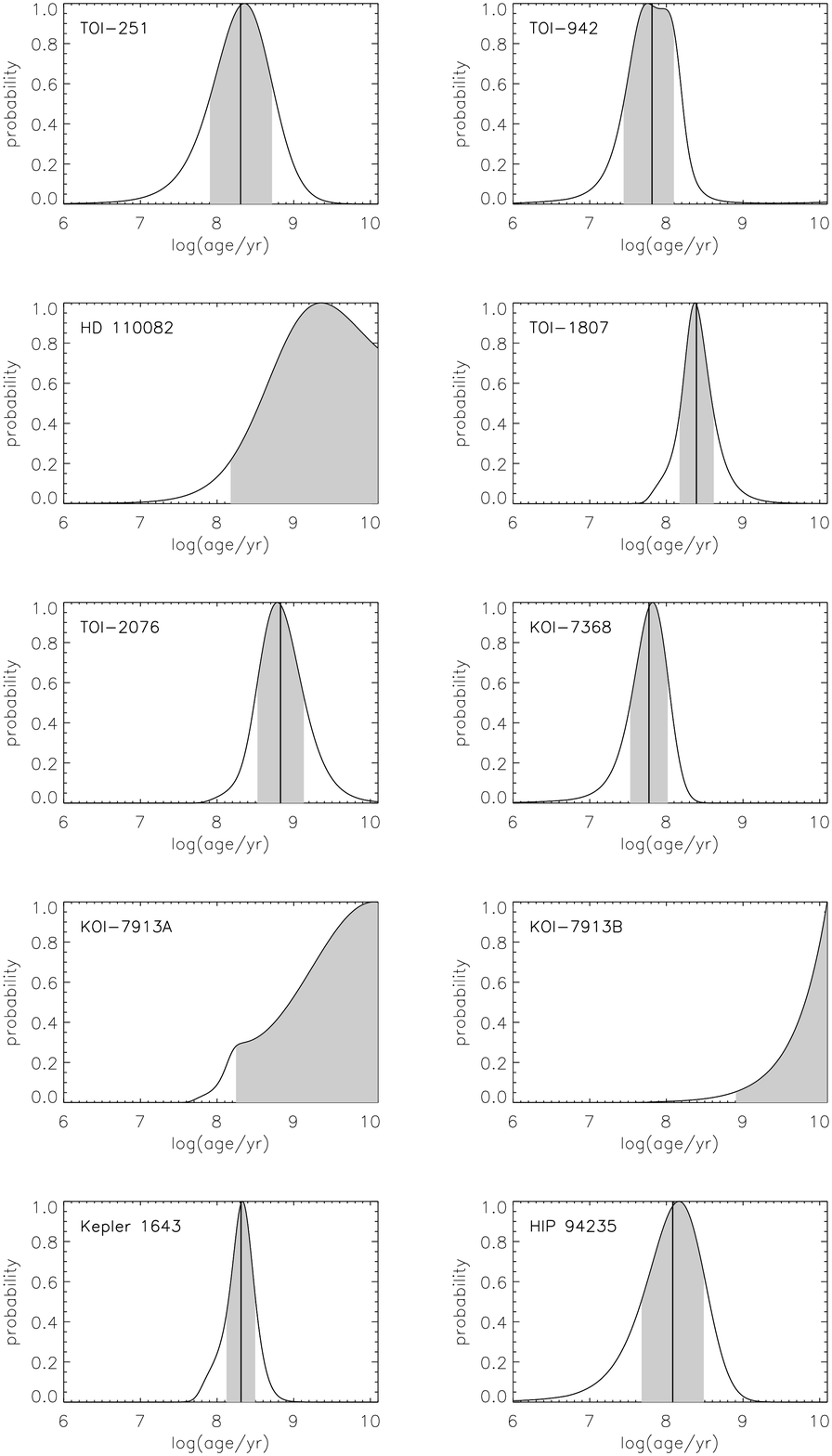} 
	\centering
	\caption{Posterior probability distributions of log age for the ten exoplanet hosts described in \S\ref{4.1}. Solid vertical lines show the median age with the grey area indicating $\pm$34 percent uncertainties around the most probable age or a 95 percent lower limit on age for targets where a most probable age could not be determined (\S\ref{3.1}).}
 \label{fieldstars_plot}
\end{figure}

Satellite surveys by Kepler and TESS have uncovered evidence for transiting exoplanets around stars in young clusters and associations as well as isolated field stars. Accurate age estimates are a crucial part of using these systems to test ideas and models for the early formation and evolution of planetary systems. The ages of isolated young field stars are usually estimated through some combination of isochronal fitting in the HR diagram, gyrochronology, magnetic activity indicators and Li depletion.
Here we provide independent Li-based age estimates and posterior age probability distributions for several examples as a demonstration of the capabilities of our model. An advantage to using our empirical model is that it can avoid some of the uncertainties in low-mass evolutionary models associated with magnetic activity and rotation (see~\S\ref{1}).  The results are summarised in Fig.~\ref{fieldstars_plot} and Table~\ref{fieldstars_tab}, where a flat prior in age is assumed. Where necessary, the published EW$_{\rm Li}$ was corrected for a blend with the Fe~{\sc i} 6707.44\AA\ line using $\Delta$EW$_{\rm Li} = 55.2 - 0.007825\,T_{\rm eff}$, derived from the relationship quoted in \cite{Soderblom1993a}.

\cite{Zhou2021a} report the discovery of transiting Neptune-sized exoplanets around TOI-251 and TOI-942. They report stellar parameters, EW$_{\rm Li}$ and ages of 40-320 Myr and 20-160 Myr respectively, based on the strengths of various magnetic activity indicators and rotation periods. The broad ranges are due to some disagreement (at the 1--3$\sigma$ level) between these diagnostics. Using a flat prior in age and no additional $T_{\rm eff}$ error, our model returns quantitative age estimates and uncertainties that are quite consistent with these, and probability distributions that are approximately normal in log age.

\cite{Tofflemire2021a} report the detection of a sub-Neptune-sized planet around the late F-star HD 110082 (TOI-1098), which itself has a co-moving wide M4 ($T_{\rm eff} = 3250$\,K) companion. The star was provisionally identified as a highly likely member of the Octans moving group (age $\sim 40$\,Myr), but its weak Li, slow rotation and the location of the M-dwarf companion in the HR diagram indicate an older age. An estimate of $250^{+50}_{-70}$ Myr was made by Tofflemire et al. based upon the rotation periods of a group of stars in the neighbourhood of HD~110082, dubbed ``MELANGE-1", that share similar space motions with HD~110082. We find that because of the high uncertainty, the EW$_{\rm Li}$ of HD~110082, after correcting for the blend with a nearby Fe~{\sc i} line, is below the threshold for assigning anything but a lower limit to its age. This lower limit is certainly incompatible with membership of the Octans moving group, but may be consistent with the gyrochronological age of MELANGE-1. Since no Li is expected to remain in M4 dwarfs at $>30$ Myr, a non-detection of Li in the M-dwarf companion would not provide tighter age constraints.

\cite{Hedges2021a} report transiting exoplanetary systems around the young K-dwarfs TOI-1807 and TOI-2076. Their ages are estimated from fitting the spectral energy distribution combined with a parallax and an age prior informed by their rotation periods. The strength of Li and magnetic activity indicators were simply said to be consistent with these estimates. Our estimates of age based on the published EW$_{\rm Li}$ are compatible in the case of TOI-1807, but significantly older for TOI-2076 where an age $<250$ Myr is ruled out with $>95$ per cent confidence. \cite{Hedges2021a} suggest that the two stars are coeval on the basis of their common space motions. Our analysis of the Li results argue this is still possible. If we treat the two stars as a ``cluster" (but with a prior that is flat in age), the combined age probability distribution gives a most probable age of $290^{+160}_{-90}$ Myr, with $\chi^2_r = 0.9$.

\cite{Bouma2022a} discuss ``mini-Neptunes" in three systems that have kinematic and spatial properties that place them within the RSG-5 and CH-2 aggregates in the young Cep-Her complex. The ages quoted arise from fitting the CMDs of these young populations and are supported by TESS-determined rotation periods and measurements of Li and chromospheric activity. For KOI-7368 our estimate of age based on its EW$_{\rm Li}$ is consistent with the published age. KOI-7913 is a wide binary with EW$_{\rm Li}$ measurements for both components. The corrected values are too small in either case to yield a well-constrained age and the flat prior leads to lower age limits that are incompatible with the ages cited by \cite{Bouma2022a}. Treating KOI-7913AB as a cluster of two stars yields a 95 per cent lower limit of $>205$ Myr. Kepler 1643 also has a low EW$_{\rm Li}$ if it were to be as young as $\sim 50$\,Myr and this was noted by \cite{Bouma2022a} who suggested their comparison samples were not large enough to fully explore the possible dispersion in EW$_{\rm Li}$ at 40-50\,Myr. Our age estimate also suggests this star is significantly older than the RSG-5 cluster and our comparison sample (the training data) is considerably larger.

\cite{Zhou2022a} presents evidence for a ``mini-Neptune" around the early-G dwarf HIP~94235. The star was kinematically assigned to the AB~Dor moving group (see \S\ref{4.2}) and given an age of 120\,Myr. The age we find from the published EW$_{\rm Li}$ (assuming no correction for any blends) is consistent with this age estimate, though the uncertainties are large.

\subsection{Estimating the age of moving groups and associations}
\label{4.2}

\begin{figure}
	\includegraphics[width = 85mm]{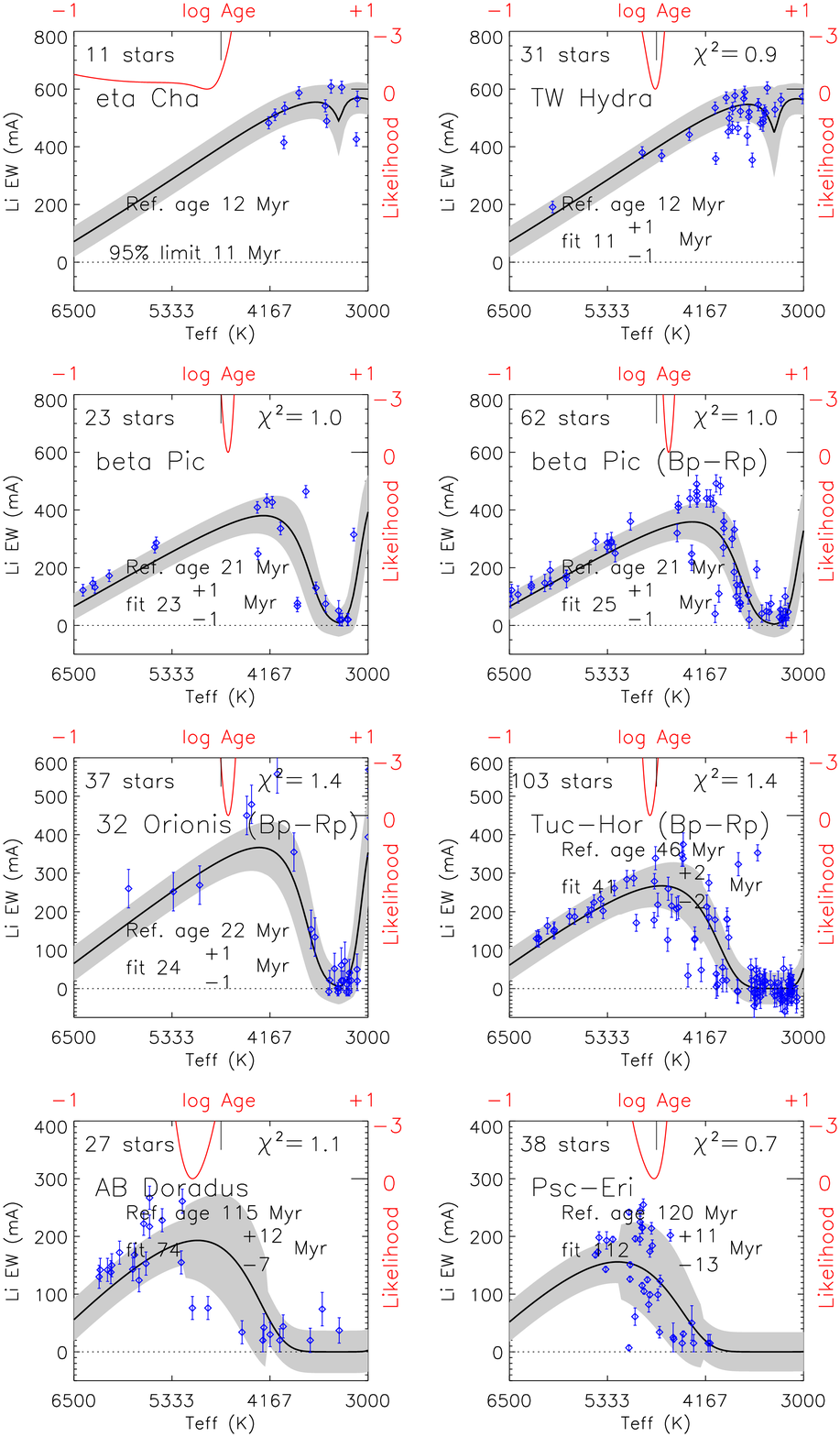} 
	\centering
	\caption{Lithium ages for moving groups and associations. Blue points show measured EW$_{\rm Li}$ as a function of $T_{\rm eff}$ with error bars indicating  the measurement errors. The black line shows the  model value EW$_{\rm m}$ at the best fit age and the grey filled region is the model estimate of the dispersion at that age. The upper red parabola shows the log likelihood as a function of log age relative to the reference age. Text on the plot shows the maximum likelihood age  and a 68 percent error bound as in Fig.~\ref{lithium_at5}. The number of data points and the reduced $\chi^2$ of the best fit isochrone to the measured data are also shown.} 
\label{moving_groups} 
\end{figure}

In the last few decades it has become well-known that the solar neighbourhood contains a large number of spatially dispersed young stars outside of obvious clusters and associations that can nevertheless be  assigned to kinematically coherent "moving groups" \citep[e.g.,][]{Zuckerman2004a, Malo2013a}. The members of many of these young groups were uncovered or confirmed by virtue of measurements of the lithium content of their members \citep[e.g.,][]{Jeffries1995a, Montes2001a, Guillout2009a, Frasca2018a} and these lithium measurements have often been used to estimate an age as an adjunct to other methods or as a semi-independent way to investigate evolutionary models. Some of the members of these moving groups are now known to host young exoplanets, adding impetus to the need for precise ages \citep{Newton2019a, Newton2021a, Zhou2022a}.

The empirical model developed here and the procedure described in \S\ref{3.2} can readily be used to constrain the age of small stellar associations with measured EW$_{\rm Li}$, placing them precisely onto the age scale defined by the GES clusters. To demonstrate how this works in practice we have used reported values of EW$_{\rm Li}$ for a non-exhaustive list of seven well-documented associations/moving groups to estimate their ages. Plots in Fig.~\ref{moving_groups} show EW$_{\rm Li}$ as a function of $T_{\rm eff}$, the best fitting empirical isochrone and the log likelihood distribution, assuming a flat prior in log age (see \S\ref{3.2}) in a similar way to Fig.~\ref{lithium_at5}.

\begin{itemize}
    \item {\bf $\eta$ Chamaeleontis moving group (Eta Cha)} is the youngest association sampled. EW$_{\rm Li}$ and T$_{\rm eff}$ data is taken from table~3 of \cite{Mentuch2008a}. Visually the  plot shows little or no evidence of lithium depletion, consequently  only an upper limit on age 11\,Myr can be resolved. This is compatible with the reference age of $12\pm$6\,Myr given my \cite{Mentuch2008a}.
    
    \item {\bf TW Hydrae association (TW Hya)} is still relatively young but maybe showing the first signs of lithium depletion amongst lower temperature targets. Data from table~2 of \cite{Mentuch2008a} yields a best fit age of 11$^{+1}_{-1}$\,Myr. This is similar to, but much more precise than  the age of $12\pm$8\,Myr given by \cite{Mentuch2008a}.
    
    \item {\bf $\beta$ Pictoris moving group (Beta Pic)} shows a fully developed ``Lithium dip" in its M-dwarfs, indicating that it is older than Eta Cha and TW Hya. Two plots are shown in Fig.~\ref{moving_groups} for Beta Pic: the first shows results using 23 targets from table~4 of \cite{Mentuch2008a}, which gave an age of $23^{+1}_{-1}$\,Myr; the second includes additional data  from table~A1 of \cite{Messina2016a} which presented EW$_{\rm Li}$ for $\sim$60 targets but no measurement uncertainties or temperature data. For the second plot $T_{\rm eff}$ data were estimated from $(G_{\rm Bp}-G_{\rm Rp})$ colour where this was available (see \S\ref{3.5} and equation \ref{BpRp2Teff}) and assuming an EW measurement uncertainty of 30\,m\AA~(where none was given) and zero reddening. This yields an age of $25^{+1}_{-1}$\,Myr which agrees within 1$\sigma$ with the age derived for the smaller sample and is comparable with the ``lithium depletion boundary" age of 17.4 to 24.3\,Myr given by \cite{Binks2014a} and \cite{Galindo-Guil2022a}. 
    
    \item {\bf 32 Oronis moving group} (32 Ori) shows a similar lithium dip to $\beta$~Pic. EW$_{\rm Li}$ data for 37 targets (including 19 upper limits) were taken from table~3 of \cite{Bell2017a}. Temperatures were estimated from $(G_{\rm Bp}-G_{\rm Rp})$ colour with zero reddening. Upper limits were assigned an EW$_{\rm Li}$ (corrected for the Fe line blend) and uncertainty equal to half the limiting value. A relatively high uncertainty of 50\,m\AA~was assumed for targets that were not upper limits, giving $\chi^2_r=2.0$ for the fit at an estimated age of $24^{+1}_{-1}$\,Myr. This matches a reported LDB age of 18--25\,Myr  \citep{Galindo-Guil2022a}. The high $\chi^2_r$ perhaps indicates that the uncertainties are even larger than 50\,m\AA, that the list of ``members" is contaminated or that the intrinsic dispersion in EW$_{\rm Li}$ is larger in this group than our model predicts.
    
    \item {\bf Tucanae-Horologium association (Tuc-Hor)} shows a broader lithium dip suggesting it is older than the previous groups. EW$_{\rm Li}$ data for 103 targets were taken from table~2 of \cite{Kraus2014a}, for which we assume a fixed uncertainty of 30\,m\AA. Temperatures were estimated from $(G_{\rm Bp}-G_{\rm Rp})$ colour with zero reddening. The estimated age of Tuc-Hor is $41\pm 2$\,Myr, which matches the reported LDB age of 41--51\, Myr given by  \cite{Galindo-Guil2022a}.
    
    \item {\bf AB Doradus moving group} (AB Dor) shows a broad lithium dip with early M-dwarfs appearing fully depleted. EW$_{\rm Li}$ and $T_{\rm eff}$ data from table~6 of \cite{Mentuch2008a} are poorly distributed, with few data points near the higher mass end of the lithium dip. Consequently the estimated age of $74^{+12}_{-7}$\,Myr is relatively imprecise compared to the other associations in Fig.~\ref{moving_groups}. The estimated age is somewhat lower than the age of 100--125Myr given by \cite{Luhman2005a} based on the similarity of the AB Dor colour-magnitude diagram to the Pleiades.
    
    \item {\bf Psc-Eri stream} (Psc-Eri) was discovered by \cite{Arancibia-Silva2020a}, who reported EW$_{\rm Li}$ and $T_{\rm eff}$ (plus other properties) of targets in the range $4000 < T_{\rm eff} < 5500$\,K. The results  shown in the final plot of Fig.~\ref{moving_groups} lie mainly to the high mass side of the lithium dip but are sufficient to provide an  estimate of age of $112^{+11}_{-13}$\,Myr, which is consistent with a reference age of 100 to 125\,Myr  based on the similarity of Psc-Eri to the Pleiades in terms of gyrochronology and the CMD of its higher mass stars \citep{Curtis2019a}.
    
\end{itemize}

\subsection{Application to clusters/data outside the Gaia-ESO Survey}
\label{4.3}

\begin{figure}
	\includegraphics[width = 85mm]{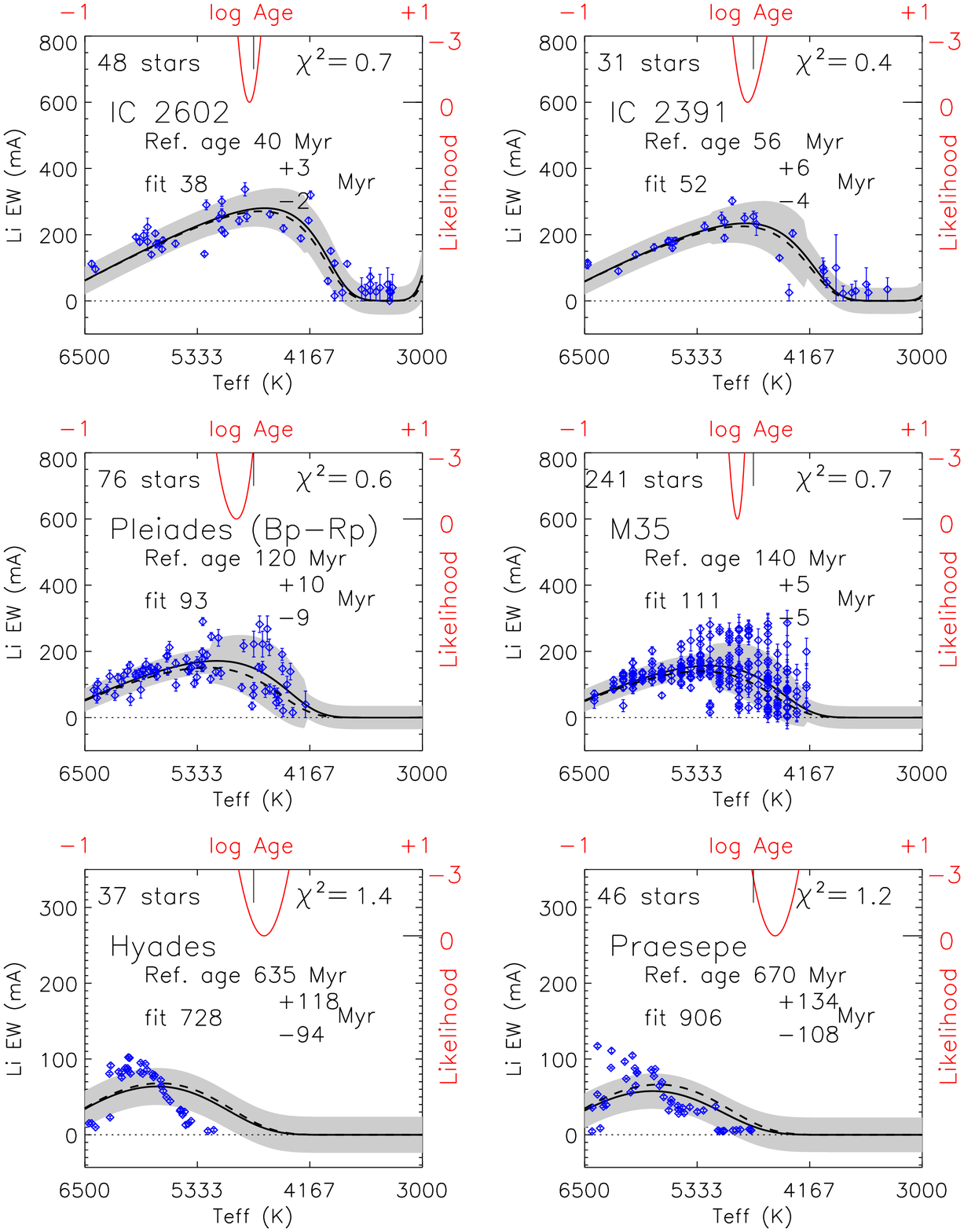} 
	\centering
	\caption{Lithium ages for clusters/data that were not part of GES or of the training dataset. Details are the same as for Figs.~\ref{lithium_at5} and~\ref{moving_groups}. A dashed line marks a model isochrone calculated at the reference ages given in each panel.} 
\label{nonges_clusters} 
\end{figure}

 A further validation of the techniques and adopted age-scale could come from applying the lithium-dating methods to data or clusters which were {\it not} part of GES. Figure~\ref{nonges_clusters} shows EW$_{\rm Li}$ as a function of $T_{\rm eff}$, the best fitting empirical isochrone and the log likelihood distribution, assuming a flat prior in log age, for six well-known clusters with data obtained from the literature.

\begin{itemize}
    \item {\bf IC~2391 and IC~2602:} This pair of clusters were part of the GES training dataset but with relatively sparse data (see Fig.~\ref{all_clusters}), particularly in the case of IC~2391. The first two panels of Fig.~\ref{nonges_clusters} show the results of fitting EW$_{\rm Li}$ and $T_{\rm eff}$ data taken from \cite{Randich2001a}. The best-fitting ages are consistent with those obtained from the GES data listed in Table~\ref{ages}, but in the case of IC~2391, considerably more precise ($63^{+27}_{-10}$ Myr vs $52^{+6}_{-4}$ Myr) thanks to the increased number of stars in an age-sensitive part of the EW$_{\rm Li}$-$T_{\rm eff}$ plane.
    
\item{\bf Pleiades and M35:} Several lithium datasets exist for the Pleiades \citep[e.g., see][]{Barrado2016a}, which has an age variously estimated between about 100\,Myr and 125\,Myr. A reference age of $\sim 120$\,Myr is cited here, based chiefly on the model-insensitive lithium depletion boundary age of $118^{+6}_{-10}$\,Myr \citep[][see \S\ref{5.1}]{Galindo-Guil2022a}. The data fitted in Fig.~\ref{nonges_clusters} are those Pleiades candidates judged to be single members by \cite{Barrado2016a}. The EW$_{\rm Li}$ values (already corrected for the Fe~{\sc i} blend) and uncertainties are from \cite{Soderblom1993a}, with $T_{\rm eff}$ calculated using Eqn.~\ref{BpRp2Teff}, and Gaia DR3 $G_{\rm Bp} - G_{\rm Rp}$ values that were dereddened according to the $E(B-V)$ values given in \cite{Soderblom1993b} and $E(G_{\rm Bp} - G_{\rm Rp}) = 1.34 E(B-V)$ \citep{Casagrande2018a}. The best-fitting age of $93^{+9}_{-10}$ Myr is in reasonable agreement with the reference age.

M35 is thought to have an age a little older than the Pleiades \citep[see][and references therein]{Anthony-Twarog2018a}, independently confirmed by 
rotation periods \citep{Jeffries2021a}, and a reference age of $140 \pm 15$ \,Myr is adopted. EW$_{\rm Li}$ and $T_{\rm eff}$ (based on spectral energy distribution fits) are from \cite{Jeffries2021a} and give a best-fitting age of $111 \pm 5$\, Myr.

The relative ages of the Pleiades and M35 are consistent with inferences from the Hertzprung-Russell diagram and the rotation periods of their low-mass stars but taken together, the results perhaps suggest a 0.1 dex discrepancy with the absolute (log) age scale defined by the GES clusters and lithium measurements.

%It is worth noting that we also tried including stars with EW$_{\rm Li}$ in \cite{Soderblom1993a} but that were classed as {\it binary} members in \cite{Barrado2016a}. The justification for this is that although a significant fraction of spectroscopic binaries will have been removed by the kinematic membership selection \citep[see discussion in][]{Jackson2022a} wider binaries will likely remain in the GES training dataset and may be responsible for some of the intrinsic dispersion in the model. With the binaries included the best-fitting Pleiades age falls to $78^{+7}_{5}$ Myr. This may indicate some issue in measuring Li in the binaries or possibly that members of binary systems in the Pleiades tend to be faster rotating \citep{Stauffer2016a} and it is known that faster rotating Pleiads have more Li than slower rotators at temperatures $T_{\rm eff}<5400$\,K \citep{Bouvier2018a}; their inclusion might therefore lead to a lower age estimate.

\item {\bf Hyades and Praesepe:} \cite{Cummings2017a} provide high quality EW$_{\rm Li}$ and $T_{\rm eff}$ data for samples of cool stars in both of these older clusters. Reference ages 
for the Hyades and Praesepe \citep[also from][]{Cummings2017a} are 635 Myr and 670\,Myr respectively. The best-fitting ages from our model are $728^{+118}_{-94}$\,Myr and $906^{+134}_{-108}$ Myr. The relatively small discrepancies between the best-fitting and reference ages suggest the model is doing reasonably well but Fig.~\ref{nonges_clusters} shows that the fits are actually quite poor in detail, with significant and systematic residuals between model and the high quality data. The cooler G-stars are more depleted than the model at either the best-fitting age or reference age. Conversely, the warmer F- and G-stars are less depleted than the model. If the cluster samples had consisted of only the warmer stars ($T_{\rm eff}>5700$\,K) then the best-fitting ages would have been significantly smaller than the reference ages by $\sim 250$ Myr. Whether this discrepancy could be attributed to differences in metallicity - both the Hyades and Praesepe are metal-rich compared with most of the GES clusters - is discussed in \S\ref{5.1}.

\end{itemize}

\section{Discussion}
\label{5}

\subsection{Systematic uncertainties}
\label{5.1}

In \S\ref{3.3} the best-fit lithium-derived ages were compared with the training ages. At young ages there was excellent agreement, but at older ages the agreement was worse, with some significant discrepancies - notably NGC~2355, M67, NGC~6253 and Berkeley 36. It is possible that other factors, besides age and $T_{\rm eff}$, might determine the photospheric Li content of a star and lead to systematic uncertainties in the ages predicted from our simple model. 

\subsubsection{Metallicity dependence}

\label{metallicity}
\begin{figure}
	\includegraphics[width = 85mm]{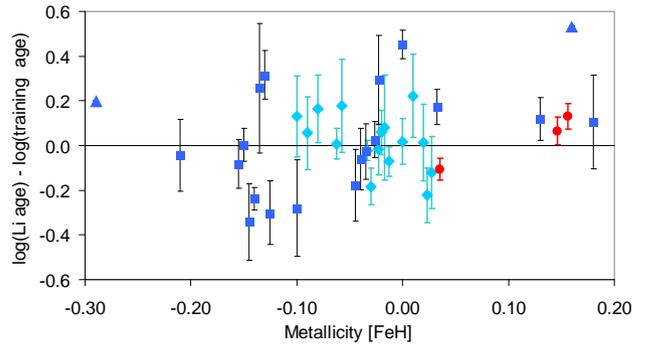} 
	\centering
	\caption{Variation of the difference between the maximum likelihood age and the training data age with metallicity. Light blue diamonds show results for clusters aged $<$200\,Myr, dark blue squares show results for older GES clusters (the triangles are clusters where we determined age lower limits). Red circles show results for three additional non-GES clusters, the Pleiades, the Hyades and Praesepe (see Fig.\ref{nonges_clusters} and \S\ref{4.3}).}  
\label{Lithium_FEH}
\end{figure}

The metallicity of a star could or should play some role in determining the amount of lithium depletion at a given age. Higher metallicity stars have deeper convection zones at a given $T_{\rm eff}$ and this should lead to more depletion during the PMS phase \citep{Pinsonneault1997a, Piau2002a, Tognelli2012a, Tognelli2021a} and possibly enhanced depletion rates during the main sequence phase, depending on which mixing mechanisms operate \citep{Chaboyer1995a}. Increased Li-depletion with metallicity would lead to over- or under-estimates of age from our model for high- and low-metallicity clusters respectively. Counterbalancing this to some extent, the adoption of a fixed template spectrum in the EW determination in \S\ref{2.2}, at some median field-star metallicity, might lead to a small over-estimate of EW$_{\rm Li}$ by a few m\AA\ and a lower age estimate for higher metallicity stars due to the blended Fe~{\sc i} line. The training dataset was drawn from clusters spanning a relatively narrow range of metallicities $-0.29<[{\rm Fe/H}]<0.18$ in the hope of minimising such effects and avoiding the added complexity of introducing a metallicity dependence into the model for EW$_{\rm m}$. 

\begin{figure}
	\includegraphics[width = 85mm]{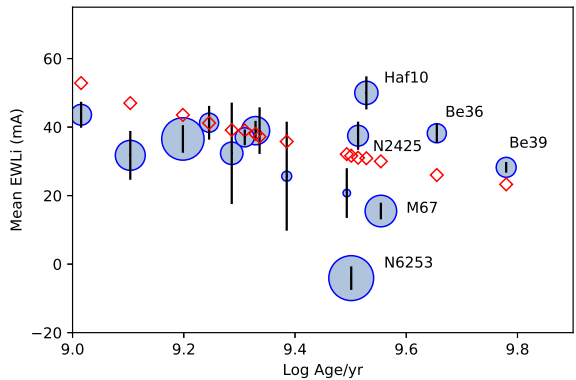} 
	\centering
	\caption{Weighted mean EW$_{\rm Li}$ for stars {\bf in the temperature range 5600 to 6200\,K} in clusters with training ages greater than 300 Myr (using mean ages in Table~\ref{clusters}). The circle sizes scale linearly with [Fe/H] and the error bars are the standard error in the weighted mean. The red diamonds show the mean EW$_{\rm Li}$ predicted by the best-fitting model, for those stars, at the training ages.}  
\label{Lithium_FEH_2}
\end{figure}

Figure~\ref{Lithium_FEH} shows the difference between the model-predicted age and training age, as a function of [Fe/H], for the clusters in Table~\ref{ages}. The majority of clusters (and all of the young, $<200$ Myr, clusters) are concentrated in a small range of metallicity ($-0.15 <$ [Fe/H] $< 0.05$) . However, the narrow distribution is resolved, since the mean cluster metallicities were determined homogeneously as part of GES and have typical internal precisions of $\sim 0.01$--$0.02$ dex \citep{Randich2022a}.

Figure~\ref{Lithium_FEH} exhibits significant scatter but no significant overall trend with metallicity; the average age difference is $0.01 \pm 0.03$\,dex. The results for the three highest metallicity (older) clusters do hint at the expected increase in the Li-determined age relative to the training age with increasing [Fe/H]. Also shown in Fig.~\ref{Lithium_FEH} are the additional results for three of the clusters considered in \S\ref{4.3}\footnote{IC~2391 and IC~2602 are already represented in the diagram using GES data. The metallicity of M35 is still uncertain \citep[see][]{Jeffries2021a}.} The first of these, the Pleiades, has [Fe/H]$=0.03 \pm 0.02$ \citep{Soderblom2009a} and sits at the metal-rich end of the $<200$ Myr cluster subset but has a model-predicted age about 0.1 dex {\it younger} than its training age. Conversely, the Hyades and Praesepe, which have a supersolar [Fe/H] of $0.146\pm 0.004$ and $0.156\pm 0.004$ respectively \citep{Cummings2017a}, demonstrate the positive residuals expected if higher metallicity led to increased Li depletion. However, Fig.~\ref{nonges_clusters} reveals a more complex picture -- the shape of the empirical model does not match the data in these clusters well. Whilst stars with $T_{\rm eff} < 5700$\,K show increased depletion with respect to the model at the reference age, which could be the signature of enhanced PMS depletion, the warmer stars have much more lithium than predicted. This might be explained if these clusters have a higher initial Li abundance commensurate with their higher metallicity, since curves of growth \citep[e.g.,][]{Franciosini2022a} indicate that an increase of 0.15 dex in Li abundance at $T_{\rm eff} \sim 6000$\,K would result in an EW$_{\rm Li}$ increase of $\sim 20$ per cent.

The lack of any strong dependence of the Li-based age estimates on metallicity is somewhat surprising given the very strong dependence of Li depletion on metallicity in theoretical models, particularly for the younger clusters where PMS Li depletion is expected to dominate. A caveat to this is the relatively narrow [Fe/H] range covered by the young training clusters and it is clear that the model is not a good fit to the EW$_{\rm Li}$ versus $T_{\rm eff}$ distributions of the metal-rich Hyades and Praesepe that are not part of the training data. An exploration of this and a detailed comparison with theoretical models is deferred to another paper, but for the purposes of empirical age estimation it appears that, at least in the metallicity range $-0.29 < [{\rm Fe/H}] < 0.18$ (or $-0.15 < $ [Fe/H] $<0.05$ for clusters with age $< 1$ Gyr), there is no compelling evidence to introduce a metallicity dependence.

\subsubsection{Systematic errors at older ages}

For clusters with training ages younger than 1 Gyr the weighted mean discrepancy between the Li-determined age (where it has been resolved) and training age is $0.00 \pm 0.02$ and with $\chi^2_{r}=1.0$,  indicating that uncertainties (in both ages) can explain the dispersion.   In contrast, the equivalent statistics for the 11 older clusters in Fig.~\ref{Lithium_FEH} are a weighted mean difference of $0.01 \pm 0.08$ with $\chi^2_{r}=8.7$. This clearly demonstrates that the scatter is much larger than the error bars and suggests that, even if metallicity is not directly implicated, there must be other parameters affecting the progress of Li depletion and thus any age determination based on Li depletion. 

Figure~\ref{Lithium_FEH_2} shows the mean EW$_{\rm Li}$ for stars with $5600 < T_{\rm eff}/K < 6200$ in clusters with age $>1$ Gyr, together with the mean EW$_{\rm m}$ that is expected at their training ages. All clusters with $\geq 5$ stars in this range are shown\footnote{This includes Pismis~15, Czernik~24 and Berkeley~22 (see Table~\ref{clusters}), which had poorly constrained ages.}. The data in this $T_{\rm eff}$ range are chiefly what determines the Li-based age for these older clusters. A single, monotonically declining relationship between EW$_{\rm Li}$ and age, as mandated by the functional form adopted in Eqn.~\ref{EW_m}, cannot match these data and comparison with Table~\ref{ages} reveals the expected trend that where the mean EW$_{\rm Li}$ is greater than the expected EW$_{\rm m}$ then the Li-based age is younger than the training age and vice-versa.
In fact, the evidence from the data is rather weak that EW$_{\rm Li}$ falls at all beyond 1 Gyr in these stars, with low Li seen only in M67 and NGC~6253. This appears to contradict work done on ``solar twin" field stars, where an order of magnitude decline in Li abundance is inferred between 2 and 8\,Gyr \citep{Carlos2019a}.

These problems were first noted by \cite{Sestito2005a} and \cite{Randich2010a} \citep[see also][]{Randich2021a} who suggested that for solar-type stars in older clusters there might be a dichotomy beyond 1 Gyr, with some showing high levels of Li depletion while in other stars the depletion might slow and plateau at $A$(Li)$\sim 2$ \citep[corresponding to EW$_{\rm Li} \simeq 20$--$30$\,m\AA\ at this $T_{\rm eff}$,][]{Franciosini2022a}. In some clusters this appears to result in a significant ($\sim$ 1 dex) dispersion and lower mean Li abundance among their solar type stars \citep[e.g.,][]{pasquini2008a, Pace2012a}, whilst in others only the less-depleted population exists and the mean Li abundance is higher. The reasons for this behaviour in the older clusters are not yet established. It appears not to be (solely) due to metallicity and it is unlikely to be due to differences in initial Li abundance -- the interstellar Li abundance in the solar vicinity seems quite uniform among young clusters \citep{randich2020a} and constant in time over at least the last 5 Gyr, since the initial solar Li abundance found in meteorites is similar to that in very young clusters. An alternative scenario might link the differences in Li depletion to differences in (internal) rotational evolution and mixing \citep{Charbonnel2005a, Eggenberger2012a, Baraffe2017a}.

\subsubsection{Comparison with lithium depletion boundary ages}
\begin{table}
\caption{Comparison of Lithium Depletion Boundary (LDB) ages of clusters and moving groups with the training and best-fit ages in Table~\ref{ages} and \S\ref{4.2}.} 
\centering
\begin{tabular}{lllll} \hline
  &  \multicolumn{2}{c}{LDB ages (ref)}  &  Training & Best fit  \\
  Clusters        & \multicolumn{4}{c}{log (age/yr)}   \\
\hline          
NGC\,2232	& $7.57^{+0.02}_{-0.03}$ (1) & --  & $7.44 \pm 0.17$ & $7.45^{+0.02}_{-0.02}$ \\	
IC\,4665	&	$7.40^{+0.06}_{-0.07}$ (2)& $7.37^{+0.01}_{-0.01}$ (3) &$7.54\pm 0.18$ & $7.76^{+0.05}_{-0.04}$\\					
NGC\,2547	&	$7.55^{+0.04}_{-0.04}$ (2) & $7.53^{+0.02}_{-0.04}$	(3)	& $7.55 \pm 0.04$ &	$7.53^{+0.02}_{-0.02}$ \\	
IC\,2602	&	$7.60^{+0.04}_{-0.04}$ (2)	& $	7.63^{+0.02}_{-0.04}$ (3) &	$7.63 \pm 0.06$ & $7.63^{+0.04}_{-0.03}$ \\	
IC\,2391	& $7.69^{+0.03}_{-0.04}$ (2) &  $7.67^{+0.01}_{-0.01}$ (3)	& $7.62\pm 0.14$ & $7.80^{+0.16}_{-0.08}$ \\	
Blanco\,1	& $8.10^{+0.07}_{-0.09}$ (2)	& $8.10^{+0.03}_{-0.13}$ (3) &	$8.04 \pm 0.05$& $7.86^{+0.07}_{-0.04}$ \\	
Pleiades    & $8.10^{+0.05}_{-0.06}$ (2)   & $8.07^{+0.03}_{-0.04}$ (3) & -- & $7.97^{+0.05}_{-0.05}$ \\  
\hline
Groups \\
\hline
Beta Pic  &   $7.31^{+0.06}_{-0.08}$ (2) & $7.24^{+0.01}_{-0.01}$ (3) & -- & $7.36^{+0.02}_{-0.02}$ \\
32 Ori    &   $7.34^{+0.07}_{-0.06}$ (4) & $7.25^{+0.02}_{-0.01}$ (3) & -- & $7.34^{+0.02}_{-0.02}$ \\
Tuc-Hor   &   $7.62^{+0.01}_{-0.02}$ (5) & $7.62^{+0.01}_{-0.01}$ (3) & -- & $7.62^{+0.03}_{-0.02}$ \\
\hline
\multicolumn{5}{l}{(1) \cite{Binks2022a}, using the \cite{Baraffe2015a} models.} \\
\multicolumn{5}{l}{(2) \cite{Soderblom2014a}, using the \cite{Chabrier1997a} models.} \\
\multicolumn{5}{l}{(3) \cite{Galindo-Guil2022a}, using the \cite{Tognelli2015a} models.}\\
\multicolumn{5}{l}{(4) \cite{Bell2017a}, using the \cite{Baraffe2015a} models.}\\
\multicolumn{5}{l}{(5) \cite{Kraus2014a}, using the \cite{Chabrier1997a} models.}
\end{tabular}
  \label{LDB}
\end{table}
Agreement between training ages and best-fit Li-age is much better for the younger clusters, with modest scatter. There is also excellent agreement (\S\ref{4.2}) between the derived and literature ages for several young moving groups. However, it is still possible that the overall age scale adopted might be offset uniformly from the truth. Whilst the older clusters have ages that are rather homogeneously determined from isochronal fitting to stars near the turn-off in the HR diagram, the ages for the younger clusters, reported by \cite{Jackson2022a} and \cite{Randich2022a}, are more heterogeneous, arising from considering both the high- and low-mass populations and there are significant model dependencies and systematic errors that might be as large as a factor of two \citep[e.g.,][]{Bell2013a, Feiden2016a, Jeffries2017a}.

Measurements of the position of the Lithium depletion boundary (LDB) -- the luminosity at which EW$_{\rm Li}$ sharply rises again at the low mass end of the lithium dip -- is a well established method of determining cluster ages and is probably the least model sensitive of all techniques \citep[][]{Jeffries2014b, Soderblom2014a, Tognelli2015a}. Its use is restricted to clusters that are old enough ($>15$ Myr) to have fully depleted lithium in mid M-dwarfs, and close enough to measure EW$_{\rm Li}$ in such intrinsically faint stars. LDB ages are known for six of the young clusters in our training set, three of the moving groups studied in \S\ref{4.2} and the Pleiades studied in \S\ref{4.3}. Table~\ref{LDB} compares the training ages and best-fit Li ages with LDB ages taken from homogeneous re-evaluations in \cite{Soderblom2014a} \citep[using the models of][]{Chabrier1997a} and \cite{Galindo-Guil2022a} \citep[using the Pisa models of][and revised Gaia-based distances and bolometric corrections]{Tognelli2015a} for {\bf six} of the clusters and the moving groups, and from \cite{Binks2022a} for NGC~2232, which was not included in the homogeneous studies. 

There is close agreement between the LDB ages and the adopted training ages. This is not altogether surprising because the ages used in \cite{Jackson2022a}, \cite{Franciosini2022a} and \cite{Randich2022a} were partly informed by the LDB ages. The best-fit ages are also reasonably consistent for both the clusters and the moving groups that were not used in training the model: IC~2391 (higher), Blanco 1 and the Pleiades (lower) have model-predicted ages that differ from their LDB ages by 1-2$\sigma$. IC~2602 and NGC~2547 agree extremely well and with very small error bars, but IC~4665 and NGC~2232 exhibit significant discrepancies in opposite directions. The discrepancy for NGC~2232 is still small ($\sim 0.1$ dex) but the best-fit age of IC~4665 is both higher than its training age and a factor of $\sim 2$ older than the precise LDB age of $28\pm 4$ Myr first quoted by \cite{Manzi2008a} and the 23--32\,Myr quoted by \cite{Galindo-Guil2022a} using a range of models. Figure~\ref{lithium_at5} shows that the older age results from attempts to explain the small EW$_{\rm Li}$ of three highly Li-depleted K-stars. This discrepancy was first highlighted by \cite{Jeffries2009a} using an independent lithium dataset and is confirmed here. Given the reasonable consistency between the training ages, best-fit Li-ages and LDB ages for the other clusters and moving groups in our sample that have these data, it is probably worth revisiting the LDB age of IC~4665 with more and better spectroscopy than obtained by \cite{Manzi2008a}.

\subsubsection{Recommendations}

In summary, any dependence of the predicted ages on metallicity appears weak, at least over the metallicity range encompassed by the training data. For young stars ($<1$ Gyr), the model should be applicable to the majority of stars in the disk population near the Sun, either for single field stars or clusters, with any systematic errors in log age limited to $<0.1$ dex. The training age scale and derived ages for young (15-120 Myr) clusters and moving groups are consistent (with one exception) with those defined by the model-insensitive LDB technique. For older stars ($>1$ Gyr), although we have used a variety of older clusters in the training sample, such that the dispersion might accurately reflect the range of possibilities and hence the age uncertainties for an individual star, there can be large (factor of two) systematic errors if the model is used to estimate the age of an older cluster, due to presently unknown factors, and our recommendation would be to avoid this.  

\subsection{Choice of prior}
\label{sec_prior}

%\begin{figure}
%	\includegraphics[width = 85mm]{lithium_at11_fit.eps} 
%	\centering
%	\caption{}
% \label{exoplanets}
%\end{figure}

In \S\ref{3.1} and~\S\ref{3.2} we introduced age prior probability distributions that were flat in age for estimating the age of single stars or flat in log age when estimating the age of a star that is a component of a cluster. These choices were largely driven by their simplicity and are the options implemented in the software provided. More complex priors could be used in cases where more is known about the star (metallicity, kinematics) or for main sequence stars with $T_{\rm eff}>6000$ K that are unlikely to be as old as 12.4 Gyr. 

The choice of prior can have a large influence on the final age estimate and its confidence interval. The systematic effect of a flat age prior compared to flat log age prior is to move the peak of the posterior probability distribution to older ages and increase the upper error bar.  This effect is more important in cases where the measurement uncertainties or intrinsic dispersion in EW$_{\rm Li}$ are large, the EW$_{\rm Li}$ is small or where the isochrones of Li depletion become bunched in Fig.~\ref{lithium_at3}, since these lead to an increased width in the age likelihood distribution and increased dominance of the prior probability. 
Examples of this problem were encountered in \S\ref{4.1} (e.g., HD~110082, KOI-7913AB) and in these cases the EW$_{\rm Li}/T_{\rm eff}$ combination is relatively uninformative and may not rule out younger ages if other information can be included (e.g. gyrochronology).
Conversely, where uncertainties are small and the likelihood sharply peaked, the choice of prior is unimportant, but this is unlikely to apply to a large fraction of stars in a general sample from the field. The likelihood only becomes sharply peaked in regions of the EW$_{\rm Li}$/$T_{\rm eff}$ plane that are rapidly traversed, so it is less likely to find stars there.   Fortunately, in a cluster, there are likely to be examples of stars in these regions and it is the likelihood distributions for these stars that yield the tight age constraints we have been able to obtain for many of the clusters and moving groups discussed in Sections~3 and 4.

\subsection{Comparison with {\sc baffles}}
\label{baffles}

\begin{figure}
	\includegraphics[width = 85mm]{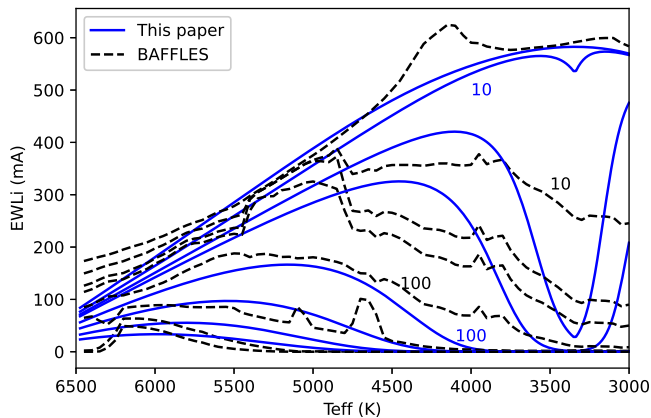} 
	\centering
	\caption{A comparison of empirical Li depletion isochrones from this paper and the {\sc baffles} model of \protect\cite{Stanford-Moore2020a}. Isochrones are plotted (from top to bottom) at 3, 10, 20, 30, 100, 300, 1000 and 3000\,Myr.}   
\label{bafflescomparison}
\end{figure}

\cite{Stanford-Moore2020a} assembled a lithium dataset for 609 stars from 10 clusters and associations with assumed ages of 5.5--4000\,Myr. This was used to make Bayesian predictions of age using a likelihood constructed from an empirical model of EW$_{\rm Li}$ and its dispersion as a function of age and $B-V$ colour, along with a flat prior in age. Their code was named {\sc baffles}. The approach here is similar in respect of philosophy and technique but offers a number of important improvements:
\begin{itemize}
\item The {\sc baffles} training data were an inhomogeneous selection from the literature, with only one cluster older than 700 Myr and only one younger than 20 Myr. The oldest cluster was M67, which may not be fully representative of stars at that age (\S\ref{5.1}). The training data here were drawn from a homogeneous analysis of 52 clusters, more densely sampling the age-$T_{\rm eff}$ plane with ten times as many stars.
\item The {\sc baffles} modelling approach made quadratic fits to EW$_{\rm Li}$ as a function of $B-V$ followed by piece-wise linear fits to the time-dependence for 64 $B-V$ slices at 10 age points plus a modelled "zeropoint" and undepleted level. The resulting model has many free parameters and abrupt, unphysical changes in EW$_{\rm m}(\log\,{\rm age}, B-V)$. The model presented here has far fewer free parameters  but can smoothly represent all the training data with improved fidelity.
\item The intrinsic dispersion in $\log$ EW$_{\rm m}$ is an empirical time- and $T_{\rm eff}$-independent non-Gaussian function in {\sc baffles}. The model for the intrinsic dispersion here is Gaussian in EW$_{\rm Li}$ but more complex, with both a time- and $T_{\rm eff}$-dependence. This does a better job of representing the dispersion at high and low EW values as well as the increased dispersion seen among cool stars at the ZAMS.
\end{itemize}

\begin{figure}
	\includegraphics[width = 85mm]{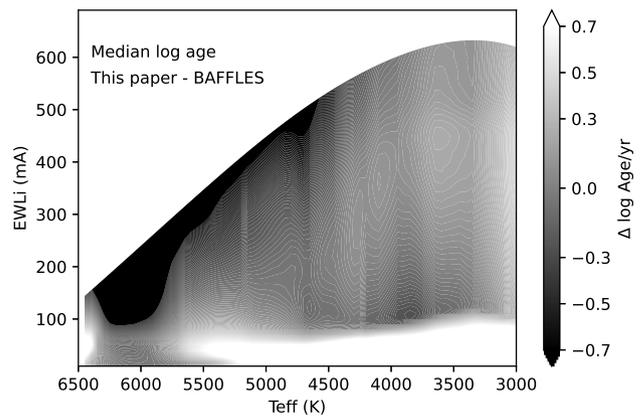} 
	\centering
	\caption{The difference in predicted median log age (in dex) between this paper and {\sc baffles} \citep{Stanford-Moore2020a} as a function of EW$_{\rm Li}$ and $T_{\rm eff}$, shown as a grayscale.}   
\label{bafflescomparison2}
\end{figure}

A quantitative comparison with the {\sc baffles} model isochrones for EW$_m$ as a function of age and $T_{\rm eff}$ is shown in Fig.~\ref{bafflescomparison}. The conversion between $B-V$ and $T_{\rm eff}$ was done using the main-sequence relationships tabulated in \cite{Pecaut2013a}.
There are clear differences, most notably at $T_{\rm eff}<4500$ K and $T_{\rm eff}>6300$ K. The former are probably due to the paucity of calibrating cool-star data in the {\sc baffles} training set, particularly beyond ages of 24\,Myr \citep[see fig.~6 in][]{Stanford-Moore2020a}. In particular there are no data in the {\sc baffles} training set with which to constrain the upturn in EW$_{\rm Li}$ at $T_{\rm eff}<3400$ K and define the lithium-chasm that opens up between 10 and 50\,Myr  The discrepancies for hotter stars are partly due to our neglect of the F-star "lithium-dip" that is seen in some older clusters \citep{Boesgaard1987a}, notably, the Hyades, which has great weight in the {\sc baffles} training data. We see very little evidence for this dip in the GES dataset for $T_{\rm eff} \leq 6500$\,K (see the final subplot in Fig.~\ref{lithium_at1_rob}), possibly because the dip is masked by observational uncertainties in EW$_{\rm Li}$ for the older clusters.

The differences in the log of {\it median} ages derived from {\sc baffles}, assuming a flat prior in age and observational uncertainties of $\pm 10$ m\AA\ in EW$_{\rm Li}$, are quantified in Fig.~\ref{bafflescomparison2} as a function of EW$_{\rm Li}$ and $T_{\rm eff}$. There is reasonable agreement over large parts of this domain, but also large discrepancies in others.  These can be explained in terms of differences in the model isochrones (Fig.~\ref{bafflescomparison}) and in the modelled intrinsic dispersion of EW$_{\rm Li}$ used to form the likelihood function. The {\sc baffles} assumption of a uniform dispersion in $\log$ EW$_{\rm Li}$ leads to very large predicted dispersions in absolute EW$_{\rm Li}$ at young ages (factors of two with significant tails out to order of magnitude deviations) but negligible dispersion in EW$_{\rm Li}$ at old ages. The combination of this with a flat prior leads to very large differences in predictions for warm stars with high EW$_{\rm Li}$, where {\sc baffles} is unable to assign them a young age. Similarly, for cool stars with small, but non-zero EW$_{\rm Li}$, the very small intrinsic dispersion in EW$_{\rm m}$ in {\sc baffles} leads to much younger predicted ages.

\section{Summary}
\label{6}

An empirical model is developed describing how photospheric lithium, represented by the equivalent width of the Li~{\sc i}~6708\AA\ line (EW$_{\rm Li}$), is depleted as a function of age and $T_{\rm eff}$, using training data for PMS and main sequence stars with $1 < {\rm age/Myr} < 6000$ in 52 open clusters observed as part of the Gaia-ESO spectroscopic survey (GES). The calibrating data cover the ranges $3000 \leq T_{\rm eff}/{\rm K} \leq 6600$ and $-0.29 < {\rm [Fe/H]} < 0.18$. The model can be used to generate a posterior probability distribution of age for stars within the scope of the training data that have both EW$_{\rm Li}$ and $T_{\rm eff}$ and hence provide age estimates, with uncertainties, for any individual star or associated group of coeval stars.

The precision of the model, applied to individual stars, is strongly $T_{\rm eff}$- and age-dependent. Precisions of 0.1 dex in log age are achievable in the best cases (M-dwarfs in the approximate age range 10--100\,Myr) but good sensitivity is also attained for K-dwarfs in the range 30--300\,Myr and G-dwarfs at 100--1000 Myr. Much better precision is obtained for coeval groups of stars, where the age probability distributions of their individual stars can be combined and the final precision depends on both the number of stars in the group and their $T_{\rm eff}$ distribution. The precision at older ages is poorer because Li is either totally depleted (K- and M-stars) or the isochrones of EW$_{\rm Li}$ versus $T_{\rm eff}$ become tightly packed (F- and G-stars) compared with typical observational uncertainties and astrophysical dispersion. 
At younger ages an undepleted EW$_{\rm Li}$ can provide only age upper limits.

The accuracy of the ages is directly linked to the accuracy of the ages assumed for the training clusters. The adopted training ages are consistent with the lithium-depletion-boundary method ages for several of the younger clusters, whilst the older clusters rely on the accuracy of main sequence turn-off ages from the literature and homogeneous determinations based on Gaia photometry and parallaxes \citep{Dias2021b}. Because of the large number of clusters used, the results are robust to errors in the assumed training age of any individual cluster. 

Any systematic errors due to the neglect of compositional effects or other confounding parameters such as the rotational history of a star or the presence of planetary systems appear limited to $<0.1$ dex in log age for stars of age $<1$ Gyr, although the training data is limited to an even narrower range of metallicity for the very youngest clusters.  At older ages ($>1$ Gyr) there are signs that our simple model is not sufficiently complex or deterministic -- there is a large dispersion in the recovered ages of the training clusters compared with their training ages that cannot be explained by uncertainties. There is little evidence that these discrepancies are linked directly to differences in metallicity  and we do not see a clear systematic dependence of lithium depletion on metallicity in the training clusters, either during PMS or main sequence evolution. As a result, we believe that the Li-based ages derived for younger stars, or young clusters/associations are robust to exact metallicities, while there is increased age uncertainty for older stars and possibly large (factor of two) systematic uncertainties if the model is applied to older clusters/associations. 
%A note of caution is the clear systematic discrepancy between the model and two metal-rich clusters (the Hyades and Praesepe) that were not part of the training data and which contain stars with $T_{\rm eff}< 5500$\,K and [Fe/H]$>0.1$ dex that are sparsely covered by the training data.

As a proof of concept and validation we have applied the model to a set of exoplanet-hosting field stars that have young ages ($< 300$\,Myr) reported in the literature. Based {\it only} on their Li content and $T_{\rm eff}$, we find that some of these stars and their exoplanets (KOI-7913, Kepler 16343) are unlikely to be as young as claimed, while the rest are quite consistent with the published ages. We have also redetermined the ages of seven well-known ``moving groups" of stars in the solar neighbourhood using only the Li content of their stars. We find excellent agreement with previous age determinations for these groups, with precisions better than 10 per cent in six cases.

We anticipate that the use of Li as a quantitative age indicator could become increasingly important in the era of large optical spectroscopic surveys, like 4MOST \citep{deJong2019a} and WEAVE \citep{Jin2022a}, that have sufficient spectral resolution ($R \geq 5000$) to measure EW$_{\rm Li}$, providing age estimates for Galactic archaeology or at least posterior probability distributions of age that are a useful adjunct to other methods. This is especially true for stars, associations and clusters at ages $<1$ Gyr, where ages from the HR diagram are either insensitive or subject to large systematic uncertainties and model-dependencies. 

Our methods are presented for wider use in the form of a Python-based software package called ``Empirical AGes from Lithium Equivalent widthS" ({\sc eagles}), that is described in Appendix~\ref{eagles}.

\section*{Acknowledgments}
Based on data products from observations made with ESO Telescopes at the La Silla Paranal Observatory under programme ID 188.B-3002. These data products have been processed by the Cambridge Astronomy Survey Unit (CASU) at the Institute of Astronomy, University of Cambridge, and by the FLAMES/UVES reduction team at INAF/Osservatorio Astrofisico di Arcetri. These data have been obtained from the Gaia-ESO Survey Data Archive, prepared and hosted by the Wide Field Astronomy Unit, Institute for Astronomy, University of Edinburgh, which is funded by the UK Science and Technology Facilities Council.

This work was partly supported by the European Union FP7 programme through ERC grant number 320360 and by the Leverhulme Trust through grant RPG-2012-541. We acknowledge the support from INAF and Ministero dell' Istruzione, dell' Universit\`a' e della Ricerca (MIUR) in the form of the grant "Premiale VLT 2012". The results presented here benefit from discussions held during the Gaia-ESO workshops and conferences supported by the ESF (European Science Foundation) through the GREAT Research Network Programme.

\section{Data availability statement}
The reduced stacked spectra underlying this article can be obtained via the European Southern Observatory (ESO) archive and is identified as the "Phase 3 release" of Gaia-ESO spectroscopic survey DR4. The full catalogue of stellar parameters derived from these spectra by the various GES working groups is also available in the "Phase 3 release" of Gaia-ESO spectroscopic survey DR5. Raw data can also be obtained from the ESO archive. 

\bibliographystyle{mnras.bst} 

\bibliography{references} 

\appendix
\section{Definition of empirical template spectra used to estimate lithium equivalent widths}
\label{AppA}
The template spectra used in the estimation of EW$_{\rm Li}$ (\S\ref{2.2}) were derived from GES spectra of dwarf field stars with membership probabilities $P_{\rm 3D} <0.1$ \citep[see Table~3 of][]{Jackson2022a}, a {\it Gaia} DR3 parallax $>1$\,mas and $v\sin i < 50$\,km\,s$^{-1}$. Stars were allocated to $\pm 100$\,K bins of $T_{\rm eff}$ in 100K steps between 3100 and 6900\,K. Spectra were offset to their rest wavelength using their GESiDR6 RV and normalised to their median values over the wavelength range 6675 to 6730\,\AA. An initial estimate of the median spectrum in each $T_{\rm eff}$ bin was calculated at each point over the wavelength range and its uncertainty estimated as 1.3 times the median absolute deviation relative to the mean.

This initial estimate was potentially biased by the presence of a small proportion of young stars in the sample. These would mostly be young cluster members with anomalous RVs due to binarity, but could include genuine young field interlopers. To minimise any bias, the initial estimate of the median spectra was used to calculate EW$_{\rm Li}$ for the set of field stars and the results used to filter out any target with $|{\rm EW}_{\rm Li}| >150$\,m\AA~for stars with $T_{\rm eff}<5700$\,K, or $|{\rm EW}_{\rm Li}|>50$\,m\AA~for hotter stars. The median spectra were then recalculated and the process repeated to produce a set of 2711 targets with EW$_{\rm Li}$ meeting the above conditions which were used to define the final set of empirical template spectra.

\begin{figure*}
    \centering
	\begin{minipage}[]{1\textwidth}
	\includegraphics[width = 170mm]{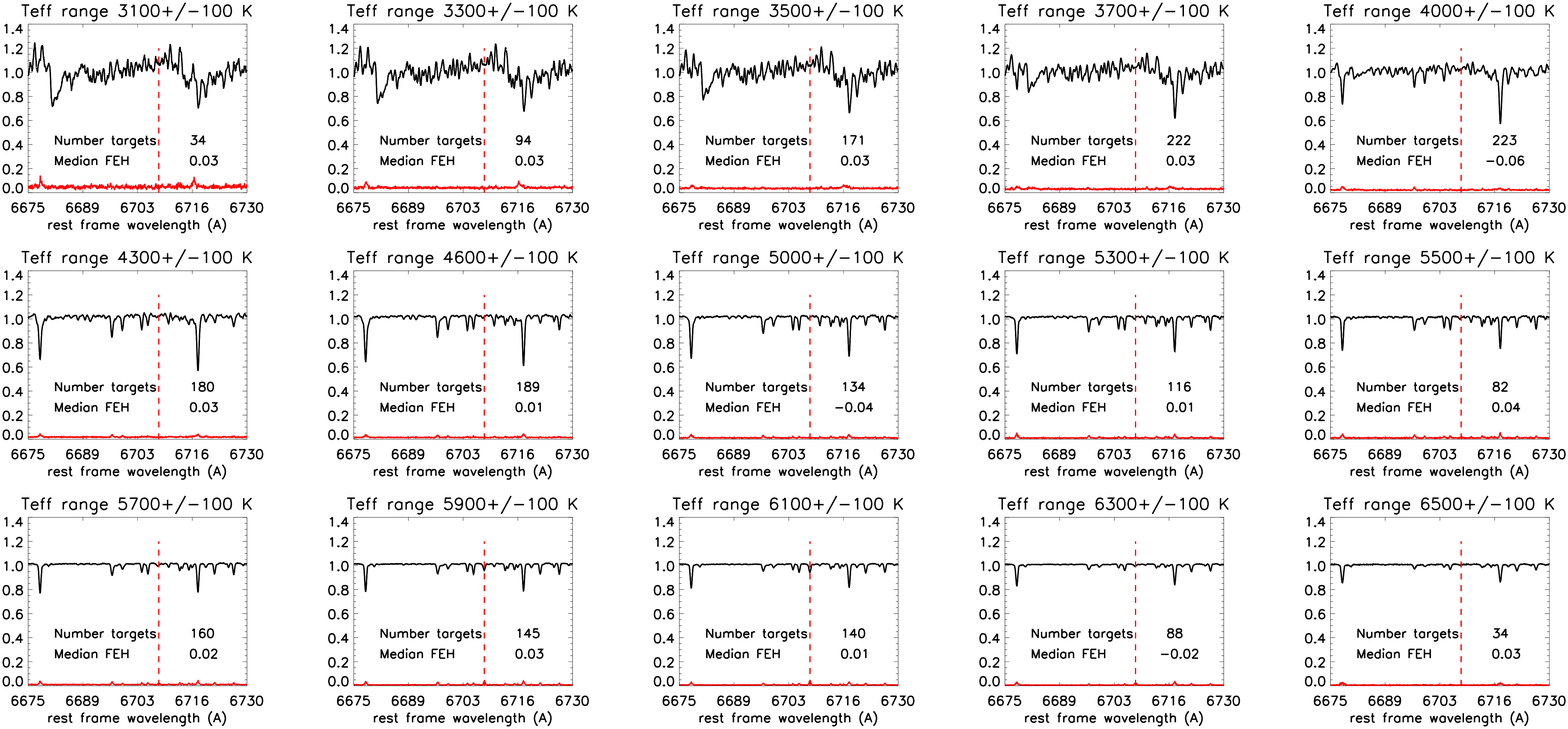}
    \end{minipage}
	\caption{Examples of median template spectra. Text on the plots show the number of spectra in each bin and their median [Fe/H]. The lower red curve shows the uncertainty in the template spectrum calculated as 1.3 times the median absolute deviation of individual spectra. The red dashed line indicates the position of the  Li~{\sc i}~6708\AA~ line. Spectra at 5900 and 6100\,K show a small but significant Lithium absorption feature indicating a non-zero EW$_{\rm Li}$ at these temperatures.}
	\label{econtinuum}	
\end{figure*}

Examples of these median spectra are shown in Fig.~\ref{econtinuum}. Median spectra for the cooler stars are expected to be fully Li-depleted. This is supported by the absence of any sign of a lithium feature in the median spectra for $T_{\rm eff} < 5500$\,K. At higher temperatures ($\sim 6000$\,K) there is a small but measurable residual EW$_{\rm Li}$. To quantify this the EW was measured in template spectra of $T_{\rm eff}>5000$\,K relative to a Li-free synthetic spectra  generated using the {\sc moog} software \citep{Sneden2012a} and \cite{Kurucz1992a} model atmospheres \citep[see][for details]{Jeffries2021a}. Interpolating these results gives the equivalent width EW$_0$ of median spectra as a function of $T_{\rm eff}$ used in Eqn.~\ref{EW_0} and listed in Table~\ref{EWo}.

\begin{table}
\caption{Equivalent width of the Li~{\sc i}~6708\AA ~of continuum spectra as a function of central bin temperature measured relative to Li-free synthetic spectra.} 
\centering

\begin{tabular}{rr}
\hline
$T_{\rm eff}$ (K) & EW$_0$ (m\AA ) \\ 
\hline          
$\leq $5600 &     0  \\ 
     5700 &     4  \\
     5800 &    14  \\
     5900 &    24  \\
     6000 &    33  \\
     6100 &    34  \\
     6200 &    20  \\
     6300 &     8  \\
     6400 &     3  \\
     6500 &     1  \\\hline
\end{tabular}
  \label{EWo}
\end{table}

\section{The {\sc eagles} code}

\label{eagles}

The model derived in \S{2} and the methods described in \S{3} have been packaged into a python code called {\sc eagles} which is made available via \url{https://github.com/robdjeff/eagles}.

{\sc eagles} is a command-line driven script that takes a simple ascii input file containing the $T_{\rm eff}$, EW$_{\rm Li}$ (corrected for any blending), $\sigma_{\rm Li}$ and (optionally) an additional $T_{\rm eff}$ uncertainty (see \S\ref{3.1}) for one or a list of stars and returns Bayesian estimates of their ages using a prior probability that is either flat in age or flat in log age (\S\ref{sec_prior}). The input stars can be treated as individuals or fitted as a coeval cluster. Outputs include the most probable age and asymmetric 68 per cent confidence interval, the median age and the full posterior age probability distribution.

Scripts are also included that will generate the isochrones shown in Fig.~\ref{lithium_at3} and a grid of estimated age as a function of EW$_{\rm Li}$ and $T_{\rm eff}$ for a given level of observational uncertainty in EW$_{\rm Li}$.

\section{Lithium versus effective temperature and likelihood plots for all clusters}

\label{clusterplots}
In \S\ref{3.3} we applied our model to all of the GES training clusters. The results were shown graphically in Fig.~\ref{lithium_at5} for several examples. This appendix (available online as supplementary material) provides plots for all of the clusters listed in Table~\ref{ages}, for which age constraints could be obtained. The Figures all conform to the same style as Fig.~\ref{lithium_at5}.

\newpage

\begin{figure*}

\includegraphics[width = 170mm]{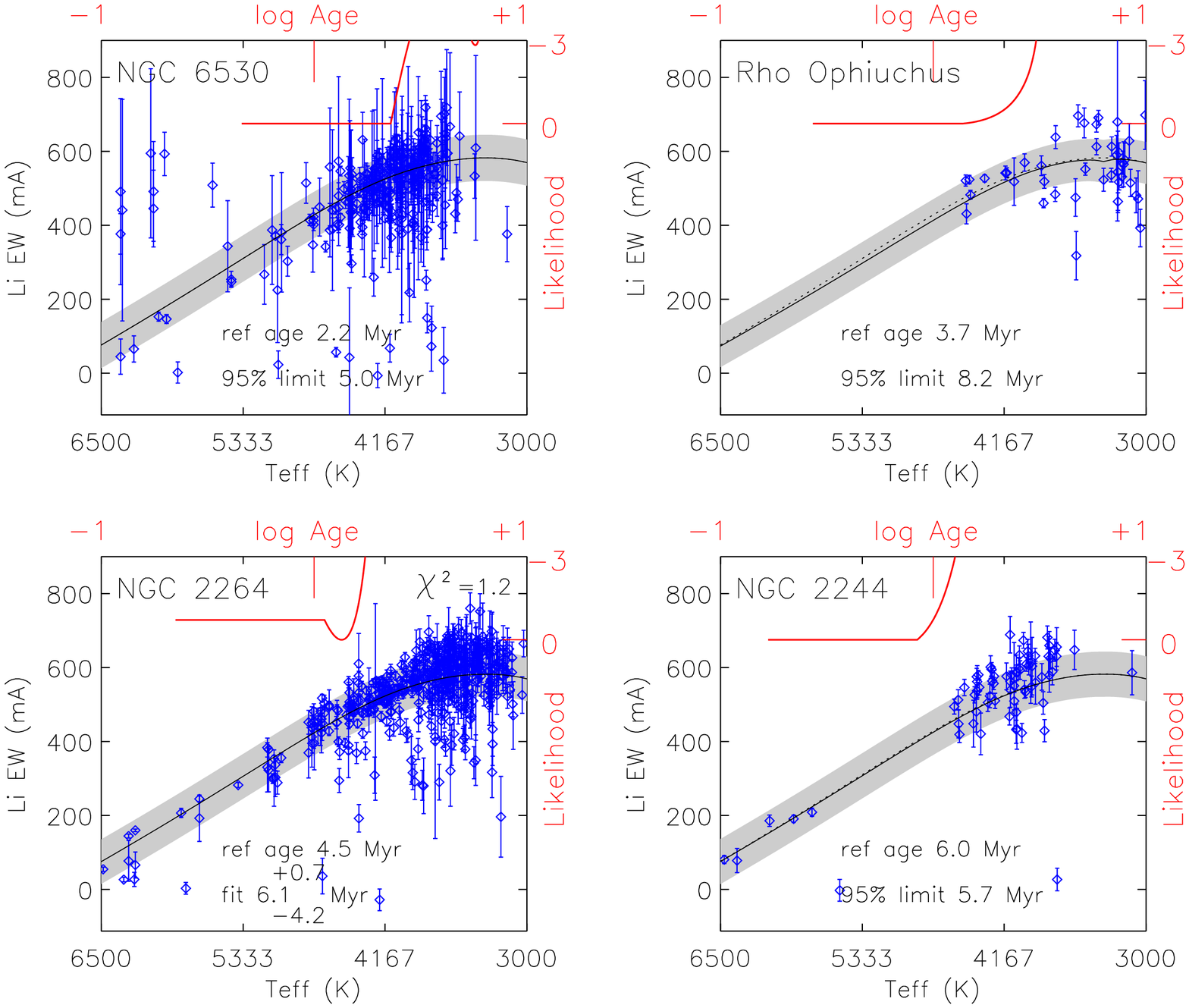}
    \caption{Best-fit Model isochrones compared with the GES training data. The clusters are presented in the order given in Table~\ref{ages} along with their reference ages and the constraints provided by the empirical model. Blue points show EW$_{\rm Li}$ as a function of $T_{\rm eff}$ (scales on the left hand and bottom axes). The solid lines show the model value of EW$_{\rm Li}$ at the fitted most probable age (shown as text on the plot) the dashed curves, where present, are a 5\,Myr isochrone. The shaded regions are the model intrinsic dispersion at the best-fit age or its upper limit. The reduced chi-squared value for the fit (where an age is determined) is shown in the top-right corner. The upper red paraboloid shows negative log likelihood (normalised to zero at the maximum-likelihood) as a function of $\log$~age relative to the age adopted in the training set (scales on the right hand and top axes).}      
\label{clusterfigures}
\end{figure*}

\addtocounter{figure}{-1}

\newpage

\begin{figure*}

\label{all_clusters}

\includegraphics[width = 170mm]{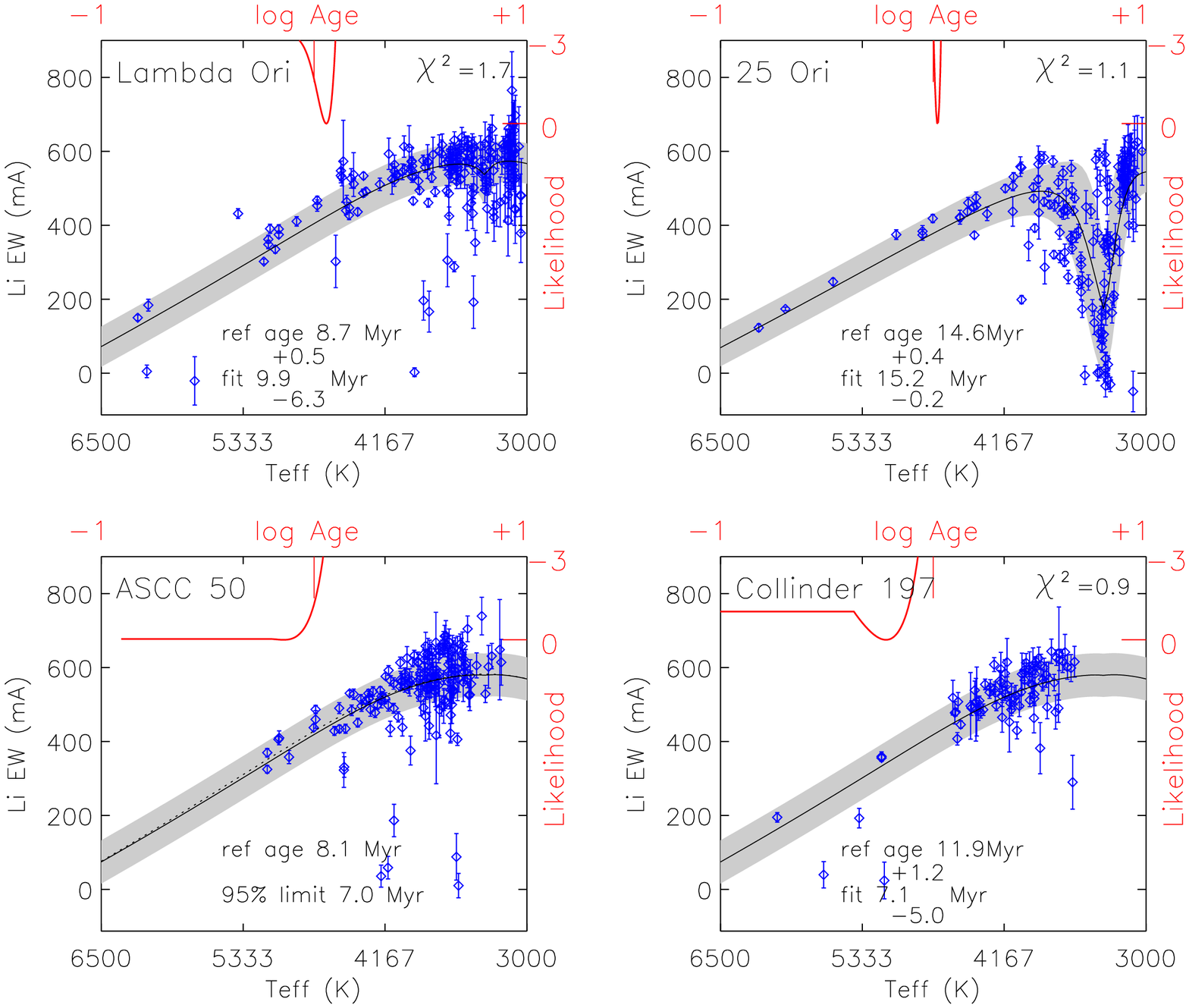}
    \caption{continued.}
\end{figure*}
\addtocounter{figure}{-1}

\newpage

\begin{figure*}

\includegraphics[width = 170mm]{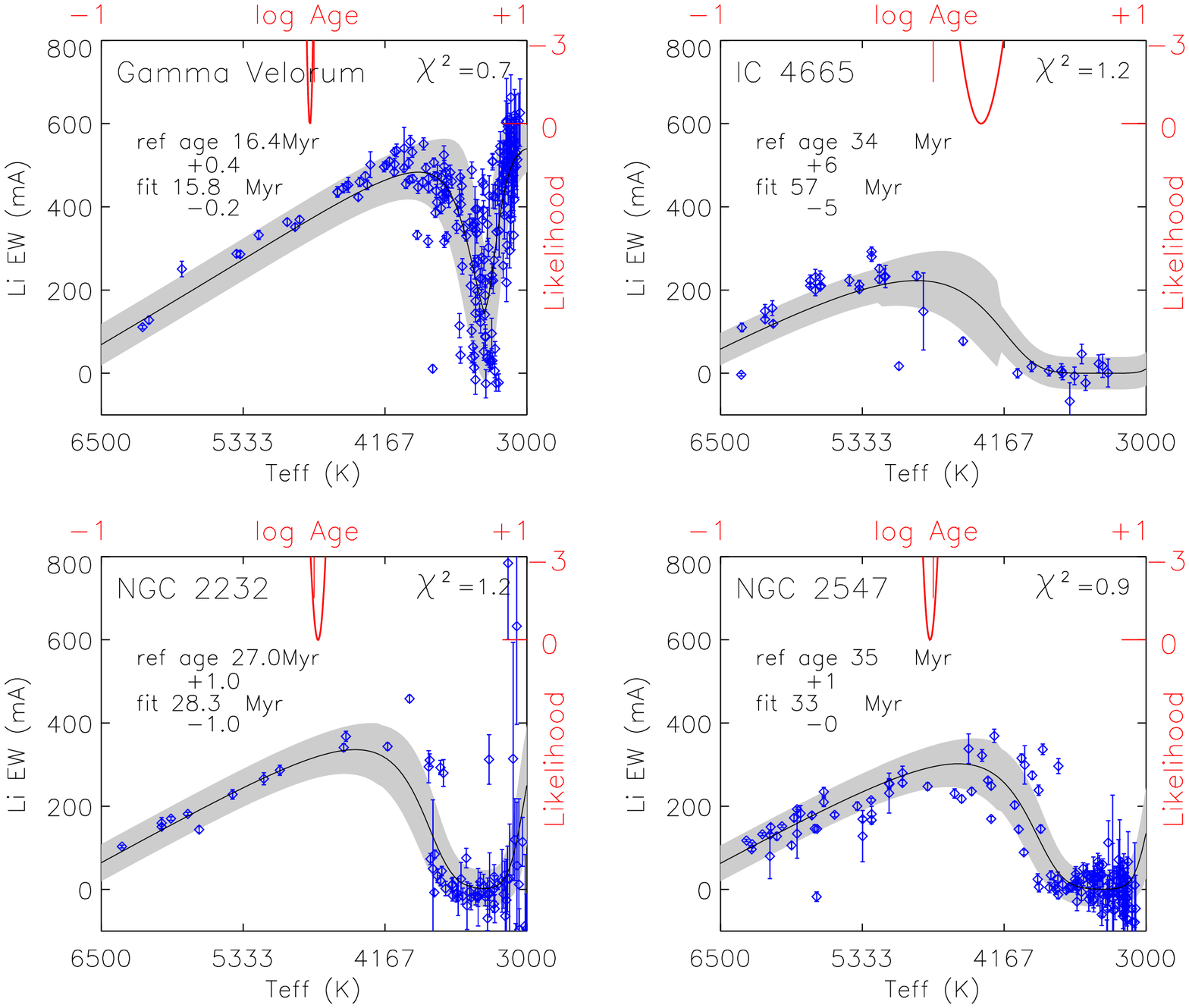}
    \caption{continued.}
\end{figure*}

\addtocounter{figure}{-1}

\newpage

\begin{figure*}

\includegraphics[width = 170mm]{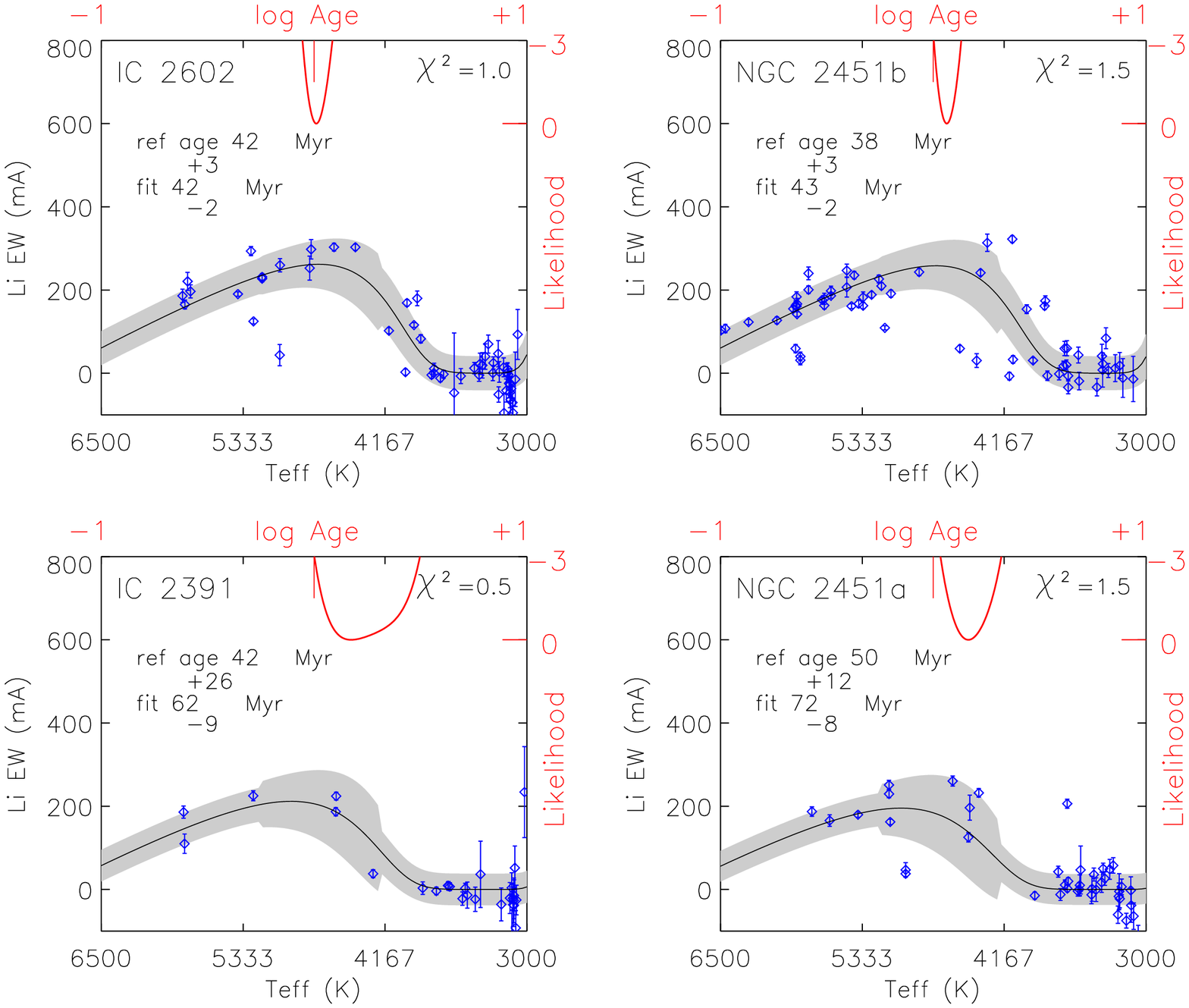}
    \caption{continued.}
\end{figure*}

\addtocounter{figure}{-1}

\newpage

\begin{figure*}

\includegraphics[width = 170mm]{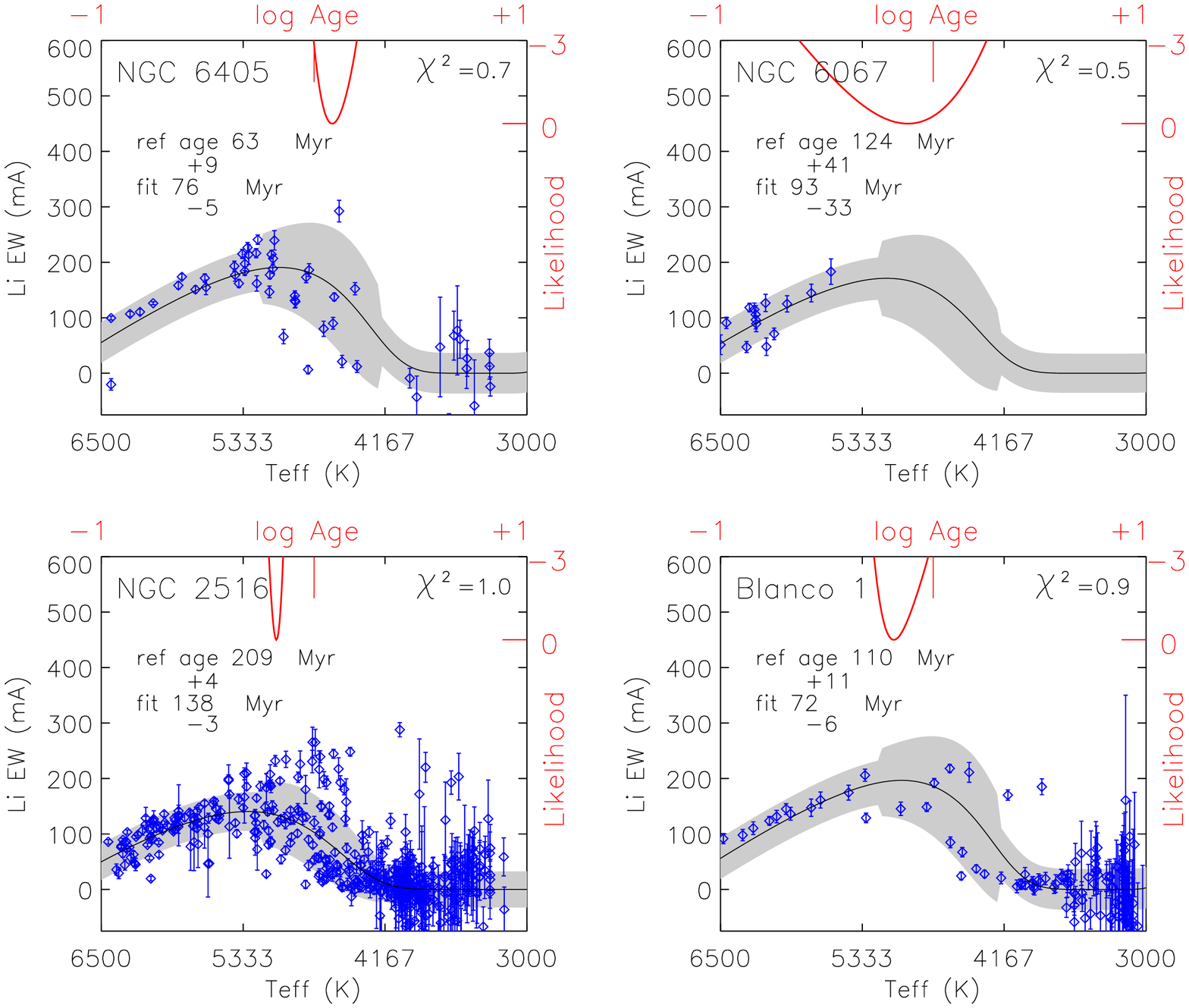}
    \caption{continued.}
\end{figure*}

\addtocounter{figure}{-1}

\newpage

\begin{figure*}

\includegraphics[width = 170mm]{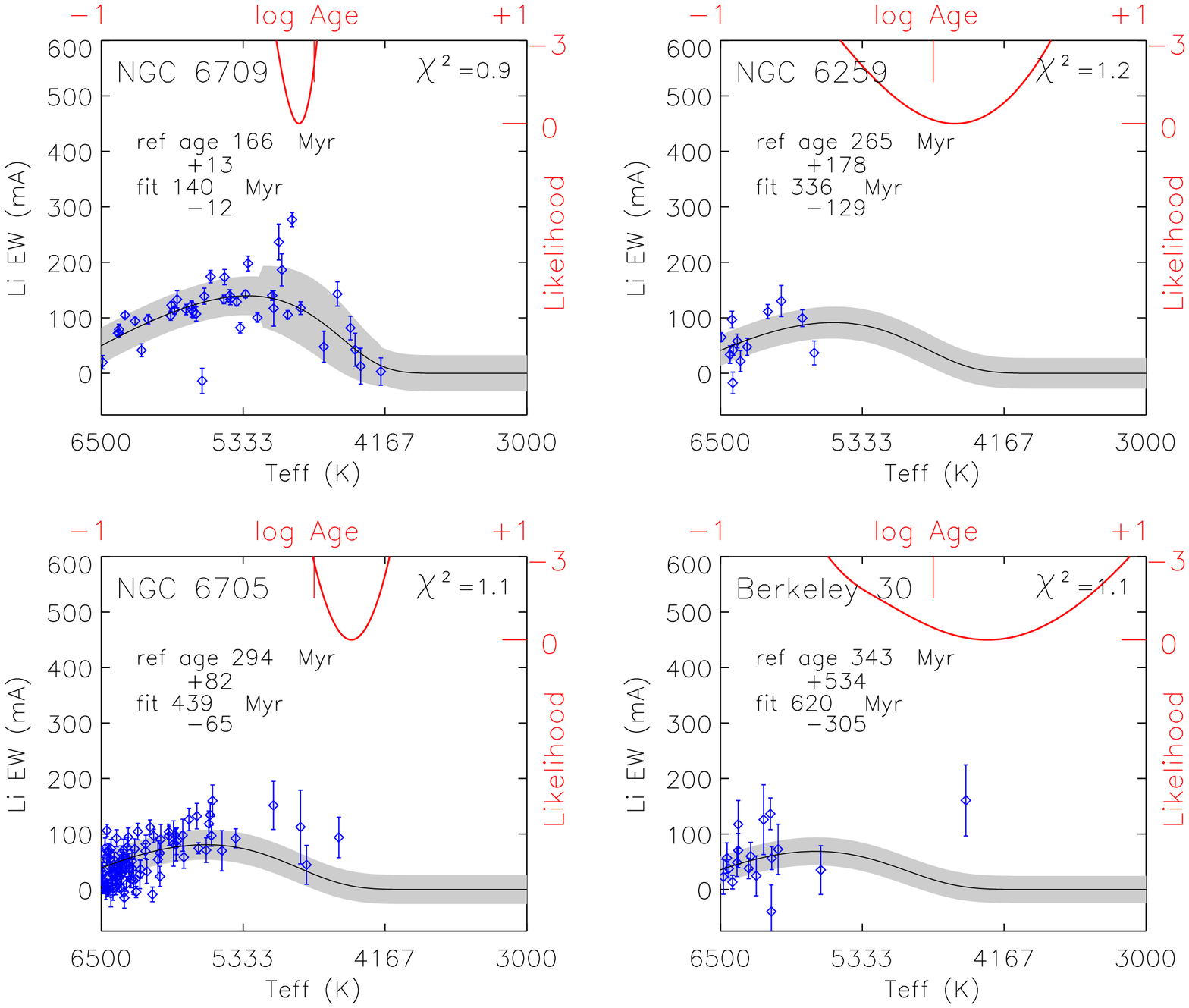}
    \caption{continued.}
\end{figure*}

\addtocounter{figure}{-1}

\newpage

\begin{figure*}

\includegraphics[width = 170mm]{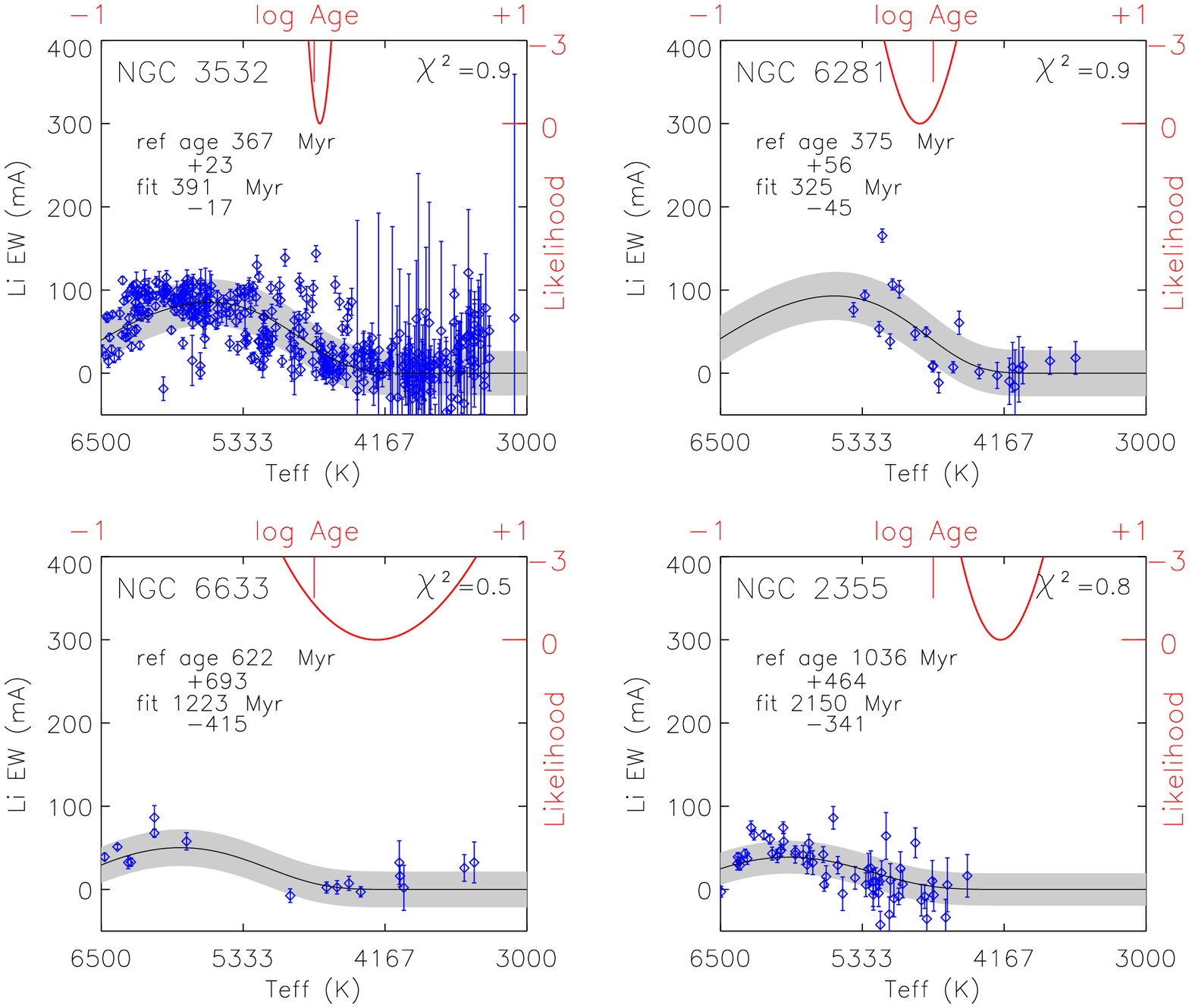}
    \caption{continued.}
\end{figure*}

\addtocounter{figure}{-1}

\newpage

\begin{figure*}

\includegraphics[width = 170mm]{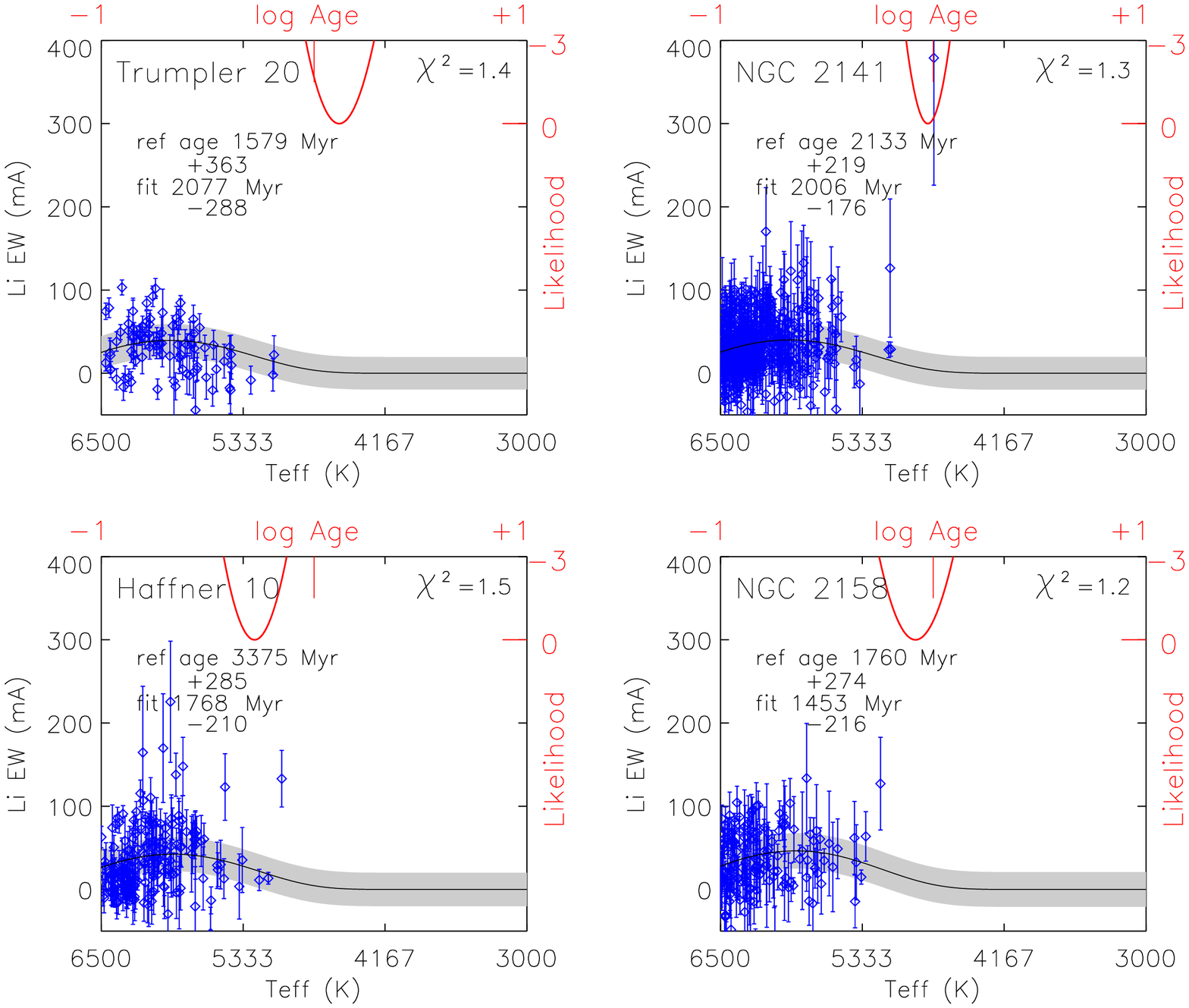}
    \caption{continued.}
\end{figure*}

\addtocounter{figure}{-1}

\newpage

\begin{figure*}

\includegraphics[width = 170mm]{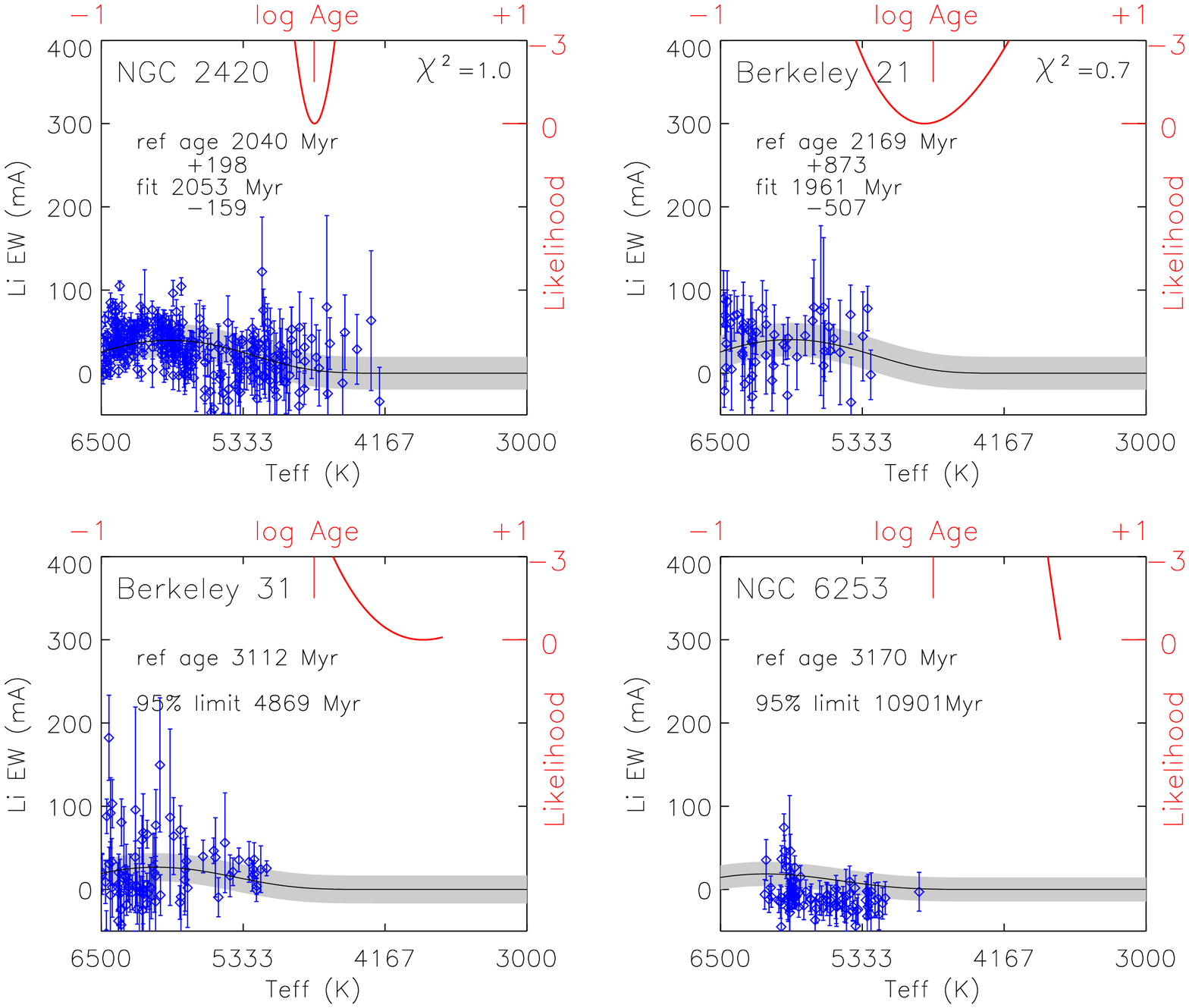}
    \caption{continued.}
\end{figure*}

\addtocounter{figure}{-1}

\newpage

\begin{figure*}

\includegraphics[width = 170mm]{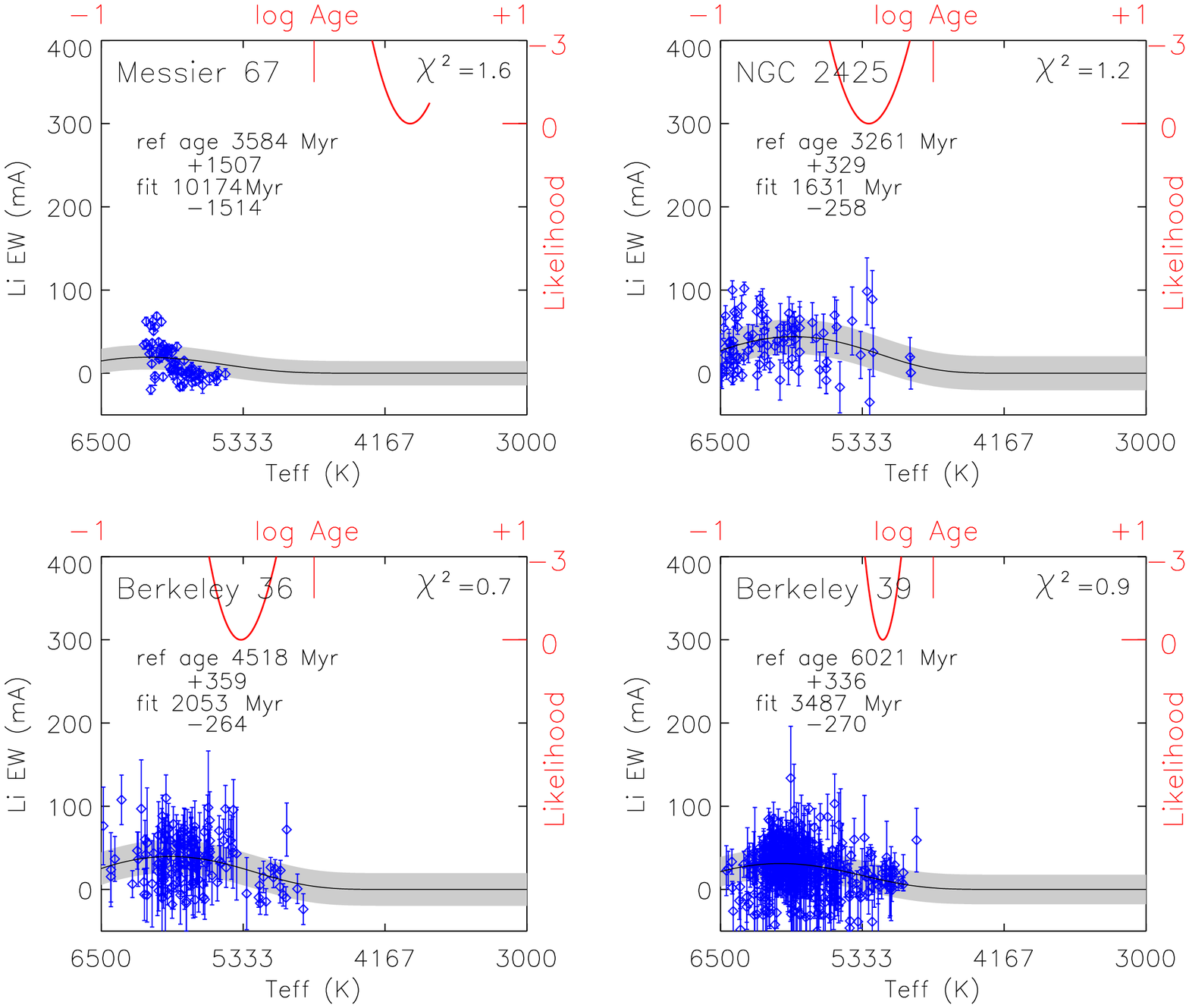}
    \caption{continued.}
\end{figure*}

\bsp % ``This paper has been produced using the ...''
\label{lastpage}
\end{document}